\documentclass[12pt]{article}
\usepackage{epsfig,amssymb,amsmath,psfrag,multirow,epstopdf,color}
\usepackage{breqn}
\usepackage{placeins}
\usepackage{array}

\numberwithin{equation}{section}

\newcommand{\be}{\begin{equation}}
\newcommand{\ee}{\end{equation}}
\newcommand{\bea}{\begin{eqnarray}}
\newcommand{\eea}{\end{eqnarray}}
\newcommand{\eqn}[1]{eq.~\eqref{#1}}

\def\sect#1{section~{\ref{#1}}}
\def\eqn#1{eq.~(\ref{#1})}
\def\Eqn#1{Equation~(\ref{#1})}
\def\eqns#1#2{eqs.~(\ref{#1}) and~(\ref{#2})}

\def\tab#1{table~{\ref{#1}}}

\def\Eqn#1{Equation~(\ref{#1})}
\def\eqn#1{eq.~(\ref{#1})}
\def\eqns#1#2{eqs.~(\ref{#1}) and~(\ref{#2})}

\def\Li{{\rm Li}}

\def\cO{{\mathcal O}}

\def\ws{{w^\ast}}

\def\to{\rightarrow}
\def\lr{\leftrightarrow}

\def\e{\epsilon}
\newcommand{\cP}{{\cal P}}

\def\Qep{Q_{\rm ep}}

\def\beq{\begin{equation}}
\def\eeq{\end{equation}}

\def\bsp#1\esp{\begin{split}#1\end{split}}

\newcommand{\Vt}{\tilde{V}}
\newcommand{\Et}{\tilde{E}}

\newcommand{\NeqFour}{{\cal N}=4}
\newcommand{\Ord}{{\cal O}}

\newfont{\scyr}{wncyr10 scaled 550}

\newcommand{\tPhi}{\tilde{\Phi}_6}
\def\beq{\begin{equation}}
\def\eeq{\end{equation}}
 
\def\bsp#1\esp{\begin{split}#1\end{split}}

\begin{document}

\catcode`\@=11
\font\manfnt=manfnt
\def\Watchout{\@ifnextchar [{\W@tchout}{\W@tchout[1]}}
\def\W@tchout[#1]{{\manfnt\@tempcnta#1\relax%
  \@whilenum\@tempcnta>\z@\do{%
    \char"7F\hskip 0.3em\advance\@tempcnta\m@ne}}}
\let\foo\W@tchout
\def\dubious{\@ifnextchar[{\@dubious}{\@dubious[1]}}
\let\enddubious\endlist
\def\@dubious[#1]{%
  \setbox\@tempboxa\hbox{\@W@tchout#1}
  \@tempdima\wd\@tempboxa
  \list{}{\leftmargin\@tempdima}\item[\hbox to 0pt{\hss\@W@tchout#1}]}
\def\@W@tchout#1{\W@tchout[#1]}
\catcode`\@=12


\thispagestyle{empty}

\null\vskip-10pt \hfill
\begin{minipage}[t]{42mm}
SLAC--PUB--16352\hskip1cm \ \ \ CALT--2015--049\\
\end{minipage}
\vspace{5mm}

\begingroup\centering
{\Large\bfseries\mathversion{bold}
The four-loop six-gluon NMHV ratio function\par}%
\vspace{7mm}

\begingroup\scshape\large
Lance~J.~Dixon$^{(1),(2)}$, Matt von Hippel$^{(3)}$ and Andrew~J.~McLeod$^{(1)}$ \\
\endgroup
\vspace{5mm}
\begingroup\small
$^{(1)}$\emph{SLAC National Accelerator Laboratory,
Stanford University, Stanford, CA 94309, USA} \\
$^{(2)}$\emph{Walter Burke Institute for Theoretical Physics,
California Institute of Technology, Pasadena, CA 91125, USA} \\
$^{(3)}$\emph{Perimeter Institute for Theoretical Physics, 
Waterloo, Ontario N2L 2Y5, Canada}
\endgroup

\vspace{0.4cm}
\begingroup\small
E-mails:\\
{\tt lance@slac.stanford.edu},\ \ \ {\tt mvonhippel@perimeterinstitute.ca},
\ \ \ {\tt ajmcleod@stanford.edu}
\endgroup
\vspace{0.7cm}

\textbf{Abstract}\vspace{5mm}\par
\begin{minipage}{14.7cm}
We use the hexagon function bootstrap to compute the ratio function
which characterizes the next-to-maximally-helicity-violating (NMHV)
six-point amplitude in planar $\NeqFour$ super-Yang-Mills theory at four loops.
A powerful constraint comes from dual superconformal invariance, in the form
of a $\bar{Q}$ differential equation, which heavily constrains the first
derivatives of the transcendental functions entering the ratio function.
At four loops, it leaves only a 34-parameter space of functions.
Constraints from the collinear limits, and from the multi-Regge limit at
the leading-logarithmic (LL) and next-to-leading-logarithmic (NLL) order,
suffice to fix these parameters and obtain a unique result.
We test the result against multi-Regge predictions at NNLL and N$^3$LL,
and against predictions from the operator product expansion involving one and
two flux-tube excitations; all cross-checks are satisfied.
We study the analytical and numerical behavior of the parity-even and parity-odd
parts on various lines and surfaces traversing the three-dimensional 
space of cross ratios.  As part of this program, we characterize all irreducible
hexagon functions through weight eight in terms of their coproduct.
We also provide representations of the ratio function
in particular kinematic regions in terms of multiple polylogarithms.
\end{minipage}\par
\endgroup

\newpage

\tableofcontents

\newpage

\section{Introduction}

Over the past few decades, the hidden simplicity of $\mathcal{N} = 4$ super-Yang-Mills (SYM) theory~\cite{Brink1976bc} has been steadily revealed.  The theory is conformally invariant for any value of the coupling~\cite{Finiteness}.  In the planar limit of a large number of colors, further simplifications take place:  the perturbative expansion has a finite radius of convergence, and the theory becomes integrable~\cite{IntegrableReview}.   Related to integrability, the theory is endowed with a dual superconformal symmetry \cite{DualConformal, FourLoopNeq4, FiveLoopNeq4, AMStrong, Drummond2008vq}, and scattering amplitudes are dual to polygonal Wilson loops with light-like edges~\cite{AMStrong, WilsonLoopWeak, Alday2008yw, Adamo2011pv}.  These features make it an ideal setting for exploring general properties of gauge theory amplitudes, especially for large numbers of external legs and high loop orders.  The infrared divergences of scattering amplitudes in planar $\mathcal{N} = 4$ SYM are captured by the BDS ansatz~\cite{BDS}.  When amplitudes are divided by this ansatz, the ratio is not only infrared-finite, but its components are functions only of dual conformally invariant cross ratios~\cite{Bern2008ap}.  This restricted set of kinematic variables simplifies dramatically the problem of determining the amplitudes.  In particular, scattering amplitudes with four or five external particles are uniquely determined, up to constants, because there are no nontrivial cross ratios in these cases.

In the six-point case, the subject of this paper, only three functions are needed to specify the scattering amplitudes.  Each function depends on three independent cross ratios, which we call $u$, $v$ and $w$. The first such function, the \emph{remainder function}, $R_6(u,v,w)$, is defined to be the maximally-helicity-violating (MHV) all-gluon amplitude divided by the BDS ansatz \cite{Bern2008ap}.  MHV amplitudes involving particles other than gluons are related to this function by the $\mathcal{N} = 4$ superalgebra, and can be combined with the all-gluon amplitude to form an MHV super-amplitude \cite{Nair1988bq,Georgiou2004by,Bianchi2008pu,ArkaniHamed2008gz}. Other helicity configurations, such as the next-to-MHV (NMHV) amplitude, are specified as \emph{ratio functions}, which are defined by dividing the super-amplitude for the chosen helicity configuration by the MHV super-amplitude~\cite{Drummond2008vq}. The NMHV ratio function can be further decomposed into two independent functions, $V$ and $\Vt$, which multiply dual superconformal $R$-invariants --- five-brackets of supersymmetric versions of momentum twistors~\cite{Hodges2009hk,Mason2009qx}.  For the six-point amplitude, the next-to-next-to-MHV amplitude is related to the MHV amplitude by parity.  Therefore, $R_6$, $V$ and $\Vt$ are the only functions that can appear in this amplitude. 

In principle, these functions could be determined at $L$ loops by direct integration of the loop integrand.  There are various approaches to computing the multi-loop integrand, see for example refs.~\cite{ArkaniHamed2010kv, Carrasco2011hw, Bourjaily2011hi, ArkaniHamed2012nw, LipsteinMason, ArkaniHamed2013jha}.  However, integrating such representations of the integrand is nontrivial.  The \emph{hexagon function bootstrap}~\cite{Dixon2011pw, Dixon2011nj, Dixon2013eka, Dixon2014voa, Dixon2014xca, Dixon2014iba} sidesteps this problem by constructing ans\"atze for the functions in the space spanned by iterated integrals~\cite{Chen} with \emph{(transcendental) weight} $2L$.  The assumption that the functions lie in this space was originally inspired by the compact analytic form found for the two-loop remainder function~\cite{Goncharov2010jf}, following earlier work~\cite{DelDuca2009au}. It can also be argued for from various ``dLog'' representations of the loop integrand~\cite{ArkaniHamed2012nw, LipsteinMason}.  Indeed, there is evidence that iterated-integral representations should exist for all scattering amplitudes with fewer than ten particles~\cite{ArkaniHamed2012nw}.  Familiar examples of iterated integrals include logarithms, polylogarithms, Riemann $\zeta$ values, and multiple polylogarithms~\cite{FBThesis,Gonch}, where the weight is given by the number of integrations. By requiring that an ansatz spanning this space of functions has the appropriate analytic properties and functional dependence, and by further matching it to known physical limits of six particle scattering, the six-point remainder and NMHV ratio functions have been uniquely determined, through four loops~\cite{Dixon2014voa} and three loops~\cite{Dixon2014iba}, respectively.  A similar {\it heptagon function} bootstrap, based on cluster variables~\cite{ClusterCoordinates,SpradlinTwoLoopMHV} has yielded the (symbol of the) seven-point remainder function --- with remarkably little input from physical limits~\cite{Drummond2014ffa}.  The main purpose of this article is to extend the hexagon function bootstrap to the NMHV six-point amplitude at four loops.

Hexagon functions are defined by two conditions~\cite{Dixon2013eka}:
\begin{enumerate}
\item Their derivatives with respect to the cross ratios can be expanded in terms of just nine hexagon functions of one lower weight, $n-1$.  Equivalently, there are nine different $\{n-1,1\}$ elements of the coproduct~\cite{Duhr2011zq,Duhr2012fh}, corresponding to nine letters in the symbol~\cite{Gonch3,Gonch2,Brown2011ik} of the function.  We also refer to these functions as \emph{final entries} (of the symbol).
\item Their branch cuts are only in $u$, $v$ and $w$, and not in any of the other six symbol letters~\cite{Gaiotto2011dt}.
\end{enumerate}
The first condition can be used to construct hexagon functions iteratively in the weight. The branch-cut condition is imposed iteratively as well, although at each order most of it is automatically obeyed, given that the first derivative obeys it by construction.
The branch-cut condition massively prunes the space of iterated integrals. For example, at weight eight --- the weight we will primarily be concerned with in this paper --- a representation of the space of iterated integrals in terms of multiple polylogarithms without imposing the branch-cut condition~\cite{Dixon2013eka} leads to 1,675,553 such functions, whereas there are only 6,916 hexagon functions. (Recently a more economical multiple-polylogarithm representation has been found which requires only 500,217 functions at weight eight~\cite{Parker2015cia}.)

In this paper, we use the hexagon function bootstrap to determine the four-loop NMHV ratio function, starting from an ansatz of weight-eight hexagon functions for each $V$ and $\Vt$. Due to the combination of $R$-invariants multiplying these functions and their permutations in the ratio function, a number of discrete symmetry constraints can be applied from the outset.  Some of the discrete symmetries are subsets of the $S_3$ group of permutations of $u$, $v$, and $w$.  There is also a ``parity'' which leaves $u,v,w$ alone but flips the sign of a square root needed to define certain symbol letters $y_i$; parity takes $y_i \lr 1/y_i$.  The function $V(u,v,w)$ must be parity-even and symmetric in the exchange $u \lr w$, while $\Vt(u,v,w)$ is parity-odd and antisymmetric under the same exchange.

A particularly powerful constraint comes from dual superconformal symmetry, which leads to a ``$\bar{Q}$'' differential equation~\cite{BullimoreSkinner,CaronHuot2011kk}.  The consequences of this equation for the first derivatives of six-point amplitudes were explored in refs.~\cite{CaronHuot2011kk,SimonSongPrivate}. It has also been studied recently in the context of the operator product expansion~\cite{BelitskyDescent}.  Here we will be interested in its global implications.  For the MHV remainder function, it implies that only six of the nine final entries are allowed.  This information was used in the hexagon function bootstrap for this function at four loops, although it still left over 100 free parameters~\cite{Dixon2014voa}. In the initial construction of the NMHV ratio function at three loops~\cite{Dixon2014iba}, a seven-final-entry condition~\cite{CaronHuot2011kk,SimonSongPrivate} was imposed on both $V$ and $\Vt$.  After the fact, it was found empirically that a function related to $V$ had only five final entries, but the connection to the $\bar{Q}$ equation was not yet clear~\cite{Dixon2014iba}.  Subsequently, we have understood that the five-final-entry condition can be derived from the $\bar{Q}$ equation, but also that this equation has much more powerful consequences~\cite{SimonSongPrivate}.  The five-final-entry condition is a restriction on just one permutation of the parity-even part of the ratio function; the full power of the $\bar{Q}$ equation comes from how it relates different permutations to each other, and also how it relates the parity-even and parity-odd functions.  Imposing the more general restrictions at the outset, along with the discrete symmetry requirements, we find only a 34-parameter family of solutions at four loops.  (The five-final-entry condition, plus a seven-final-entry condition on $\Vt$, together with the same discrete symmetry constraints, would have left 808 parameters at four loops.)

To this 34-parameter ansatz we apply the same physical constraints used at three loops~\cite{Dixon2014iba}.  In the collinear limit, in which two external legs of the amplitude become parallel, the six-point amplitude must reduce to a splitting function times a five point amplitude.  Because the five-point ratio function is trivial, loop corrections to the six-point ratio function must vanish in this limit.  This constraint fixes all but five of the 34 parameters.  Furthermore, while the hexagon functions are free of unphysical singularities, some of the $R$-invariants have spurious poles. Therefore, any linear combination of $V$ and $\Vt$ that multiplies an $R$-invariant that has a spurious pole must vanish as that pole is approached.  Previously, this condition provided a useful constraint~\cite{Dixon2011nj,Dixon2014iba}.  Now, however, the combination of the $\bar{Q}$ and collinear constraints is so powerful that no additional parameters are fixed by the spurious-pole constraint (at least through four loops).

To fix the five remaining parameters at four loops, we turn to the multi-Regge limit.  There has been considerable study of the remainder function in this limit~\cite{Bartels2008ce, Bartels2008sc, Lipatov2010qg, Lipatov2010ad, Bartels2010tx, BLPCollRegge, Dixon2011pw, Fadin2011we, Dixon2012yy, Pennington2012zj, CaronHuot2013fea, Dixon2014voa, Hatsuda2014oza, BCHS, JamesYorgos}.  In the NMHV case, a factorization was proposed at the leading-logarithmic level by Lipatov, Prygarin and Schnitzer~\cite{Lipatov2012gk}, and later extended to all orders~\cite{BCHS,Dixon2014iba}.  The quantities entering the multi-Regge factorization --- the BFKL eigenvalue and the impact factor --- can either be determined order-by-order~\cite{Dixon2014iba}, or all at once using integrability and a continuation from the near-collinear limit~\cite{BCHS} (see also ref.~\cite{JamesYorgos}).  The three-loop ratio function suffices to determine the multi-Regge limit to next-to-leading-logarithmic (NLL) accuracy. Matching the five-parameter ansatz at four loops to the NLL result, we fix all five parameters remaining in the ansatz.

Once we have uniquely determined the solution, we can check it against further boundary data.  It predicts the next-to-next-to-leading-logarithmic (NNLL) terms in the multi-Regge limit, and even the N$^3$LL impact factor. All of these results agree with previous predictions~\cite{Dixon2014voa,Dixon2014iba,BCHS}.  Many further checks come from the operator product expansion (OPE) controlling the near-collinear limit~\cite{Alday2010ku, Gaiotto2010fk, Gaiotto2011dt, Sever2011da}, by virtue of the representation of the (super)amplitude as a light-like polygonal Wilson (super)loop.  The Wilson loop OPE can be calculated nonperturbatively in the coupling, using technology first developed by Basso, Sever and Vieira (BSV), wherein the expansion is carried out in the number of flux tube excitations \cite{Basso2013vsa, Basso2013aha, Basso2014koa, BSVIV}.  This expansion corresponds to the number of powers of $\sqrt{w}$ in the series expansion around the collinear limit $w\to0$, $u+v\to1$.  More recently, this flux-tube approach has been extended to all helicity configurations~\cite{BDM, BelitskyI, BelitskyII, AllHelicityOPE, POPE}.  Previously, we used some of this information in the construction of the three-loop ratio function.  With the additional $\bar{Q}$ constraints imposed, the OPE comparison becomes purely a cross-check, at least through four loops.  We have compared the series expansion of our results to both the single and double flux-tube excitation OPE predictions, and all are in agreement.\footnote{We thank Andrei Belitsky for assistance with this comparison.}

Another interesting limit is that of multi-particle factorization, where the six-point amplitude splits into two four-point amplitudes connected by a single-particle exchange (at tree level).  In this limit, two cross ratios get large at the same rate: $u,w\to\infty$ with $u/w$ and $v$ fixed.  At three loops, it was found that the behavior of the even part of the ratio function in this limit was extremely simple, and could be expressed just in terms of a polynomial in one kinematic combination, $\ln(uw/v)$, with constant ($\zeta$-valued) coefficients.  We find that this pattern persists at four loops.

In order to gain some insight into the structure of the NMHV amplitude, we explore the analytic and numerical features of $V$ and $\Vt$ through four loops in a number of kinematic regions.  We give (relatively) compact formulas for $V$ and $\Vt$ on particular lines through the space of cross ratios where they simplify.  We obtain numerical values and plot them on these lines, and on various two-dimensional surfaces.  From the finite radius of convergence of the perturbative expansion of planar $\mathcal{N}=4$ SYM, we expect the ratios of perturbative coefficients at successive loop orders to eventually approach the same negative constant.  However, the rate at which this happens can depend on the location within the space of cross ratios.  In many limits, there are logarithmic divergences, where the power of the logarithm increases with the loop order.  Sufficiently close to these limits, the generic asymptotic behavior does not hold.  However, we observe that away from these singular regions, the ratios between successive loop orders do become increasingly flat as the loop order increases.

Another aspect of this work is to improve our knowledge of the space of hexagon functions at higher weight, not only to help with the four-loop construction performed in this article, but also as a platform for going to higher loops in the future.  We have constructed a basis for this space now through weight eight, whereas previously only a weight-five basis had been constructed~\cite{Dixon2013eka}. The weight-six part of the basis allows us to write the three-loop quantities $R_6^{(3)}$, $V^{(3)}$ and $\Vt^{(3)}$ as single functions, whereas previously we had to describe them for generic $(u,v,w)$ in terms of their first derivatives, or equivalently their $\{5,1\}$ coproduct elements.  Similarly, we can express the four-loop quantities for generic $(u,v,w)$ in terms of the weight-eight basis, although the expressions do start to become rather lengthy.

The structure of this paper is as follows.  In \sect{setupoverview} we describe the setup and give an overview of the constraints we impose.  We also outline the iterative construction of a basis of hexagon functions. In \sect{qbar} we discuss the constraints coming from the $\bar{Q}$ equation, which does the bulk of the work in fixing parameters.  In \sect{mrk} we discuss the multi-Regge constraint, which fixes the final five parameters in our four-loop ansatz.  In \sect{ope} we analyze the near-collinear limit and compare it to the OPE predictions.  In \sect{multiparticle} we study the multi-particle factorization limit.  In \sect{quant} we study the quantitative behavior of the result on various lines and surfaces in the space of cross ratios.  Finally, in \sect{conclusions} we conclude and provide our outlook for the future.  There are four appendices.  Appendix~\ref{hexagon_basis} gives more details on the construction of a hexagon function basis.  Appendix~\ref{RVVt3inwt6basis} gives the three-loop quantities $R_6^{(3)}$, $V^{(3)}$ and $\Vt^{(3)}$ in terms of the weight-six basis, while appendix~\ref{RVVt4inwt8basis} gives parts of the expressions of the corresponding four-loop quantities in terms of the weight-eight basis.  Finally, appendix~\ref{SP_basis} describes the basis of functions of $(u,v)$ to which the hexagon functions collapse on the surface $w=1$.  This function space is useful for implementing the spurious-pole constraint.

Many of the analytic results in this paper are too lengthy to present in the manuscript.  Instead we provide a webpage containing a set of ancillary files in computer-readable format~\cite{V4website}.  The files describe: functional integrability constraints, the ratio function and remainder function through four loops in terms of the the weight-eight basis, a coproduct-based definition of the basis, expansions of the ratio function in the near-collinear limit and in the multi-Regge limit, multiple polylog representations in other ``bulk'' regions, harmonic polylog representations on particular lines, a basis of functions for the surface $w=1$ through weight seven, and the ratio function and remainder function on $w=1$ through three loops in terms of this basis.


\section{Setup and overview of constraints}
\label{setupoverview}

\subsection{Decomposition of ratio function in terms of $R$-invariants}
\label{onshellsuper}

As in past work at one, two, and three loops~\cite{Drummond2008vq,Drummond2008bq,Dixon2011nj,Dixon2014iba}, we describe the six-point amplitude using an on-shell superspace~\cite{Nair1988bq,Georgiou2004by,Bianchi2008pu,ArkaniHamed2008gz}.  We package the on-shell states of the theory into a superfield $\Phi$ depending on Grassmann variables $\eta^A$, $A=1,2,3,4$, transforming in the fundamental representation of $SU(4)$:
\be
\Phi\ =\ G^+ + \eta^A \Gamma_A + \tfrac{1}{2!} \eta^A \eta^B S_{AB}
+ \tfrac{1}{3!} \eta^A \eta^B \eta^C \epsilon_{ABCD} \overline{\Gamma}^D
+ \tfrac{1}{4!} \eta^A \eta^B \eta^C \eta^D \epsilon_{ABCD} G^-.
\label{onshellmultiplet}
\ee
Here $G^+$, $\Gamma_A$,
$S_{AB}=\tfrac{1}{2}\epsilon_{ABCD}\overline{S}^{CD}$,
$\overline{\Gamma}^A$, and $G^-$ are the positive-helicity gluon,
gluino, scalar, anti-gluino, and negative-helicity gluon states,
respectively.

The superamplitude $\mathcal{A}(\Phi_1,\Phi_2,\ldots,\Phi_n)$ contains all the information about the component helicity amplitudes, which can be extracted as particular terms in the expansion in the Grassmann variables.  The superamplitude can be factored into the product of the MHV superamplitude and the ratio function $\cP$~\cite{Drummond2008vq},
\be
\mathcal{A}\ =\ \mathcal{A}_{\rm MHV} \, \times \, \cP\,.
\label{ratiodef}
\ee
The ratio function is infrared finite.  Expanding it in the $\eta$ variables for six-particle scattering yields three terms,
\be
\cP\ =\ 1 + \cP_{\rm NMHV}
+ \cP_{\overline{\rm MHV}}\,.
\ee
Because $\mathcal{A}_{\rm MHV} \, \times \, \cP_{\overline{\rm MHV}}$ is just the parity conjugate of the MHV superamplitude $\mathcal{A}_{\rm MHV}$, the only quantity not determined by the MHV expression is $\cP_{\rm NMHV}$, which we compute.

We represent the kinematic variables in terms of dual coordinates $(x_i, \theta_i)$. (For a full discussion see e.g.~ref.~\cite{ElvangHuangReview}.)   The momenta $k_i^{\alpha\dot\alpha} = k_i^\mu \sigma_\mu^{\alpha\dot\alpha}$ and supermomenta $q_i^{\alpha A}$ are expressed in terms of the dual coordinates as,
\be
k_i^{\alpha \dot\alpha} = 
\lambda_i^\alpha \tilde{\lambda}_i^{\dot\alpha}
= x_i^{\alpha \dot\alpha} - x_{i+1}^{\alpha \dot\alpha}, \qquad
q_i^{\alpha A} = \lambda_i^{\alpha} \eta_i^A
= \theta_i^{\alpha A} - \theta_{i+1}^{\alpha A}\,.
\label{xthetadef}
\ee
The dual coordinates appear in the amplitude either through the three dual conformal cross ratios, or (in the $R$-invariants) through the momentum supertwistors.

The three cross ratios are given by,
\beq\label{uvw_def}
u = u_1 = \frac{x_{13}^2\,x_{46}^2}{x_{14}^2\,x_{36}^2}\,, 
\qquad v = u_2 = \frac{x_{24}^2\,x_{51}^2}{x_{25}^2\,x_{41}^2}\,, \qquad
w = u_3 = \frac{x_{35}^2\,x_{62}^2}{x_{36}^2\,x_{52}^2}\,,
\eeq
where $x_{ij}^2 \equiv (x_i^\mu - x_j^\mu)^2$.  The momentum supertwistors~\cite{Hodges2009hk,Mason2009qx} are
\be
\mathcal{Z}_i = (Z_i \, | \, \chi_i), \qquad 
Z_i^{R=\alpha,\dot\alpha} = 
(\lambda_i^\alpha , x_i^{\beta \dot\alpha}\lambda_{i\beta}),
\qquad
\chi_i^A= \theta_i^{\alpha A}\lambda_{i \alpha} \,.
\ee
The momentum twistors $Z_i$ transform linearly under dual conformal symmetry, so that the four-bracket $\langle abcd\rangle \equiv \epsilon_{RSTU} Z_a^R Z_b^S Z_c^T Z_d^U$
is a dual conformal invariant (although it is not invariant under projective transformations of the $Z_i$).  To construct dual superconformal invariants we can package the four-brackets, along with the $\chi_i$, into five-brackets of momentum supertwistors called $R$-invariants as follows:
\be
(f)\ \equiv\ [abcde]\ =\ 
\frac{\delta^4\bigl(\chi_a \langle bcde\rangle + {\rm cyclic}\bigr)}
{\langle abcd\rangle\langle bcde\rangle
\langle cdea\rangle\langle deab\rangle\langle eabc\rangle}\,.
\label{five_bracket_def}
\ee
Here the six external lines are labeled $\{a,b,c,d,e,f\}$, and we use shorthand notation to represent the five-bracket of $\mathcal{Z}_a,\mathcal{Z}_b,\mathcal{Z}_c,\mathcal{Z}_d,$ and $\mathcal{Z}_e$ by the remaining leg $f$.

For higher-point amplitudes these $R$-invariants obey many identities; however, here it is sufficient to only consider one~\cite{Drummond2008vq}:
\be
(1) - (2) + (3) - (4) + (5) - (6)\ =\ 0.
\label{5bracketidentity}
\ee
Using this identity the tree-level ratio function can be represented in two equivalent ways:
\be
\cP^{(0)}_{\rm NMHV}\  =\ (2) + (4) + (6)\ =\ (1) + (3) + (5).
\label{NewPtree}
\ee

At loop level, the $R$-invariants are dressed by two functions of the cross ratios: a parity-even function $V(u,v,w)$ and a parity-odd function $\Vt(y_u,y_v,y_w)$~\cite{Drummond2008vq,Dixon2011nj}:
\be
\bsp
&\cP_{\rm NMHV}\ =\ \frac{1}{2}\Bigl[
 [(1) + (4)] V(u,v,w) + [(2) + (5)] V(v,w,u) + [(3) + (6)] V(w,u,v)  \\
&\hskip1.8cm 
+ [(1) - (4)] \Vt(y_u,y_v,y_w) - [(2)-(5)] \Vt(y_v,y_w,y_u)
  + [(3) - (6)] \Vt(y_w,y_u,y_v) \Bigr] \,.
\label{PVform}
\esp
\ee
The $y_i$ are dual conformally invariant parity-odd variables; indeed the definition of parity is the inversion $y_i \lr 1/y_i$.  The $y_i$ variables can be defined in terms of $(u,v,w)$ as follows:
\be
y_u = \frac{u-z_+}{u-z_-}\,, \qquad y_v = \frac{v-z_+}{v-z_-}\,, 
\qquad y_w = \frac{w - z_+}{w - z_-}\,,
\label{yfromu}
\ee
where
\be
z_\pm = \frac{1}{2}\Bigl[-1+u+v+w \pm \sqrt{\Delta}\Bigr]\,, 
\qquad \Delta = (1-u-v-w)^2 - 4 uvw\,.
\label{z_definition}
\ee
So alternatively, parity can be defined as $\sqrt{\Delta} \lr -\sqrt{\Delta}$, while leaving $(u,v,w)$ invariant.  Each point $(u,v,w)$ corresponds to two points in the $y_i$ variables, $(y_u,y_v,y_w)$ and $(1/y_u,1/y_v,1/y_w)$.  Parity-even functions have the same values at both $y_i$ points, whereas the values of parity-odd functions flip sign between the two $y_i$ points.

\subsection{$V$, $\Vt$, $E$, $\Et$ and $U$}
\label{VVtEEtU}

The functions $V(u,v,w)$ and $\Vt(y_u,y_v,y_w)$ can be expanded perturbatively.
At tree level, the function $V(u,v,w)$ is equal to unity, while $\Vt(y_u,y_v,y_w)$ vanishes.
Their full loop expansions are
\bea
V &=& 1 + \sum_{L=1}^\infty a^L V^{(L)} \,, \qquad \label{Vexpand}\\
\Vt &=& \sum_{L=2}^\infty a^L \Vt^{(L)} \,, 
\label{Vtexpand}
\eea
where $a=g_{\textrm{YM}}^2N_c/(8\pi^2)$ is our loop expansion parameter,
in terms of the Yang-Mills coupling constant $g_{\textrm{YM}}$ and
the number of colors $N_c$.  (The one-loop quantity $\Vt^{(1)}$ vanishes
because there is no parity-odd weight-two hexagon function.)

It is convenient to introduce some other functions $E$ and $\Et$, which are closely related to $V$ and $\Vt$, but defined more directly in terms of the NMHV amplitude, rather than its ratio to the MHV amplitude.  The $\bar{Q}$ equation will be simplest when expressed in terms of these functions.  First recall that the MHV amplitude can be expressed in terms of two quantities, the BDS ansatz~\cite{BDS}
and the remainder function $R_6$~\cite{Bern2008ap}:
\be
\mathcal{A}_{\rm MHV}\ =\ \mathcal{A}_{\rm BDS} \, \times \, \exp(R_6)\,.
\label{remainderdef}
\ee
Therefore if we divide the NMHV superamplitude by the BDS ansatz $\mathcal{A}_{\rm BDS}$, rather than by the MHV amplitude, that ratio will have the same expansion~(\ref{PVform}), but with $V\to V\exp(R_6)$ and $\Vt\to \Vt\exp(R_6)$.  In fact, we are going to divide the NMHV amplitude by a slightly-different, ``BDS-like'' function.  Such a quantity has already been considered in the analysis of the strong-coupling behavior of amplitudes~\cite{AGM}, as well as in the study of the multi-particle factorization limit of the NMHV amplitude~\cite{Dixon2014iba}.

Before describing the BDS-like ansatz, we recall that the BDS ansatz can be written as~\cite{BDS},
\be
\frac{\mathcal{A}_n^{\rm BDS}}{\mathcal{A}_n^{{\rm MHV}\,(0)}}\ =\ 
\exp\biggl[ 
\sum_{L=1}^\infty a^L \Bigl( f^{(L)}(\e) \frac{1}{2} M_n^{\rm 1-loop}(L\e) + C^{(L)} \Bigr)
\biggr] \,,
\label{BDSAnsatz}
\ee
where $\mathcal{A}_n^{{\rm MHV}\,(0)}$ is the MHV tree-level super-amplitude, and
\be
f^{(L)}(\e) \equiv f_0^{(L)} + \e \, f_1^{(L)} + \e^2 \, f_2^{(L)} \,.
\label{fepsdef}
\ee
Two of the constants,
\be
f_0^{(L)}\ =\ \frac{1}{4} \, \gamma_K^{(L)} \,, \qquad
f_1^{(L)}\ =\ \frac{L}{2} \, {\cal G}_0^{(L)} \,,
\label{cuspG}
\ee
are given in terms of the cusp anomalous dimension $\gamma_K$
and the ``collinear'' anomalous dimension ${\cal G}_0$, 
while $f_2^{(L)}$ and $C^{(L)}$ are other (zeta-valued) constants. We won't need the specific values of any of these constants except for the cusp anomalous dimension.  This quantity is known to all orders~\cite{Beisert2006ez}; its expansion to four loops is given by
\be
\gamma_K(a) = \sum_{L=1}^\infty a^L\,\gamma_K^{(L)} = 4a - 4 \zeta_2\,a^2
+ 22\zeta_4\,a^3
- 4 \biggl( \frac{219}{8} \zeta_6 + (\zeta_3)^2 \biggr) a^4 
+ \mathcal{O}(a^5)\,.
\label{cuspdef}
\ee
The function $M_n^{\rm 1-loop}(L\e)$ is the one-loop amplitude, normalized by the tree amplitude $\mathcal{A}_n^{{\rm MHV}\,(0)}$, and evaluated in dimensional regularization with $D=4-2\e$, but letting $\e\to L\e$.

The normalized six-point one-loop amplitude is given by~\cite{Unitarity}
\bea
M_6^{\rm 1-loop}(\e) &=&
  \sum_{i=1}^6 \biggl[
  -\frac{1}{\e^2} (-s_{i,i+1})^{-\e}
 - \ln\Bigl(\frac{-s_{i,i+1}}{-s_{i,i+1,i+2}}\Bigr)
   \ln\Bigl(\frac{-s_{i+1,i+2}}{-s_{i,i+1,i+2}}\Bigr)
 + \frac{1}{4} \ln^2\Bigl(\frac{-s_{i,i+1,i+2}}{-s_{i+1,i+2,i+3}}\Bigr) \biggl]
\nonumber\\ &&\null\hskip0.5cm
 - \Li_2(1-u)  - \Li_2(1-v) - \Li_2(1-w) + 6 \, \zeta_2 \,,
\label{M6}
\eea
where $s_{i,i+1} = (k_i+k_{i+1})^2$ and $s_{i,i+1,i+2} = (k_i+k_{i+1}+k_{i+2})^2$.  Notice that $M_6^{\rm 1-loop}$ has non-trivial dependence on the three-particle momentum invariants $s_{i,i+1,i+2}$, both explicitly and implicitly through the three cross ratios. However, this dependence can be removed by shifting $M_6^{\rm 1-loop}$ by a particular totally symmetric function of the cross ratios,
\be
Y(u,v,w)\ \equiv\ \Li_2(1-u) + \Li_2(1-v) + \Li_2(1-w) 
         + \frac{1}{2} \Bigl( \ln^2u + \ln^2v + \ln^2w \Bigr) \,.
\label{Ydef}
\ee
We let
\bea
\hat{M}_6(\e) &=& M_6^{\rm 1-loop} + Y(u,v,w) \nonumber\\
&=&  \sum_{i=1}^6 \biggl[
  - \frac{1}{\e^2} \Bigl( 1 - \e \ln(-s_{i,i+1}) \Bigr)
  - \ln(-s_{i,i+1})\ln(-s_{i+1,i+2})
  + \frac{1}{2} \ln(-s_{i,i+1})\ln(-s_{i+3,i+4}) \biggr]\nonumber\\
&&\hskip0cm\null + 6 \, \zeta_2 \,,
\label{M6hat}
\eea
which contains only the two-particle invariants $s_{i,i+1}$.

Then we can define the BDS-like function by
\be
\frac{\mathcal{A}_6^{\rm BDS-like}}{\mathcal{A}_6^{{\rm MHV}\,(0)}}\ =\ 
\exp\biggl[ 
\sum_{L=1}^\infty a^L \Bigl( f^{(L)}(\e) \frac{1}{2} \hat{M}_6(L\e) + C^{(L)} \Bigr)
\biggr] \,.
\label{BDSlikeAnsatz}
\ee
Using \eqn{M6hat}, it is related to the BDS ansatz by
\be
\mathcal{A}_6^{\rm BDS-like}
= \mathcal{A}_6^{\rm BDS} \, \exp\biggl[ \frac{\gamma_K}{8} Y(u,v,w) \biggr] \,.
\label{BDSlike}
\ee
Finally, we normalize the NMHV superamplitude by the BDS-like ansatz, and define new functions
$E(u,v,w)$ and $\Et(u,v,w)$ as the coefficients of the $R$-invariants:
\be
\bsp
&\frac{\mathcal{A}_{\rm NMHV}}{\mathcal{A}_6^{\rm BDS-like}}
\ =\ \frac{1}{2}\Bigl[
 [(1) + (4)] E(u,v,w) + [(2) + (5)] E(v,w,u) + [(3) + (6)] E(w,u,v)  \\
&\hskip1.8cm 
+ [(1) - (4)] \Et(y_u,y_v,y_w) - [(2)-(5)] \Et(y_v,y_w,y_u)
  + [(3) - (6)] \Et(y_w,y_u,y_v) \Bigr] \,.
\label{Eform}
\esp
\ee

The relations between the new expansion coefficients, $E$ and $\Et$, and the old ones,
$V$ and $\Vt$, are:
\bea
E(u,v,w) &=& V(u,v,w) \, \exp\biggl[ R_6(u,v,w) - \frac{\gamma_K}{8} Y(u,v,w) \biggr] \,, 
\label{EVrelation}\\
\Et(u,v,w) &=& \Vt(u,v,w) \, \exp\biggl[ R_6(u,v,w) - \frac{\gamma_K}{8} Y(u,v,w) \biggr] \,.
\label{EtVtrelation}
\eea
As long as the remainder function $R_6$ is known to the same loop order, it is straightforward
to pass back and forth between $(E,\Et)$ and $(V,\Vt)$.  The consequences of the $\bar{Q}$
equations, which hold globally in $(u,v,w)$, are simplest to describe in terms of $E$ and $\Et$.
On the other hand, the boundary data is often described in terms of $V$ and $\Vt$.

One exception is the limit of multi-particle factorization,
in which the perturbative simplicity of $E$, or rather its logarithm $U$,
was first noticed.  We define
\be
U(u,v,w) = \ln E(u,v,w), \qquad  E(u,v,w) = \exp\Bigl[ U(u,v,w) \Bigr] \,.
\label{Udef}
\ee
In \sect{multiparticle} we will see that this function has the same simple behavior
at four loops that it has through three loops~\cite{Dixon2014iba}.


\subsection{Hexagon functions}
\label{hexfnsubsection}

In order to construct the NMHV amplitude at four loops, we build on the observation that through three loops $V^{(L)}$ and $\Vt^{(L)}$ have been found to belong to the space of hexagon functions of weight $2L$~\cite{Dixon2014iba}. A hexagon function is defined to be any function whose symbol is constructed from letters drawn from the set
\be
\mathcal{S}_u = \{u, v, w, 1-u, 1-v, 1-w, y_u, y_v, y_w\} \,,
\label{uLetters}
\ee
and which has only physical branch cuts~\cite{Dixon2013eka}. The latter condition implies that hexagon functions can only have discontinuities when either $u$, $v$, or $w$ approaches zero or infinity.  This condition can be enforced at the level of the symbol by only allowing the variables $u$, $v$, and $w$ to appear in the first entry of the symbol. Hexagon functions in which none of the variables $y_u$, $y_v$, or $y_w$ appear can be factored into functions whose symbols have letters drawn from $\{u,1-u\}$, or $\{v,1-v\}$, or $\{w,1-w\}$.  Such functions can be expressed as (products of) harmonic polylogarithms (HPLs) of a single variable~\cite{Remiddi1999ew}.  Functions whose symbols contain $y_u$, $y_v$, or $y_w$ are more complex.  They can be defined iteratively in terms of lower-weight hexagon functions by means of their derivatives. They can also be represented in terms of multiple polylogarithms in particular regions. In ref.~\cite{Dixon2013eka}, the space of hexagon functions was explored through weight six and a basis of irreducible hexagon functions through weight five was introduced.  Irreducible hexagon functions are those that cannot be written as products of lower-weight hexagon functions.

The derivatives of a weight-$n$ hexagon function $F$ are given by~\cite{Dixon2013eka}
\bea
&&\hskip0.7cm
\frac{\partial F}{\partial u} \bigg |_{v,w} =
\frac{F^u}{u} - \frac{F^{1-u}}{1-u} + \frac{1 - u - v - w}{u \sqrt{\Delta}} F^{y_u}
+ \frac{1 - u - v + w}{(1-u) \sqrt{\Delta}} F^{y_v}
+ \frac{1 - u + v - w}{(1-u) \sqrt{\Delta}} F^{y_w} \,, \nonumber\\
&~& \label{dFu}\\
&&\hskip-0.7cm\sqrt{\Delta} y_u \frac{\partial F}{\partial y_u} \bigg|_{y_v, y_w}
= (1 - u) (1 - v - w) F^u - u (1 - v) F^v - u (1 - w) F^w - u (1 - v - w) F^{1-u}
\nonumber\\ 
&&\hskip2.6cm\null
+ u v F^{1-v} + u w F^{1-w} + \sqrt{\Delta} F^{y_u} \,, \label{dFyu}
\eea
as well as the cyclic permutations of these formulae under $u\to v\to w\to u$, $y_u \to 1/y_v \to y_w \to 1/y_u$.  Each of the rational prefactors in \eqn{dFu} is $[\partial (\ln x)/\partial u]|_{v,w}$ for some $x\in \mathcal{S}_u$, while in \eqn{dFyu} the corresponding rational prefactor is $[\partial (\ln x)/\partial y_u]|_{y_v,y_w}$.  The $F^x$ for $x\in \mathcal{S}_u$ denote nine weight-$(n-1)$ hexagon functions.  These functions are also referred to as elements of the $\{n-1,1\}$ coproduct component of $F$~\cite{Duhr2012fh}:
\be
\Delta_{n-1,1}(F) \equiv \sum_{i=1}^3 \Bigl[ F^{u_i} \otimes \ln u_i + F^{1-u_i} \otimes \ln (1-u_i) + F^{y_i} \otimes \ln y_i \Bigr] \,.
\ee
The $\{n-1,1\}$ coproduct component specifies all the first derivatives of $F$.  Hence it completely specifies $F$, up to an additive constant.

To fix the additive constant, we will typically require that basis functions vanish at the point $(u,v,w)=(1,1,1)$.  Physical constraints are imposed elsewhere, so we need to transfer information about the value of functions at other points to the point $(1,1,1)$.  We can transfer the information along special lines that cut through the $(u,v,w)$ space.  For example, the line $(1,v,v)$ connects $(1,1,1)$ to $(1,0,0)$.  The latter point corresponds to a soft limit (a special case of two collinear limits), where there are physical constraints. On the line $(1,v,v)$, all hexagon functions collapse to HPLs in the single variable $v$.  The standard notation for such functions is $H_{\vec{w}}(v)$, where $\vec{w}=w_1,w_2,\ldots,w_n$ is a list of $n$ elements (at weight $n$), all of which are either 0 or 1.  We can use shuffle identities to always choose $w_n=1$ for $n>1$, and it is convenient to have the argument be $1-v$ so that the function is regular at $v=1$.  Furthermore we use a compressed notation in which $(m-1)$ 0's followed by a 1 is written as $m$.  Thus we define $H_{3,1,1}^v = H_{0,0,1,1,1}(1-v)$, and so forth.  The function $\Li_2(1-v)$ entering the definition of $Y(u,v,w)$ is $H_2^v$ in this notation. 

\Eqn{dFu} and its cyclic permutations form the cornerstone for the construction of a basis of hexagon functions, iteratively in the weight.  Suppose one knows all hexagon functions at weight $(n-1)$.  One can define a candidate set of weight $n$ hexagon functions by introducing arbitrary linear combinations of the weight $(n-1)$ functions for each of the $\{n-1,1\}$ coproduct elements $F^x$, $x\in \mathcal{S}_u$.  This construction is naturally graded by parity.  That is, if $F$ is parity-even, then the six coproducts $F^{u_i}$ and $F^{1-u_i}$ are parity-even and should be drawn from the parity-even subspace at weight $(n-1)$, while the three coproducts $F^{y_i}$ are parity-odd.  If $F$ is parity-odd, the reverse is true.

Not all combinations of $\{n-1,1\}$ coproduct elements $F^x$ correspond to actual functions.
First of all, they should obey the functional integrability conditions,
\be
\frac{\partial^2F}{\partial u_i \partial u_j} = 
\frac{\partial^2F}{\partial u_j \partial u_i} \,, \qquad i \neq j.
\label{integr}
\ee
These conditions can be recast as linear constraints on the $\{n-2,1,1\}$ coproduct elements of $F$,
namely $F^{y,x}$, where $F^{y,x}$ is defined as the $y$ coproduct element for $F^x$, i.e.
\be
\Delta_{n-2,1}(F^x) \equiv \sum_{i=1}^3 \Bigl[ F^{u_i,x} \otimes \ln u_i + F^{1-u_i,x} \otimes \ln (1-u_i) + F^{y_i,x} \otimes \ln y_i \Bigr] \,.
\label{Fxcoprod}
\ee
In fact, the functional integrability conditions~(\ref{integr}) only involve the antisymmetric combination $F^{[x,y]} \equiv F^{x,y} - F^{y,x}$.  The constraints are given by:
\be
\bsp
F^{[u_i,u_j]} &= - F^{[y_i,y_j]} \,, \\
F^{[1-u_i,1-u_j]} &= F^{[y_i,y_j]} + F^{[y_j,y_k]} + F^{[y_k,y_i]} \,, \\
F^{[u_i,1-u_j]} &= - F^{[y_k,y_i]} \,, \\
F^{[u_i,y_i]} &= 0 \,, \\
F^{[u_i,y_j]} &= F^{[u_j,y_i]} \,, \\
F^{[1-u_i,y_i]} &= F^{[1-u_j,y_j]} - F^{[u_j,y_k]} + F^{[u_k,y_i]} \,, \\
F^{[1-u_i,y_j]} &= - F^{[u_k,y_j]} \,,
\label{integrabilityc}
\esp
\ee
for all $i\neq j \neq k \in \{1,2,3\}$.  There are a total of 12 independent parity-even relations (if $F$ is even) and 14 parity-odd ones.  We list them all explicitly in an ancillary file.

One can solve the system of linear equations~(\ref{integrabilityc}) to obtain a set of functions $F$, which is almost the set of hexagon functions at weight $n$.  There is one more branch-cut condition that has to be satisfied~\cite{Dixon2013eka}:  The derivative $\partial_uF$ in \eqn{dFu} has a $1/(1-u)$ singularity as $u\to1$, which will lead to a $\ln(1-u)$ branch-cut unless we require,
\be
 \Bigl[ F^{1-u} + F^{y_v} - F^{y_w} \Bigr] \Big|_{u \to 1}\ =\ 0.
\label{Fomuvanish_original}
\ee
Although this condition appears to be a strong one, holding for any $v$ and $w$, for $u=1$ the combination $F^{1-u} + F^{y_v} - F^{y_w}$ turns out to be independent of $v$ and $w$, once the integrability conditions~(\ref{integrabilityc}) are satisfied.  This constancy can be verified using the basis of functions described in appendix~\ref{SP_basis}.  Thus \eqn{Fomuvanish_original} only fixes weight $(n-1)$ (zeta-valued) constants in $F^{1-u}$, if $F$ is parity-even.  The constants can be fixed in the corner of the $u=1$ plane where $v$ and $w$ both vanish, namely the Euclidean multi-Regge kinematics (EMRK), which is also known as the soft limit~\cite{Dixon2013eka}.  This limit can also be reached by taking $y_u\to1$ with $y_v$ and $y_w$ fixed. In this limit, $\Delta=0$ and the parity-odd functions $F^{y_i}$ vanish, so the condition~(\ref{Fomuvanish_original}) and its permutations reduce to the three conditions
\be
F^{1-u_i} |_{u_i \to 1,\ u_j,u_k\to 0}\ =\ F^{1-u_i}(y_i=1,y_j,y_k)\ =\ 0,
\qquad i \neq j \neq k,
\label{Fomuvanish}
\ee
for even $F$.  If $F$ is parity-odd, then \eqn{Fomuvanish_original} involves the constant part of the parity-even functions $F^{y_v}$ and $F^{y_w}$.  However, such constant terms are forbidden by the requirement that $F$ vanishes when $y_i\to1$, independently of $y_j$ and $y_k$.  This is equivalent to the conditions,
\be
F^{y_j}(y_i=1,y_j,y_k)\ =\ 0,
\qquad i \neq j \neq k,
\label{Fyuvanish}
\ee
for odd $F$.

The combined solution to eqs.~(\ref{integrabilityc}), (\ref{Fomuvanish}) and (\ref{Fyuvanish}), for otherwise arbitrary hexagon functions as $\{n-1,1\}$ coproduct elements, generates the space of weight-$n$ hexagon functions $F$, apart from a few constants.  These constants are the linear combinations of the independent multiple zeta values (MZVs) at weight $n$.  Most of the weight-$n$ functions are reducible, i.e.~they are products of lower-weight hexagon functions.  In order to identify the irreducible subspace, one can generate the vector space of reducible hexagon functions, and remove them from the complete space of solutions.  This procedure was carried out in ref.~\cite{Dixon2013eka}, and a basis of hexagon functions was constructed through weight five.


\subsection{A basis at weight six, seven and eight}
\label{wt678basis}

Our calculation of the four-loop ratio function was facilitated by extending this basis of hexagon functions to weight six and seven.  We also constructed a weight-eight basis, but only after obtaining the four-loop result.  The extension of the basis beyond weight five was not strictly necessary; indeed, the four-loop remainder function was determined without such a basis~\cite{Dixon2014voa}.  In this case, the weight-five basis was used repeatedly to generate all of the $\{5,1,1,1\}$ elements of the coproduct of a generic (parity-even) weight-eight function.  From these functions all of the $\{6,1,1\}$ coproduct elements were constructed, then all of the $\{7,1\}$ coproduct elements, and finally all of the weight-eight functions.  The integrability and branch-cut conditions were imposed at each step, but there was no attempt to construct a basis beyond weight five.  However, the present approach provides a more direct route to the weight-eight four-loop ratio function.  It will also be a platform for going to five loops in the future, starting with the $\{8,1,1\}$ coproduct elements.  (Or one could extend the basis to weight nine and work with the $\{9,1\}$ coproduct elements.)

The basis at weight six also allows us to present results for $R_6$, $V$ and $\Vt$ at three loops that are significantly more compact than previous representations in terms of the $\{5,1\}$ coproducts (see appendix~\ref{RVVt3inwt6basis}).  Similarly, the weight-eight basis lets us write each of the four-loop functions as a single weight-eight function, although of course the four-loop results are not as compact as the three-loop ones.  In ancillary files, we provide $R_6^{(L)}$, $V^{(L)}$ and $\Vt^{(L)}$ for $L=3,4$.  We also provide a coproduct description of the hexagon function basis at weight six, seven and eight; this basis is described further in appendix \ref{hexagon_basis}.

There is a certain arbitrariness in defining a basis of irreducible functions; in principle, one can make an arbitrary linear transformation on the basis, and one can add any linear combination of reducible functions to any candidate basis function.  However, in the course of constructing the higher-weight basis functions, we found that some care in the construction leads to much simpler representations for physical quantities such as $R_6^{(3)}$, $V^{(3)}$, and $\Vt^{(3)}$.  One can generate a ``random'' basis by asking {\sc Maple} or {\sc Mathematica} to provide a null space ``orthogonal'' to the reducible function space. However, when $R_6^{(3)}$, $V^{(3)}$, or $\Vt^{(3)}$ are expressed in terms of such a basis, the rational numbers multiplying the basis functions in the expressions for these quantities have quite large numerators and denominators, with sometimes as many as 13 digits.  A better way to select the basis for irreducible hexagon functions at weight $n$ is to require that their weight $\{n-1,1\}$ coproduct elements collectively contain exactly one of the weight $(n-1)$ basis elements, and with unit coefficient.  One cannot require this for all weight $n$ irreducible functions; there are too many of them, compared with the number of weight $(n-1)$ ones.  We start by imposing this criterion on the $y_i$ coproduct entries, and preferentially for the functions with the most $y_i$ entries in their symbol, as these typically have the most complicated coproducts.  When we run out of weight $(n-1)$ irreducible functions, we impose the criterion using products of logarithms and weight $(n-2)$ irreducible functions instead. It is usually possible to further reduce the number of terms appearing in the coproducts of the basis functions by adding suitable linear combinations of reducible functions to them. Finally, as in ref.~\cite{Dixon2013eka}, we constructed our basis functions so that they form orbits under the permutation group $S_3$ acting on the variables $u$, $v$, and $w$, either singlets, three-cycles or six-cycles. 

The basis we have constructed in this way through weight eight leads to quite parsimonious rational number coefficients when $R_6$, $V$, and $\Vt$ (or their coproduct elements) are expanded in terms of the basis functions. For instance, the rational numbers multiplying the weight-six irreducible functions in $R_6^{(3)}$, $V^{(3)}$, and $\Vt^{(3)}$ have denominators that are all powers of 2, up to an occasional factor of 3.  The largest denominator is 128, while the largest numerator is 149.  (The coefficients in front of the pure-HPL terms don't boast the same level of simplicity, but this is unsurprising since the above prescription for choosing irreducible hexagon functions only constrains each function up to the addition of reducible functions.)  We also constructed a set of weight-five basis functions without the degeneracy of the basis defined in ref.~\cite{Dixon2013eka}, by organizing the $S_3$ orbits differently.  Even so, converting $R_6^{(3)}$, $V^{(3)}$, and $\Vt^{(3)}$ to the weight-five basis of ref.~\cite{Dixon2013eka} (which was selected with slightly different criteria in mind) only gives rise to slightly more complicated rational-number coefficients.  So we will continue to use the weight-five basis of ref.~\cite{Dixon2013eka}.

Using the basis through weight six, we give the results for the three-loop functions $R_6^{(3)}$, $V^{(3)}$, and $\Vt^{(3)}$ in eqs.~(\ref{R63A}), (\ref{V3A}) and (\ref{Vt3A}) of appendix~\ref{RVVt3inwt6basis} and in an ancillary file.  Continuing the construction to weight eight, we give a similar representation for the the four-loop functions $R_6^{(4)}$, $V^{(4)}$, and $\Vt^{(4)}$ in appendix~\ref{RVVt4inwt8basis}.  In this case, we only give the terms containing the irreducible weight-eight basis functions in the text; the remaining terms, which are products of lower-weight functions, are very lengthy and can be found in the same ancillary file.


\subsection{Overview of the constraints}
\label{constraintsoverview}

Our goal is to find a unique pair of functions $E(u,v,w)$ and $\Et(u,v,w)$ at four loops. We begin with an ansatz for the $\{7,1\}$ coproduct of a generic weight 8 hexagon function. There are 5153 such functions with even parity, which are candidates for $E^{(4)}$, and 1763 with odd parity, which are candidates for $\Et^{(4)}$. We then apply a succession of constraints to our ansatz in order to arrive at a unique result.

We largely follow the methodology of ref.~\cite{Dixon2014iba}, with some refinements. In particular, we apply the following constraints:
\begin{itemize}
\item \textbf{Symmetry:} Under the exchange of $u$ and $w$, $E$ is symmetric, while $\Et$ is antisymmetric:
\be
E(w,v,u)\ =\ E(u,v,w),\qquad
\Et(y_w,y_v,y_u)\ =\ -\Et(y_u,y_v,y_w).
\label{EEtsym}
\ee
\item \textbf{$\bar{Q}$ Equation:} Caron-Huot and He predicted~\cite{CaronHuot2011kk,SimonSongPrivate} that the final entries of the hexagon functions that make up $V(u,v,w)$ should belong to a seven-element set. At lower loop orders, two of us observed~\cite{Dixon2014iba} that the function $U(u,v,w)$ has final entries from a more constrained five-element set.  This relation can now be derived from the $\bar{Q}$ equation, but there are a host of other relations, which we describe further below.  Together they are very powerful and do the bulk of the work in reducing the number of parameters in the ansatz, at four loops as well as at lower loops.
\item \textbf{Collinear Vanishing:} In the collinear limit, the six-point ratio function should approach the five-point ratio function, multiplied by some splitting function.  Because the only non-vanishing components of the five-point super-amplitude are MHV and NMHV, which are related by parity, and because there are no dual conformally invariant cross ratios at five points, the five-point ratio function is trivial; it vanishes at loop level. As such, the loop level six-point ratio function must vanish in the collinear limit. We take this limit by sending $w\rightarrow 0$ and $v\rightarrow 1-u$. In this limit, all of the $R$-invariants vanish except for $(1)$ and $(6)$, which become equal. Taking into account that parity-odd functions such as $\Vt$ always vanish in this limit, we have the constraint,
\be
[ V(u,v,w) + V(w,u,v) ]_{w\rightarrow 0,\ v\rightarrow 1-u}\ =\ 0.
\label{collvanish}
\ee
\item \textbf{Spurious Pole Vanishing:} Physical states give rise to poles in scattering amplitudes when the sums of color-adjacent momenta vanish, when $(k_i+k_{i+1}+\ldots+k_{j-1})^2\equiv x_{ij}^2=0$. These sums come from four-brackets of the form $\langle i-1,i,j-1,j\rangle$. Poles of any other form, in particular poles arising from other four-brackets, should not appear. Individual $R$-invariants have such spurious poles, so these must cancel between $R$-invariants at tree level. At loop level, the corresponding condition is that the relevant combination of $V$ and $\Vt$ must vanish on any spurious pole. As it happens, examining one of these spurious poles is sufficient to guarantee vanishing on the others, by Bose symmetry of the super-amplitude. If we choose to fix behavior on the pole $\langle2456\rangle\rightarrow 0$, we need to cancel potential poles from $R$-invariants $(1)$ and $(3)$ with equal and opposite residues.  This leads to the condition,
\be
[ V(u,v,w) - V(w,u,v) + \Vt(y_u,y_v,y_w) - \Vt(y_w,y_u,y_v)
]_{\langle 2456\rangle \rightarrow 0} \ =\ 0.
\label{spurious}
\ee
where the $\langle 2456\rangle\rightarrow 0$ limit can be implemented by taking
$w\to1$ with $u$ and $v$ held fixed; more precisely,
\be
w\rightarrow 1 \,,
\quad y_u\rightarrow(1-w)\frac{u(1-v)}{(u-v)^2} \,,
\quad y_v\rightarrow\frac{1}{(1-w)}\frac{(u-v)^2}{v(1-u)} \,,
\quad y_w\rightarrow\frac{1-u}{1-v} \,.
\ee
We have used a basis of irreducible two-variable functions, discussed in appendix~\ref{SP_basis}, to impose this constraint.
\item \textbf{Multi-Regge Limit:} The multi-Regge limit is a generalization of the Regge limit for $2\rightarrow n$ scattering, where the outgoing particles are strongly ordered in rapidity. We build on our three-loop results, using our generalization of the work of Lipatov, Prygarin, and Schnitzer~\cite{Lipatov2012gk} to subleading logarithmic order.  We also compare our results to a recent all-orders proposal~\cite{BCHS}. 
\item \textbf{Near-Collinear Limit:} As at three loops, we employ the pentagon decomposition of the NMHV Wilson loop OPE developed by Basso, Sever, and Vieira~\cite{Basso2013aha}. Their calculation uses integrability to compute the OPE nonperturbatively in the coupling, in an expansion in the number of flux-tube excitations, corresponding to powers of $\sqrt{w}$ in the near-collinear limit. Actually, our new understanding of the $\bar{Q}$ equation is such a powerful constraint that our ansatz is completely fixed before comparing with the OPE constraints, so the OPE results serve as a pure cross check of our assumptions (and theirs).  We perform these checks at the first order of the OPE, corresponding to one state propagating across the Wilson loop~\cite{Basso2013aha}, and then at second order (two flux excitations)~\cite{Basso2014koa} using explicit results of Belitsky~\cite{BelitskyI,BelitskyII,BelitskyPrivate}.  In an ancillary file, we provide limits of $V$ and $\Vt$ to third order, making possible comparisons to the OPE terms involving three flux-tube excitations (we leave these checks to the intrepid reader).
\end{itemize}

In addition to these constraints, we should point out a residual freedom in our definition of $\Vt$, first noticed in ref.~\cite{Dixon2014iba}. If we add an arbitrary cyclicly symmetric function $\tilde{f}$ to $\Vt$, we find that it vanishes in the full ratio function:
\be
\bsp
&\hskip0.7cm \frac{1}{2}\left[ [(1)-(4)] \tilde{f}(u,v,w)
  -[(2)-(5)] \tilde{f}(u,v,w) + [(3)-(6)] \tilde{f}(u,v,w) \right]\\
&=\ \frac{1}{2}\Bigl[ [(1)+(3)+(5)]-[(2)+(4)+(6)] \Bigr]\ \tilde{f}(u,v,w)\\
&=\ 0,
\label{cyclicvanish}
\esp
\ee
and thus remains unfixed by any physically meaningful limits.

This ``gauge freedom'' was used in ref.~\cite{Dixon2014iba} to set the sum of the cyclic permutations of $\Vt$ to zero, essentially as an arbitrary choice of gauge.  We make the same choice here.  However, when presenting numerical results we usually present ``gauge invariant'' quantities:  Instead of $\Vt$, we use the difference of two cyclic permutations, such as $\Vt(v,w,u)-\Vt(w,u,v)$. Any cyclicly-symmetric contribution vanishes in such linear combinations, while the physical information is still preserved. Whenever $\Vt$ appears in physical limits, it does so in these linear combinations.


\section{$\bar{Q}$ Equation}
\label{qbar}

In refs.~\cite{BullimoreSkinner,CaronHuot2011kk}, an equation was presented describing the
action of the dual superconformal generator $\bar{Q}$ on a generic amplitude.
In terms of the dual Grassmann variables $\chi_i$ and momentum twistors $Z_i$,
the dual superconformal generator for an $n$-point amplitude is a first-order
differential operator,
\be
\bar{Q}^A_a\ =\ (S_\alpha^A, \bar{Q}^A_{\dot\alpha})
  \ =\ \sum_{i=1}^n \chi_i^A \frac{\partial}{\partial Z_i^a} \,.
\label{Qbardef}
\ee
The reason it does not annihilate the amplitude is because of a collinear anomaly,
and so its action on an $L$-loop N$^k$MHV amplitude can be expressed
in terms of the integral over an $(L-1)$-loop N$^{k+1}$MHV amplitude with one more external leg.
For the NMHV six-point amplitude we need the N$^2$MHV seven-point amplitude, but
by parity this amplitude is equivalent to the NMHV amplitude.
The $\bar{Q}$ equation for the NMHV six-point amplitude
takes the form~\cite{BullimoreSkinner,CaronHuot2011kk},
\be
\bar{Q} {\cal R}_{6,1}\ =\ \frac{\gamma_K}{8} \int d^{2|3} {\cal Z}_7
\Bigl[ {\cal R}_{7,2} - {\cal R}_{6,1} {\cal R}_{7,1}^{\rm tree} \Bigr]
\ +\ {\rm cyclic},
\label{Qbarstart}
\ee
where
\be
{\cal R}_{6,1}\ =\ \frac{\mathcal{A}_{\rm NMHV}}{\mathcal{A}_6^{\rm BDS}} \,. 
\label{R61def}
\ee
Similarly, ${\cal R}_{7,2}$ is the BDS-normalized N$^2$MHV 7-point amplitude, and ${\cal R}_{7,1}^{\rm tree}$ is the ratio of NMHV to MHV 7-point tree super-amplitudes. The integration is over a super-momentum-twistor ${\cal Z}_7$ along a collinear limit corresponding to one edge of the hexagon.  The ``$+$~cyclic'' terms correspond to the other edges.

An analysis of the leading singularities of
${\cal R}_{7,2}$~\cite{SimonSongPrivate} shows that
there are only four linearly independent residues
from the edge shown,
\be
(1) \, \bar{Q} \ln \frac{\langle5612\rangle}{\langle5614\rangle} \,, \qquad
(2) \, \bar{Q} \ln \frac{\langle5612\rangle}{\langle5614\rangle} \,, \qquad
(4) \, \bar{Q} \ln \frac{\langle5612\rangle}{\langle5614\rangle} \,, \qquad
(5) \, \bar{Q} \ln \frac{\langle5612\rangle}{\langle5614\rangle} \,, \label{R72residues}
\ee
where $(1),(2),(4),(5)$ are the $R$-invariants $(f)$ defined in \eqn{five_bracket_def}.  However, integration of the seven-point tree amplitude in
the second, collinear subtraction term in \eqn{Qbarstart} would seem
to give more possible residues.  Using eq. (3.7) of ref.~\cite{CaronHuot2011kk},
one finds a term~\cite{SimonSongPrivate}
\be
\int d^{2|3} {\cal Z}_7 {\cal R}_{7,1}^{\rm tree}
\ =\ \ln\frac{\langle6134\rangle\langle6523\rangle}
    {\langle6123\rangle\langle6534\rangle}
  \, \bar{Q} \ln \frac{\langle5612\rangle}{\langle5613\rangle}  \,.
\label{QbarR71tree}
\ee
The unwanted $\langle5613\rangle$ term can be removed by considering
the action of $\bar{Q}$ on $\hat{\cal R}_{6,1}$ rather than ${\cal R}_{6,1}$,
where
\be
\hat{\cal R}_{6,1}\ =\ \frac{\mathcal{A}_{\rm NMHV}}{\mathcal{A}_6^{\rm BDS-like}}
\ =\ {\cal R}_{6,1} \times \exp\biggl[ - \frac{\gamma_K}{8} Y(u,v,w) \biggr] \,.
\label{hatR61def}
\ee
Here $\hat{\cal R}_{6,1}$ is the quantity expanded in terms of $E$ and $\Et$ in \eqn{Eform}. The extra factor of $\exp[-\frac{\gamma_K}{8} Y]$ in $\hat{\cal R}_{6,1}$ leads to an additional contribution from the action of $\bar{Q}$ on $Y$.

Note from \eqn{Ydef} that
\be
\partial_u Y\ =\ \frac{\ln u}{u(1-u)}
\ =\ \ln u \ \partial_u \ln\biggl(\frac{u}{1-u}\biggr) \,.
\label{dYdu}
\ee
Using the cyclic symmetry of $Y$ and rewriting $u,v,w$ in terms of momentum-twistors, we have for $\bar{Q}Y$ (as for any first-order differential operator acting on $Y$),
\be
\bar{Q} \, Y\ =\ \ln\frac{\langle3456\rangle\langle6123\rangle}
    {\langle6134\rangle\langle5623\rangle}
    \ \bar{Q} \ln \frac{\langle6123\rangle\langle3456\rangle}
            {\langle5613\rangle\langle2346\rangle} \ + (2\ {\rm cyclic}) \,.
\label{QbarY}
\ee
From this form, it is apparent that in $\bar{Q}\hat{\cal R}_{6,1}$
the $\bar{Q} \ln\langle5613\rangle$ term cancels between the $\bar{Q}Y$
contribution and \eqn{QbarR71tree}.

As a result, the residues in $\bar{Q}\hat{\cal R}_{6,1}$
are given by \eqn{R72residues} plus cyclic permutations.
Taking into account the identity~\cite{CaronHuot2011kk}
\be
(6) \ \bar{Q} \ln \frac{\langle1234\rangle}{\langle1235\rangle}\ =\ 0,
\label{Qbarident}
\ee
and all of its permutations, and completing the momentum twistors into
the projectively invariant variables in $\mathcal{S}_u$ in \eqn{uLetters},
one finds that \eqn{R72residues} is equivalent to the following set of final
entries~\cite{SimonSongPrivate}:
\bea
&&(1) \, d\ln(uw/v)\,, \qquad (1) \, d \ln\biggl(\frac{(1-w)u}{y_v\,w(1-u)}\biggr) \,,
\label{dlnfinalentries}\\
&&\Bigl[ (2) + (5) + (3) + (6) \Bigr] d \ln\biggl(\frac{v}{1-v}\biggr)
+ (1) \, d\ln\biggl(\frac{w}{y_u\,(1-w)}\biggr)
+ (4) \, d \ln\biggl(\frac{u}{y_w\,(1-u)}\biggr) \,,
\nonumber
\eea
plus cyclic rotations, for a total of $3\times 6=18$ linear combinations.
This number should be compared with a naive count of $6\times9=54$ possible $R$-invariants times final entries, or $5\times9=45$ independent functions if we take into account the tree identity~(\ref{5bracketidentity}).

Next we impose the $\bar{Q}$ relations~(\ref{dlnfinalentries})
as constraints on the $\{n-1,1\}$ coproducts of the
functions $E$ and $\Et$ defined by \eqn{Eform}.
We do this in the cyclic-vanishing gauge for $\Et$:
\be
\Et(u,v,w) + \Et(v,w,u) + \Et(w,u,v)\ =\ 0.
\label{Etcycvanish}
\ee
We can rewrite the derivatives of this condition in terms of the $\{n-1,1\}$ coproducts
of $\Et$:
\bea
\Et^{u}(u,v,w) + \Et^u(v,w,u) + \Et^u(w,u,v) &=& 0, \label{Etucycvanish}\\
\Et^{1-u}(u,v,w) + \Et^{1-u}(v,w,u) + \Et^{1-u}(w,u,v) &=& 0, \label{Etomucycvanish}\\
\Et^{y_u}(u,v,w) + \Et^{y_u}(v,w,u) + \Et^{y_u}(w,u,v) &=& 0, \label{Etyucycvanish}
\eea
as well as the cyclic images of these equations.

Then the $\bar{Q}$ relations that involve parity-even functions
(except the first, which we group here for convenience) are
\bea
E^{y_u}(u,v,w) &=& E^{y_w}(u,v,w),   \label{Qbareven1}\\
E^{1-v}(u,v,w) &=& 0,               \label{Qbareven2}\\
E^{1-u}(u,v,w) &=& - E^u(u,v,w) - E^v(u,v,w),   \label{Qbareven3}\\
E^{1-u}(u,v,w) + E^{1-w}(u,v,w) &=& E^{1-v}(v,w,u) + E^{1-u}(v,w,u),
\label{Qbareven4}\\
3 \, [ \Et^{y_u}(u,v,w) - \Et^{y_v}(u,v,w) ] &=& 2 \, E^{1-w}(u,v,w) - E^{1-w}(w,u,v),
\label{Qbareven5}
\eea
while the remaining ones, which involve parity-odd functions, are
\bea
3 \, [ \Et^{u}(w,u,v) + \Et^{1-u}(w,u,v) ] &=&
\Et^{v}(u,v,w) + \Et^{w}(v,w,u) - \Et^{v}(w,u,v) - \Et^{w}(w,u,v), \nonumber\\
&~& \label{Qbarodd1}\\
3 \, \Et^{1-u}(u,v,w) &=& \Et^{v}(u,v,w) + \Et^{w}(u,v,w) - \Et^{v}(v,w,u) - \Et^{w}(v,w,u)\nonumber\\
&&\hskip0.2cm\null
- E^{y_u}(u,v,w) + E^{y_v}(u,v,w),  \label{Qbarodd2}\\
2 \, [ E^{y_u}(u,v,w) - E^{y_v}(u,v,w) ] &=&
3 \, [ \Et^{w}(v,w,u) - \Et^{u}(w,u,v) ] + \Et^{v}(v,w,u) - \Et^{v}(w,u,v). \nonumber\\
&~& \label{Qbarodd3}
\eea
All permutations of these equations are implied.
The first three of the above equations do not mix different
permutations of $E$.  They are equivalent to the five-final-entry conditions
found for $U=\ln E$~\cite{Dixon2014iba}.  These relations are also manifest
from the form~(\ref{dlnfinalentries}).

We have used the symmetry relations~(\ref{EEtsym}) in writing
these equations.
Using this symmetry, the arguments of $E$ and $\Et$
can be restricted to $(u,v,w)$, $(v,w,u)$, $(w,u,v)$.
At the outset there are nine final entries, for a total
of $2\times3\times9=54$ independent functions (not counting
how they are related to each other by permutations).
Altogether there are 18 independent even relations and 18 odd relations
(including the cyclic vanishing conditions)
which leads to 9 linearly independent even functions and 9 odd ones.
This agrees with the 18 linear combinations of final entries described in \eqn{dlnfinalentries}.

In practice, we use the $\bar{Q}$ relations to write all of
the other $\{n-1,1\}$ coproducts in terms of just six functions:
$E^{u}(u,v,w)$, $E^{v}(u,v,w)$ (symmetric in $(u\lr w)$),
$E^{y_v}(u,v,w)$ (symmetric in $(u\lr w)$),
$\Et^{u}(u,v,w)$, $\Et^{v}(u,v,w)$ (antisymmetric in $(u\lr w)$),
and $\Et^{y_v}(u,v,w)$ (antisymmetric in $(u\lr w)$).
For these six functions, we insert the most general linear combination of
weight $(2L-1)$ hexagon functions with the right symmetry.
Then we use the $\bar{Q}$ relations to generate the
rest of the $\{n-1,1\}$ coproducts of $E$ and $\Et$,
and also as further constraints on the ansatz.
At the same time, we impose the functional integrability
constraints~(\ref{integrabilityc}), as well as
the branch-cut conditions~(\ref{Fomuvanish}) and (\ref{Fyuvanish}).
Solving all these equations simultaneously leads to
the remaining number of parameters in the line labelled ``$\bar{Q}$ equation''
in \tab{tab:NMHV4_constr_fullQbar}.

We never need to construct the full space
of weight $2L$ functions directly.  The number of
initial parameters is dictated by the number of weight $(2L-1)$
functions.  At four loops, there are 1,801 parity-even weight 7 functions,
and 474 parity-odd weight 7 functions.
We start with 4,550 unknown parameters, from $E^u$ (1,801), $E^v$ (996),
$E^{y_v}$ (272), $\Et^u$ (474), $\Et^v$ (202) and $\Et^{y_v}$ (805).
This is just twice the total number of weight 7 functions.
One implementation of the combined equations gives 28,569
equations for the 4,550 parameters --- obviously with a great deal of redundancy.
This linear system can be solved by {\sc Maple} in under an hour
on a single processor, in terms of just 30 remaining parameters.
(There are four more parameters, corresponding to the weight 8 constants
$\zeta_8$, $\zeta_3 \zeta_5$, $\zeta_2 (\zeta_3)^2$ and $\zeta_{5,3}$.
These parameters are invisible at the level of the $\{7,1\}$ coproducts,
but they are fixed in the next step by the collinear vanishing condition.)

\renewcommand{\arraystretch}{1.25}
\begin{table}[!t]
\centering
\begin{tabular}[t]{l c c c c c c c}
\hline
Constraint & $L=1$ & \multicolumn{2}{c}{$L=2$} 
                   & \multicolumn{2}{c}{$L=3$}
                   & \multicolumn{2}{c}{$L=4$}\\\hline
       & even & even & odd & even & odd & even & odd\\\hline
0. Integrable functions          & 10 & 82 & 6 & 639 & 122 & 5153 & 1763\\\hline
1. (Anti)symmetry in $u$ and $w$  & 7 & 50 & 2 & 363 & 49  & 2797 & 786 \\\hline
2. Cyclic vanishing of $\tilde{V}$ & 7 & 50 & 2 & 363 & 39 & 2797 & 583\\\hline
3. $\bar{Q}$ equation       & 2 & \multicolumn{2}{c}{ 5 }
        & \multicolumn{2}{c}{ 12 } & \multicolumn{2}{c}{ 34 }\\\hline
4. Collinear vanishing             & 0 & \multicolumn{2}{c}{ 0 }
      & \multicolumn{2}{c}{ 1 } & \multicolumn{2}{c}{ 5 }\\\hline
5. Spurious-pole vanishing         & 0 & \multicolumn{2}{c}{ 0 }
      & \multicolumn{2}{c}{ 1 } & \multicolumn{2}{c}{ 5 }\\\hline
6. LL multi-Regge kinematics  & 0 & \multicolumn{2}{c}{ 0 }
      & \multicolumn{2}{c}{ 0 } & \multicolumn{2}{c}{ 1 }\\\hline
7. NLL multi-Regge kinematics  & 0 & \multicolumn{2}{c}{ 0 }
      & \multicolumn{2}{c}{ 0 } & \multicolumn{2}{c}{ 0 }\\\hline
\end{tabular}
\caption{Remaining parameters in the ans\"{a}tze for
$V^{(L)}$ and $\tilde{V}^{(L)}$ after each constraint is applied,
at each loop order.  Here we use the full $\bar{Q}$
equation, which together with symmetry
and functional integrability fixes almost all of the parameters at the
outset.}
\label{tab:NMHV4_constr_fullQbar}
\end{table}

The collinear vanishing condition~(\ref{collvanish}) is simple to implement
and it fixes all of the remaining parameters at one and two loops.
At three and four loops it leaves only one and five parameters, respectively.

It might seem counterintuitive at first sight that the combination of the $\bar{Q}$ and collinear constraints could fix all of the parameters through two loops, because each constraint appears to be homogeneous, i.e.~the right-hand side of the constraint is zero.  A homogeneous constraint should always allow for at least one free parameter, from rescaling any solution by an overall multiplicative constant.  The catch, of course, is that the $\bar{Q}$ constraint is on $E$ and $\Et$, while the collinear constraint is on $V$ and $\Vt$, and these are related to each other inhomogeneously, by a known additive function at a given loop order.  In other words, in terms of $E$ and $\Et$, the collinear vanishing constraint is inhomogeneous.

Next we examine the spurious-pole condition~(\ref{spurious}).  It depends on two variables, $u$ and $v$. We impose it by making use of the function space described in appendix \ref{SP_basis}, for which we have a basis through weight seven.  At four loops, in order to use the weight-seven basis, we first take the derivative of \eqn{spurious} with respect to $u$, using \eqn{uvderiv} to write it in terms of the $\{7,1\}$ coproduct components.  (The condition is antisymmetric in $(u\lr v)$, so it is sufficient to inspect the $u$ derivative.)  However, we find that the full $\bar{Q}$ relations seem to almost completely subsume the spurious-pole condition.  That is, when we impose the spurious-pole condition after the collinear vanishing condition, {\it no} additional parameters are fixed by it, at least through four loops.

In order to see how much the $\bar{Q}$ relations cover the spurious-pole condition, we also tried imposing this condition {\it before} the collinear vanishing condition.  In this case, a few parameters can be fixed, exclusively those that multiply very simple functions in the parity-even part $E$,
of the form
\be
c \, \ln^k(uw/v)
\label{spurambig}
\ee
for odd values of $k$.  Here $c$ is a weight-$(2L-k)$ zeta-value that gives
the correct total weight to the function~(\ref{spurambig}), namely $2L$ at $L$ loops.  It is easy to see that functions of the form~(\ref{spurambig}) cannot be fixed by $\bar{Q}$ for either even or odd $k$.  The only $\bar{Q}$ relation to which these functions contribute at all is \eqn{Qbareven3}, and they cancel trivially between the two terms on the right-hand side, $E^u$ and $E^v$.  For even values of $k$, the functions~(\ref{spurambig}) are still unfixed by the $\bar{Q}$ relations, but they drop out of the spurious-pole condition~(\ref{spurious}), simply because $\ln^k(uw/v) - \ln^k(vw/u) \to 0$ as $w\to1$.

At three and four loops, we need to impose constraints from the multi-Regge limit to fix the final few parameters.  That is the subject of the next section.

Before we appreciated the full power of the $\bar{Q}$ relations, we carried
out a similar analysis, but only imposing the five final-entry condition on $U$
and a seven final-entry condition on $\Vt$.  In order to impose the latter
condition at four loops, we needed to leave the cyclic-vanishing gauge for $\Vt$.  This introduced a number of unphysical, gauge parameters. In \tab{tab:NMHV4_constr_OLD} we tabulate the remaining parameters at different loop orders under these conditions.  It is remarkable how much more power there is in the full $\bar{Q}$ relations, namely the ones that relate $\{n-1,1\}$ coproducts of $E$ and $\Et$ with different permutations.  Whereas in \tab{tab:NMHV4_constr_fullQbar}
there are only 34 parameters left after imposing the $\bar{Q}$ constraint, at the same level in \tab{tab:NMHV4_constr_OLD}, after imposing the 7 final-entry condition on $\Vt$ there are still $487+321=808$ physical parameters!

It is clear that this kind of massive parameter reduction at the outset will make it much more feasible to go to higher loops.  It also drastically reduces the amount of boundary data required. In \tab{tab:NMHV4_constr_OLD} we see that at four loops we needed to use the NNLL multi-Regge information. (Information
at this accuracy is available~\cite{Dixon2014voa,Dixon2014iba} without relying on integrability-based predictions~\cite{BCHS}.) We also needed to use the ${\cal O}(T^1)$ terms in the OPE limit to fix the final two parameters. In contrast, in \tab{tab:NMHV4_constr_fullQbar} all parameters are fixed without any use of the OPE limit, and only the NLL approximation for multi-Regge-kinematics.

\renewcommand{\arraystretch}{1.25}
\begin{table}[!t]
\centering
\begin{tabular}[t]{l c c c c c c c}
\hline
Constraint & $L=1$ & \multicolumn{2}{c}{$L=2$} 
                   & \multicolumn{2}{c}{$L=3$}
                   & \multicolumn{2}{c}{$L=4$}\\\hline
       & even & even & odd & even & odd & even & odd\\\hline
0. Integrable functions          & 10 & 82 & 6 & 639 & 122 & 5153 & 1763\\\hline
1. (Anti)symmetry in $u$ and $w$  & 7 & 50 & 2 & 363 & 39\,+\,10 &
                                             2797    & 583\,+\,203 \\\hline
2. 5 final-entry condition (even only) & 3 & 14 & 2 &  78 & 39\,+\,10 &
                                                487 & 583\,+\,203\\\hline
3. 7 final-entry condition (odd only) & 3 & 14 & 1 &  78 & 21\,+\,3 &
                                                  487 & 321\,+\,64\\\hline
4. Collinear vanishing             & 0 &  2 & 1 &  28 & 21\,+\,3 &
                                                  284 & 321\,+\,64\\\hline
5. $\Ord(T^1)$ 6134 OPE            & 0 &  0 & 1 &   0 & 21\,+\,3 &
                                                  110 & 321\,+\,64\\\hline
6. NNLL multi-Regge kinematics     & 0 &  0 & 0 &   0 & 3\,+\,3 &
                                                    0 & 219\,+\,64\\\hline 
7. Spurious-pole vanishing & 0
& \multicolumn{2}{c}{ 0 } & \multicolumn{2}{c}{ 0\,+\,3 }
                          & \multicolumn{2}{c}{ 2\,+\,64 }\\\hline
8. $\Ord(T^1)$ 1111 OPE & 0 
& \multicolumn{2}{c}{ 0 } & \multicolumn{2}{c}{ 0\,+\,3 }
                          & \multicolumn{2}{c}{ 0\,+\,64 }\\\hline
9. $\Ord(T^{1,2})$ 1114 OPE & 0 
& \multicolumn{2}{c}{ 0 } & \multicolumn{2}{c}{ 0\,+\,3 }
                          & \multicolumn{2}{c}{ 0\,+\,64 }\\
\hline
\end{tabular}
\caption{Remaining parameters in the ans\"{a}tze for $V^{(L)}$ and $\Vt^{(L)}$ after each constraint is applied, at each loop order.  In this version we do not use the full $\bar{Q}$ equation, but only the 5 (7) final-entry condition in the parity even (odd) sector.  The first six constraints do not mix the parity-even and parity-odd function spaces, so we can count the number of even and odd parameters separately until we reach the spurious-pole constraint.  The 7 final-entry condition can only be satisfied if we abandon the cyclic-vanishing condition, which leaves some unphysical ``gauge'' parameters. We split the number of odd parameters into ``physical\,+\,gauge''; only the former number is relevant.}
\label{tab:NMHV4_constr_OLD}
\end{table}


\section{Multi-Regge kinematics}
\label{mrk}

In order to fix the last few parameters at four loops, we analyze the limit of multi-Regge kinematics (MRK) for the NMHV amplitude, following closely ref.~\cite{Dixon2014iba}. The multi-Regge limit in this context refers to $2\rightarrow 4$ scattering, with the four outgoing particles strongly ordered in rapidity. In particular, it involves the all-gluon amplitude, with helicities
\be
3^+ 6^+ \, \to \, 2^+ 4^- 5^+ 1^+ \,,
\label{NMHVconfig}
\ee
where the cross ratios become
\beq
u_1 \to 1\,,\qquad u_2,u_3 \to 0\,,
\eeq
with the ratios
\beq
\frac{u_2}{1-u_1}\equiv \frac{1}{(1+w)\,(1+\ws)} {\rm~~~~and~~~~} 
\frac{u_3}{1-u_1}\equiv \frac{w\ws}{(1+w)\,(1+\ws)}
\label{wdef}
\eeq
held fixed. Here we use $(u_1,u_2,u_3)$ instead of $(u,v,w)$ for the cross ratios, to avoid confusion with the traditional MRK variable $w$.

In ref.~\cite{Dixon2014iba} two of us extended the NMHV leading-logarithmic MRK ansatz of Lipatov, Prygarin, and Schnitzer~\cite{Lipatov2012gk} along the lines of the MHV MRK factorization described by Fadin and Lipatov~\cite{Fadin2011we}.  We proposed the following ansatz:
\be
\bsp
\cP_{\textrm{NMHV}}\times e^{R_6+i\pi\delta}|_{\textrm{MRK}}
\ =\ \cos\pi\omega_{ab} - i  \frac{a}{2} \sum_{n=-\infty}^\infty
(-1)^n\left(\frac{w}{\ws}\right)^{\frac{n}{2}}
&\int_{-\infty}^{+\infty}
\frac{d\nu}{(i\nu+\frac{n}{2})^2}|w|^{2i\nu}
\, \Phi^{\textrm{NMHV}}_{\textrm{Reg}}(\nu,n)\\
&\hskip0.5cm\times\left(-\frac{1}{1-u_1}\frac{|1+w|^2}{|w|}\right)^{\omega(\nu,n)}
\,.
\esp
\label{extMRK}
\ee
where
\be
\bsp
\omega_{ab} &= \frac{1}{8}\,\gamma_K(a)\,\log|w|^2\,,\\
\delta &= \frac{1}{8}\,\gamma_K(a)\,\log\frac{|w|^2}{|1+w|^4}\,,
\esp
\ee
and $\gamma_K(a)$ is the cusp anomalous dimension, given in \eqn{cuspdef}. Here $\omega(\nu,n)$ is known as the BFKL eigenvalue, and is the same for MHV and NMHV, while $\Phi^{\textrm{NMHV}}_{\textrm{Reg}}(\nu,n)$ is the NMHV impact factor. Both may be expanded perturbatively in $a$:
\beq\bsp
\omega(\nu,n) &\,= 
- a \left(E_{\nu,n} + a\,E_{\nu,n}^{(1)}+ a^2\,E_{\nu,n}^{(2)}+\cO(a^3)\right)\,,\\
\Phi^{\textrm{NMHV}}_{\textrm{Reg}}(\nu,n)&\, = 1 + a \, \Phi^{\textrm{NMHV},(1)}_{\textrm{Reg}}(\nu,n)
 + a^2 \, \Phi^{\textrm{NMHV},(2)}_{\textrm{Reg}}(\nu,n)
 + a^3 \, \Phi^{\textrm{NMHV},(3)}_{\textrm{Reg}}(\nu,n)+\cO(a^4)\,.
\label{expandomegaPhi}
\esp\eeq
By expanding \eqn{extMRK} in $a$ and performing the summation and integration, we are left with functions of $w$ and $\ws$ that we can compare to the MRK limit of the ratio function.

The configuration~(\ref{NMHVconfig}) corresponds to the $(\chi_4)^4$ component of the ratio function. Taking the MRK limit of this component, the $R$-invariants reduce to functions of $\ws$:
\be
(1)\rightarrow\frac{1}{1+\ws} \,, \qquad (5)\rightarrow \frac{\ws}{1+\ws} \,,
\qquad (6)\rightarrow  1,
\label{RinvsMRK}
\ee
while the other $R$-invariants vanish.

Parity symmetry of the ratio function leads, in this limit, to a symmetry under $(w,\ws)\rightarrow(1/w,1/\ws)$. Taking advantage of this symmetry, we break up the ratio function as follows:
\bea
\cP_{\textrm{MRK}}^{(L)} &=& 2\pi i \, \sum_{r=0}^{L-1}\ln^r(1-u_1)
\biggl\{
\frac{1}{1+\ws} \Bigl[ p_r^{(L)}(w,\ws) + 2\pi i\, q_r^{(L)}(w,\ws) \Bigr]
\nonumber\\
&&\hskip3.2cm\null +\frac{\ws}{1+\ws}
\Bigl[ p_r^{(L)}(w,\ws) + 2\pi i\, q_r^{(L)}(w,\ws) \Bigr] 
\Big|_{(w,w^*)\rightarrow(1/w,1/w^*)} \biggr\}
\nonumber\\
&&\hskip0.1cm\null + \cO(1-u_1) \,.
\label{NMHVMRKgeneral}
\eea
Here the $p_r^{(L)}(w,\ws)$ and $q_r^{(L)}(w,\ws)$ are composed of functions known as single-valued harmonic polylogarithms (SVHPLs)~\cite{BrownSVHPLs,Dixon2012yy}. In general, $p_{r}^{(L)}$ and $q_{r-1}^{(L)}$ are closely related to each other.  They are determined by the BFKL eigenvalue and impact factor evaluated to the same subleading order in $a$.  Essentially, $q_{r-1}^{(L)}$ is generated by taking the log of $(-1)$ out of the last factor of \eqn{extMRK} instead of a $\ln(1-u_1)$.  For this reason, $q_{L-1}^{(L)}$ vanishes, and we will refer to both $p_{L-1}^{(L)}$ and $q_{L-2}^{(L)}$ as leading-log (LL), $p_{L-2}^{(L)}$ and $q_{L-3}^{(L)}$ as next-to-leading-log (NLL), and so on.

The relations between $p_{r}^{(L)}$ and $q_{r-1}^{(L)}$ that we quote below involve the coefficients appearing in the MRK expansion of the remainder function,
\be
[R_6]_{\textrm{MRK}}^{(L)} = 2\pi i \, \sum_{r=0}^{L-1}\ln^r(1-u_1)
\Bigl[ g_r^{(L)}(w,\ws) + 2\pi i\, h_r^{(L)}(w,\ws) \Bigr] \,,
\label{MHVMRK}
\ee
which can be found through four loops in refs.~\cite{Dixon2012yy,Dixon2014voa}.
They also involve the lower-loop $p_r^{(L)}$ functions, given in
ref.~\cite{Dixon2014iba}.

After imposing collinear vanishing, we fix the five remaining parameters in our four-loop ansatz by matching to the functions $p_r^{(4)}$ and $q_r^{(4)}$. Four of the five parameters are fixed merely by matching to the LL expressions $p^{(4)}_3$ and $q^{(4)}_2$.  We remark that when we perform the same analysis at three loops, there is a single undetermined parameter at this stage, which is fixed by the LL coefficient $p^{(3)}_2$.

At four loops, the one parameter remaining after LL matching is fixed by matching to the NLL coefficients $p^{(4)}_2$ and $q^{(4)}_1$.  The NLL BFKL eigenvalue and NMHV impact factor needed to compute these functions were already fixed at lower loops. The four-loop coefficient functions through NLL are presented below.  We express them in terms of functions $L_{\vec{w}}^\pm$ defined in ref.~\cite{Dixon2012yy}, which are combinations of SVHPLs having definite symmetry properties under complex conjugation ($w\lr \ws$) and inversion ($w\lr1/w$, $\ws\lr1/\ws$):
\bea
q^{(4)}_3 &=& 0 \,, \\
p^{(4)}_3 &=& \frac{1}{768} \Bigl[ - 120 \, L_4^- + 192 \, L_{2,1,1}^-
 - 4 \, ( L_0^- - 20 \, L_1^+ ) \, L_3^+ + 96 \, L_1^+ \, L_{2,1}^-
 + 8 \, (L_2^-)^2 + 8 \, (L_0^-)^2 \, L_2^-
\nonumber\\ &&\hskip0.8cm\null
 - 5 \, (L_0^-)^3 \, L_1^+ - 10 \, (L_0^-)^2 \, (L_1^+)^2
 - 8 \, L_0^- \, (L_1^+)^3 - 16 \, (L_1^+)^4
 + 96 \, \zeta_3 \, L_1^+ \Bigr] \,, \\
q^{(4)}_2 &=& \frac{3}{2} \, p^{(4)}_3 - \frac{1}{2} L_1^+ \, p^{(3)}_2
 - g^{(2)}_1 \, p^{(2)}_1 - g^{(3)}_2 \, p^{(1)}_0 \,, \\
p^{(4)}_2 &=& \frac{1}{64} \biggl\{
 - 87 \, L_5^+ + 14 \, L_{4,1}^- + 32 \, L_{3,1,1}^+ + 8 \, L_{2,2,1}^+
 - 96 \, L_{2,1,1,1}^-
 - \frac{1}{2} \, ( 11 \, L_0^- + 46 \, L_1^+ ) \, L_4^-
\nonumber\\ &&\hskip0.8cm\null
 - ( L_0^- - 4 \, L_1^+ ) \, L_{3,1}^+
 + 12 \, L_0^- \, L_{2,1,1}^-
 + \Bigl[ 12 \, (L_0^-)^2 - 11 \, L_0^- \, L_1^+ + 20 \, (L_1^+)^2 \Bigr] \, L_3^+
\nonumber\\ &&\hskip0.8cm\null
 + 2 \, \Bigl[ (L_0^-)^2 - 2 \, L_0^- \, L_1^+ + 12 \, (L_1^+)^2 \Bigr] \, L_{2,1}^-
\nonumber\\ &&\hskip0.8cm\null
 + \biggl[ \frac{5}{24} \, (L_0^-)^3 + \frac{13}{4} \, (L_0^-)^2 \, L_1^+
   - L_0^- \, (L_1^+)^2 + 4 \, (L_1^+)^3 \biggr] \, L_2^-
 - \frac{13}{240} \, (L_0^-)^5
\nonumber\\ &&\hskip0.8cm\null
 - \frac{11}{8} \, (L_0^-)^4 \, L_1^+
 + \frac{5}{4} \, (L_0^-)^3 \, (L_1^+)^2
 - \frac{7}{3} \, (L_0^-)^2 \, (L_1^+)^3
 - 2 \, L_0^- \, (L_1^+)^4 - \frac{12}{5} \, (L_1^+)^5
\nonumber\\ &&\hskip0.8cm\null
 + \zeta_2 \, \Bigl[ - 48 \, L_3^+ - 48 \, L_{2,1}^- - 24 \, L_1^+ \, L_2^-
               + 3 \, (L_0^-)^3 + 6 \, (L_0^-)^2 \, L_1^+ + 16 \, (L_1^+)^3 \Bigr]
\nonumber\\ &&\hskip0.8cm\null
 + \zeta_3 \, \Bigl[ 2 \, L_2^- + (L_0^-)^2 + 28 \, L_0^- \, L_1^+
 + 8 \, (L_1^+)^2 \Bigr]
 - 102 \, \zeta_5 - 48 \, \zeta_2 \, \zeta_3 \biggr\} \,, \\
q^{(4)}_1 &=& p^{(4)}_2
  - \frac{1}{2} L_1^+ \Bigl[ p^{(3)}_1 - \zeta_2 \, p^{(2)}_1 \Bigr]
  - g^{(2)}_1 \, p^{(2)}_0 - g^{(2)}_0 \, p^{(2)}_1 - g^{(3)}_1 \, p^{(1)}_0 \,.
\label{pqLLNLL}
\eea

Once the final five parameters are fixed, we can obtain the NNLL and N$^3$LL coefficients $p^{(4)}_1$, $q^{(4)}_0$ and $p^{(4)}_0$ with no ambiguity.  We obtain:
\bea
p^{(4)}_1 &=& \frac{1}{64} \biggl\{
  96  \, L_{6}^- + 58  \, L_{5,1}^+ + 16  \, L_{4,2}^+ - 12  \, L_{4,1,1}^-
  - 24  \, L_{3,1,1,1}^+ + 240  \, L_{2,1,1,1,1}^-
\nonumber\\ &&\hskip0.8cm\null
 - \frac{1}{2}  \, ( 3  \, L_{0}^- + 450  \, L_{1}^+ )  \, L_{5}^+
 - ( 9  \, L_{0}^- - 22  \, L_{1}^+ )  \, L_{4,1}^-
 + 4  \, ( L_{0}^- + 5  \, L_{1}^+ )  \, L_{3,1,1}^+
\nonumber\\ &&\hskip0.8cm\null
 + 16  \, L_{1}^+  \, L_{2,2,1}^+
 - 12  \, ( L_{0}^- + 6  \, L_{1}^+ )  \, L_{2,1,1,1}^-
 - \Bigl[ 13  \, (L_{0}^-)^2 + 25  \, L_{0}^-  \, L_{1}^+
         + 16  \, (L_{1}^+)^2 \Bigr] \, L_{4}^-
\nonumber\\ &&\hskip0.8cm\null
 - \Bigl[ 5  \, (L_{0}^-)^2 - 18  \, L_{0}^-  \, L_{1}^+
         - 8  \, (L_{1}^+)^2 \Bigr] \, L_{3,1}^+
 - 4  \, \Bigl[ 2  \, (L_{0}^-)^2 - 3  \, L_{0}^-  \, L_{1}^+
              + 12  \, (L_{1}^+)^2 \Bigr] \, L_{2,1,1}^-
\nonumber\\ &&\hskip0.8cm\null
  + \biggl[ \frac{3}{8}  \, (L_{0}^-)^3
     + \frac{67}{2}  \, (L_{0}^-)^2  \, L_{1}^+
     - 12  \, L_{0}^-  \, (L_{1}^+)^2
     + \frac{71}{3}  \, (L_{1}^+)^3 \biggr]  \, L_{3}^+
\nonumber\\ &&\hskip0.8cm\null
  + \Bigl[ 2  \, (L_{0}^-)^3 - (L_{0}^-)^2  \, L_{1}^+
     + 5  \, L_{0}^-  \, (L_{1}^+)^2 + 14  \, (L_{1}^+)^3 \Bigr] \, L_{2,1}^-
  - 7  \, (L_{3}^+)^2 - 4  \, (L_{2,1}^-)^2
\nonumber\\ &&\hskip0.8cm\null
  + 8  \, L_{2,1,1}^-  \, L_{2}^-
  - \frac{1}{4}  \, \Bigl[ (L_{0}^-)^2 + 12  \, (L_{1}^+)^2 \Bigr] \, (L_{2}^-)^2
\nonumber\\ &&\hskip0.8cm\null
  - \biggl[ 4  \, L_{0}^-  \, L_{3}^+ - \frac{13}{8}  \, (L_{0}^-)^4
    - \frac{25}{6}  \, (L_{0}^-)^3  \, L_{1}^+
    + \frac{1}{2}  \, (L_{0}^-)^2  \, (L_{1}^+)^2
    + L_{0}^-  \, (L_{1}^+)^3 - 8  \, (L_{1}^+)^4 \biggr] \, L_{2}^-
\nonumber\\ &&\hskip0.8cm\null
  - \frac{37}{720}  \, (L_{0}^-)^6 - \frac{1}{48}  \, (L_{0}^-)^5  \, L_{1}^+
  - \frac{97}{24}  \, (L_{0}^-)^4  \, (L_{1}^+)^2
  + 2  \, (L_{0}^-)^3  \, (L_{1}^+)^3
\nonumber\\ &&\hskip0.8cm\null
  - \frac{13}{3}  \, (L_{0}^-)^2  \, (L_{1}^+)^4
  - L_{0}^-  \, (L_{1}^+)^5 - \frac{22}{15}  \, (L_{1}^+)^6
\nonumber\\ &&\hskip0.8cm\null
  + \zeta_2  \, \biggl[ 180  \, L_{4}^- - 8  \, L_{3,1}^+ - 144  \, L_{2,1,1}^- 
  - 4  \, \Bigl[ 8  \, (L_{0}^-)^2 + 3  \, L_{0}^-  \, L_{1}^+
          - 12  \, (L_{1}^+)^2 \Bigr]  \, L_{2}^-
\nonumber\\ &&\hskip1.6cm\null
      - 44  \, ( L_{0}^- - L_{1}^+ )  \, L_{3}^+
      - 4  \, ( L_{0}^- - 6  \, L_{1}^+ )  \, L_{2,1}^- - 4  \, (L_{2}^-)^2
      + \frac{1}{6}  \, (L_{0}^-)^4 + 16  \, (L_{0}^-)^3  \, L_{1}^+
\nonumber\\ &&\hskip1.6cm\null
      - 26  \, (L_{0}^-)^2  \, (L_{1}^+)^2
      - 58  \, L_{0}^-  \, (L_{1}^+)^3 + 108  \, (L_{1}^+)^4 \biggr]
\nonumber\\ &&\hskip0.8cm\null
  + \zeta_3  \, \biggl[ 22  \, L_{3}^+ - 4  \, L_{2,1}^-
      + 4  \, ( 6  \, L_{0}^- - L_{1}^+ )  \, L_{2}^-
      - \frac{5}{3}  \, (L_{0}^-)^3 + 3  \, (L_{0}^-)^2  \, L_{1}^+
\nonumber\\ &&\hskip1.6cm\null
      + 35  \, L_{0}^-  \, (L_{1}^+)^2
              - 10  \, (L_{1}^+)^3 \biggr]
  + \zeta_4  \, \Bigl[ 216  \, L_{2}^-
              + 108  \, ( L_{0}^- - 2  \, L_{1}^+ )  \, L_{1}^+ \Bigr]
\nonumber\\ &&\hskip0.8cm\null
  - \zeta_5  \, ( 21  \, L_{0}^- + 54  \, L_{1}^+ )
  - 4  \, \zeta_2  \, \zeta_3  \, ( 3  \, L_{0}^- - 10 \, L_{1}^+ )
\biggr\} \,,
\label{pNNLL}
\eea
\bea
q^{(4)}_0 &=& \frac{1}{2} \, p^{(4)}_1
- \frac{1}{2} L_1^+
     \biggl[ p^{(3)}_0 - \zeta_2 \, p^{(2)}_0
          + \frac{11}{2} \, \zeta_4 \, p^{(1)}_0 \biggr]
+ \pi^2 \Bigl[ p^{(4)}_3 - g^{(2)}_1 p^{(2)}_1 - 2 \, g^{(3)}_2 p^{(1)}_0 \Bigr]
\nonumber\\
&&\hskip0cm\null
- \pi^2 \, L_1^+ \, \Bigl[ p^{(3)}_2 - 2 \, g^{(2)}_1  p^{(1)}_0 \Bigr]
+ \frac{\pi^2}{2} \, (L_1^+)^2 \, p^{(2)}_1 - \zeta_2 \, (L_1^+)^3 \, p^{(1)}_0
- g^{(2)}_0 p^{(2)}_0 - g^{(3)}_0 p^{(1)}_0 \,,~~~~
\label{qNNLL}
\eea
and
\bea
p^{(4)}_0 &=& \frac{1}{64} \biggl\{
1718 \, L_{7}^+ - 96 \, L_{6,1}^- - 42 \, L_{5,1,1}^+ - 72 \, L_{4,2,1}^+
+ 12 \, L_{4,1,1,1}^- - 24 \, L_{3,3,1}^+ - 8 \, L_{3,1,1,1,1}^+
\nonumber\\ &&\hskip0.8cm\null
- 48 \, L_{2,2,1,1,1}^+ - 16 \, L_{2,1,2,1,1}^+ - 240 \, L_{2,1,1,1,1,1}^-
+ 2 \, ( 43 \, L_{0}^- + 24 \, L_{1}^+ ) \, L_{6}^-
\nonumber\\ &&\hskip0.8cm\null
+ \frac{1}{2} \, ( 3 \, L_{0}^- + 122 \, L_{1}^+ ) \, L_{5,1}^+
+ 16 \, L_{1}^+ \, L_{4,2}^+
+ ( 17 \, L_{0}^- - 6 \, L_{1}^+ ) \, L_{4,1,1}^-
\nonumber\\ &&\hskip0.8cm\null
- 4 \, ( L_{0}^- + 3 \, L_{1}^+ ) \, L_{3,1,1,1}^+
+ 12 \, ( 3 \, L_{0}^- + 10 \, L_{1}^+ ) \, L_{2,1,1,1,1}^-
\nonumber\\ &&\hskip0.8cm\null
- \frac{1}{4} \, \Bigl[ 849 \, (L_0^-)^2 - 132 \, L_{0}^- \, L_{1}^+
         + 552 \, (L_1^+)^2 \Bigr] \, L_{5}^+
\nonumber\\ &&\hskip0.8cm\null
+ \Bigl[ 13 \, (L_0^-)^2 - 19 \, L_{0}^- \, L_{1}^+
       + 8 \, (L_1^+)^2 \Bigr] \, L_{4,1}^-
- \Bigl[ 3 \, (L_0^-)^2 + 16 \, L_{0}^- \, L_{1}^+
       + 12 \, (L_1^+)^2 \Bigr] \, L_{3,1,1}^+
\nonumber\\ &&\hskip0.8cm\null
+ 2 \, \Bigl[ 3 \, (L_0^-)^2 + 4 \, (L_1^+)^2 \Bigr] \, L_{2,2,1}^+
+ 8 \, \Bigl[ (L_0^-)^2 - 4 \, L_{0}^- \, L_{1}^+
            + 3 \, (L_1^+)^2 \Bigr] \, L_{2,1,1,1}^-
\nonumber\\ &&\hskip0.8cm\null
+ 4 \, L_{0}^- \, L_{2,1}^- \, L_{3}^+
+ 2 \, ( 3 \, L_{0}^- \, L_{3,1}^+ + 4 \, L_{1}^+ \, L_{2,1,1}^- ) \, L_{2}^-
\nonumber\\ &&\hskip0.8cm\null
+ \frac{1}{16} \, \Bigl[ 128 \, L_{2,1}^- - 163 \, (L_0^-)^3
         - 118 \, (L_0^-)^2 \, L_{1}^+
         - 332 \, L_{0}^- \, (L_1^+)^2 - 56 \, (L_1^+)^3 \Bigr] \, L_{4}^-
\nonumber\\ &&\hskip0.8cm\null
- \frac{1}{8} \, \Bigl[ 3 \, (L_0^-)^3 + 52 \, (L_0^-)^2 \, L_{1}^+
         - 80 \, L_{0}^- \, (L_1^+)^2 - 24 \, (L_1^+)^3 \Bigr] \, L_{3,1}^+
\nonumber\\ &&\hskip0.8cm\null
- \frac{1}{6} \, \Bigl[ 23 \, (L_0^-)^3 + 18 \, (L_0^-)^2 \, L_{1}^+
         + 18 \, L_{0}^- \, (L_1^+)^2 + 132 \, (L_1^+)^3 \Bigr] \, L_{2,1,1}^-
\nonumber\\ &&\hskip0.8cm\null
+ \frac{1}{48} \Bigl[ 1041 \, (L_0^-)^4 - 312 \, (L_0^-)^3 L_{1}^+
           + 996 \, (L_0^-)^2 (L_1^+)^2 + 16  \, L_{0}^- (L_1^+)^3
           + 496 \, (L_1^+)^4 \Bigr] L_{3}^+
\nonumber\\ &&\hskip0.8cm\null
- \frac{1}{8} \, \Bigl[ 13 \, (L_0^-)^4 - 38 \, (L_0^-)^3 \, L_{1}^+
           - 16 \, (L_0^-)^2 \, (L_1^+)^2
           -  80 \, L_{0}^- \, (L_1^+)^3 + 16 \, (L_1^+)^4 \Bigr] \, L_{2,1}^-
\nonumber\\ &&\hskip0.8cm\null
- \frac{1}{2} \, \Bigl[ 3 \, (L_0^-)^2 \, L_{1}^+
                      + 4 \, (L_1^+)^3 \Bigr] \, (L_2^-)^2
- \frac{1}{8} \, \biggl[ 64 \, L_{4,1}^- + 16 \, L_{0}^- \, L_{1}^+ \, L_{3}^+
          - \frac{43}{5} \, (L_0^-)^5
\nonumber\\ &&\hskip1.6cm\null
         - 8 \, (L_0^-)^4 \, L_{1}^+
          - \frac{97}{3} \, (L_0^-)^3 \, (L_1^+)^2 + 10 \, (L_0^-)^2 \, (L_1^+)^3
    - 4 \, L_{0}^- \, (L_1^+)^4 - 24 \, (L_1^+)^5 \biggr] \, L_{2}^-
\nonumber\\ &&\hskip0.8cm\null
+ \frac{83}{2016} \, (L_0^-)^7 - \frac{1691}{720} \, (L_0^-)^6 \, L_{1}^+
+ \frac{223}{240} \, (L_0^-)^5 \, (L_1^+)^2
- \frac{109}{36} \, (L_0^-)^4 \, (L_1^+)^3
\nonumber\\ &&\hskip0.8cm\null
- \frac{1}{2} \, (L_0^-)^3 \, (L_1^+)^4
- \frac{44}{15} \, (L_0^-)^2 \, (L_1^+)^5
+ \frac{1}{3} \, L_{0}^- \, (L_1^+)^6 - \frac{32}{105} \, (L_1^+)^7
\nonumber
\eea
\bea
&&\hskip0.8cm\null
+ \zeta_2 \, \biggl[ 542 \, L_{5}^+ - 84 \, L_{4,1}^- - 72 \, L_{3,1,1}^+
   - 16 \, L_{2,2,1}^+ + 432 \, L_{2,1,1,1}^-
   + ( 65 \, L_{0}^- + 42 \, L_{1}^+ ) \, L_{4}^-
\nonumber\\ &&\hskip1.6cm\null
   + 4 \, ( 2 \, L_{0}^- + 5 \, L_{1}^+ ) \, L_{3,1}^+
   - 4 \, ( 11 \, L_{0}^- + 54 \, L_{1}^+ ) \, L_{2,1,1}^-
\nonumber\\ &&\hskip1.6cm\null
   - \Bigl[ 81 \, (L_0^-)^2 - 212 \, L_{0}^- \, L_{1}^+
          + 436 \, (L_1^+)^2 \Bigr] \, L_{3}^+
\nonumber\\ &&\hskip1.6cm\null
   + 4 \, \Bigl[ 5 \, (L_0^-)^2 - 6 \, L_{0}^- \, L_{1}^+
               - 6 \, (L_1^+)^2 \Bigr] \, L_{2,1}^-
\nonumber\\ &&\hskip1.6cm\null
   - \Bigl[ 192 \, L_{3}^+ + \frac{49}{6} \, (L_0^-)^3
         - 24 \, (L_0^-)^2 \, L_{1}^+ + 19 \, L_{0}^- \, (L_1^+)^2
         - 62 \, (L_1^+)^3 \Bigr] \, L_{2}^-
\nonumber\\ &&\hskip1.6cm\null
   - \frac{43}{40} \, (L_0^-)^5 + \frac{131}{12} \, (L_0^-)^4 \, L_{1}^+
   - 33 \, (L_0^-)^3 \, (L_1^+)^2
   + \frac{176}{3} \, (L_0^-)^2 \, (L_1^+)^3
\nonumber\\ &&\hskip1.6cm\null
   - 34 \, L_{0}^- \, (L_1^+)^4
   + \frac{344}{5} \, (L_1^+)^5 \biggr]
\nonumber\\
&&\hskip0.8cm\null
+ \zeta_3 \, \biggl[ 4 \, L_{4}^- - 26 \, L_{3,1}^+ + 4 \, L_{2,1,1}^-
   - ( 47 \, L_{0}^- - 70 \, L_{1}^+ ) \, L_{3}^+
   + 4 \, ( 2 \, L_{0}^- + L_{1}^+ ) \, L_{2,1}^-
\nonumber\\ &&\hskip1.6cm\null
   + \frac{1}{4} \, \Bigl[ 7 \, (L_0^-)^2 + 124 \, L_{1}^+ \, L_{0}^-
            - 12 \, (L_1^+)^2 \Bigr] \, L_{2}^-
   - \frac{3}{8} \, (L_0^-)^4 + \frac{119}{12} \, (L_0^-)^3 \, L_{1}^+
\nonumber\\ &&\hskip1.6cm\null
   - \frac{21}{2} \, (L_0^-)^2 \, (L_1^+)^2
   + 3 \, L_{0}^- \, (L_1^+)^3 - \frac{34}{3} \, (L_1^+)^4 \biggr]
\nonumber\\ &&\hskip0.8cm\null
+ \zeta_4 \, \biggl[ 804 \, L_{3}^+ + 504 \, L_{2,1}^-
   + 14 \, ( 7 \, L_{0}^- - 18 \, L_{1}^+ ) \, L_{2}^-
   - 23 \, (L_0^-)^3 - 130 \, (L_0^-)^2 \, L_{1}^+
\nonumber\\ &&\hskip1.6cm\null
   + 108 \, L_{0}^- \, (L_1^+)^2 - 384 \, (L_1^+)^3 \biggr]
\nonumber\\ &&\hskip0.8cm\null
+ \frac{1}{2} \, \zeta_5 \biggl[
                  16 \, L_{2}^- - 125 \, (L_0^-)^2 - 84 \, L_{0}^- \, L_{1}^+
                  + 224 \, (L_1^+)^2 \biggr]
+ 4 \, \zeta_2 \, \zeta_3
    \Bigl[ L_{2}^- - 21 \, L_{0}^- \, L_{1}^+ - 6 \, (L_1^+)^2 \Bigr]
\nonumber\\ &&\hskip0.8cm\null
- 438 \, \zeta_6 \, ( L_{0}^- - 2 \, L_{1}^+ )
- 2 \, (\zeta_3)^2 \, ( 13 \, L_{0}^- + 30 \, L_{1}^+ )
- 720 \, \zeta_7 + 504 \, \zeta_3 \, \zeta_4 + 396 \, \zeta_2 \, \zeta_5
\biggr\} \,. \label{pNNNLL}
\eea

In addition to presenting these functions here in the main text, we also include them, alongside their lower-loop analogues, in computer-readable format in an ancillary file.

These functions are also predicted by the recent all-orders proposal~\cite{BCHS} for the BFKL eigenvalue and impact factor.  In particular, the NNNLL NMHV impact factor $\Phi^{\textrm{NMHV}, (3)}_{\textrm{Reg}}(\nu,n)$ enters the computation of $p^{(4)}_0$.  It can be extracted from the MHV impact factor (computed through NNNLL in ref.~\cite{Dixon2014voa}) and the relation~\cite{BCHS}
\be
\Phi^{\textrm{NMHV}}_{\textrm{Reg}}(\nu,n) = \Phi^{\textrm{MHV}}_{\textrm{Reg}}(\nu,n)
\times \frac{\nu-\frac{in}{2}}{\nu+\frac{in}{2}}
\, \frac{x(u+\frac{in}{2})}{x(u-\frac{in}{2})} \,,
\label{BCHSallorders}
\ee
where
\be
x(u) = \frac{1}{2} \Bigl[ u + \sqrt{ u^2 - 2 a } \Bigr]
\label{Zhukovsky}
\ee
is the Zhukovsky variable.  The rapidity $u$ entering this formula is related to the variable $\nu$ by an integral expression~\cite{BCHS}; its expansion to the relevant order in our notation is
\bea
u &=& \nu - \frac{i}{2} \, a \, V
+ \frac{i}{8} \, a^2 \, V \, ( N^2 + 4 \, \zeta_2 )\nonumber\\
&&\hskip0cm\null
- \frac{i}{32} \, a^3 \biggl\{
  V \, \biggl[ 3 \, N^2 \, V^2 + \frac{5}{4} \, N^4
           - 2 \, \zeta_2 \, ( 4 \, V^2 - N^2 ) + 88 \, \zeta_4 \biggr]
  - 8 \, \zeta_3 \, \Bigl[ -i\partial_\nu E_{\nu,n} \Bigr]
  \biggr\} \nonumber\\
&&\hskip0cm\null
+ {\cal O}(a^4), 
\label{u_nu}
\eea
where $V=i\nu/(\nu^2+n^2/4)$, $N=n/(\nu^2+n^2/4)$, and $E_{\nu,n}$ is the
LL BFKL eigenvalue,
\be
E_{\nu,n} =
  \psi\left(1+i\nu+\frac{|n|}{2}\right)
+ \psi\left(1-i\nu+\frac{|n|}{2}\right) 
- 2\psi(1) - \frac{N}{2} \,.
\label{E_0}
\ee
Expanding \eqn{BCHSallorders} to ${\cal O}(a^3)$, we see that the relation between the NMHV and MHV impact factors becomes non-rational in $\nu$ and $n$ at NNNLL, due to the $\psi$ function appearing in \eqn{u_nu} for $u$, via \eqn{E_0}.

When we compute $p^{(4)}_1$, $q^{(4)}_0$ and $p^{(4)}_0$ from the master formula~(\ref{extMRK}), using \eqn{BCHSallorders} for $\Phi^{\textrm{NMHV}}_{\textrm{Reg}}$, we find precise agreement with the above values extracted from our unique solution.  Given the complexity of eqs.~(\ref{pNNLL}), (\ref{qNNLL}) and (\ref{pNNNLL}), this is already a rather stringent cross-check.


\section{Near-collinear limit}
\label{ope}

By examining the near-collinear limit of the ratio function, we can make contact with the Pentagon OPE approach of Basso, Sever, and Vieira, allowing for a rich array of further cross-checks.  The duality between amplitudes and Wilson loops relates NMHV amplitudes to Wilson loops with states inserted on the boundary, with different choices of states corresponding to different NMHV components~\cite{Mason2010yk,CaronHuot2010ek}.  Through four loops, we have compared our limits with BSV's calculation of the $\chi_6\chi_1\chi_3\chi_4$ and $(\chi_1)^4$ components of the super-Wilson loop~\cite{Basso2013aha}, as well as Belitsky's computation of the $\chi_1^3\chi_4$ component~\cite{BelitskyII} and  $\chi_1^2\chi_4^2$ component~\cite{BelitskyI,BelitskyPrivate}.

To approach the $w\rightarrow 0$ collinear limit, we convert from the cross ratios $(u,v,w)$ to the variables $(F,S,T) \equiv (e^{i\phi},e^\sigma,e^{-\tau})$ used by BSV in ref.~\cite{Basso2013vsa}:
\be
\bsp
u &= \frac{F}{F+FS^2+ST+F^2ST+FT^2} \,,\\
v &= \frac{FS^2}{(1+T^2)(F+FS^2+ST+F^2ST+FT^2)} \,,\\
w &= \frac{T^2}{1+T^2} \,,\\
y_u &= \frac{F+ST+FT^2}{F(1+FST+T^2)} \,,\\
y_v &= \frac{FS+T}{F(S+FT)} \,,\\
y_w &= \frac{(S+FT)(1+FST+T^2)}{(FS+T)(F+ST+FT^2)} \,.
\esp
\label{NMHVBSVparam}
\ee
In these variables, the collinear limit corresponds to $\tau\rightarrow\infty$, or $T\rightarrow 0$.

BSV investigate the $(\chi_1)^4$ component of the NMHV amplitude by inserting a gluonic state on the bottom cusp of the Wilson loop. Up to first order in $T$, the $R$-invariants in this component become
\be
\bsp
&(1)\rightarrow0,\qquad 
(2)\rightarrow \frac{F T}{S(1+S^2)}+\cO\left(T^2\right),\qquad 
(3)\rightarrow 1 - F S T +\cO\left(T^2\right),\\
&(4)\rightarrow 1 - \frac{F T}{S} + \cO\left(T^2\right), \qquad
(5)\rightarrow \frac{F S^3 T}{1+S^2}+\cO\left(T^2\right),\qquad
(6)\rightarrow 0 + \cO(T^4) \,.
\esp
\ee
As in ref.~\cite{Dixon2014iba}, we find that the ratio function in this limit can be expressed as:
\be
\bsp
\cP^{(1111)}\ =&\ \frac{1}{2} \biggl\{
V(u,v,w) + V(w,u,v) - \Vt(u,v,w) + \Vt(w,u,v) \\
&\hskip0.5cm
 + F T \biggl[ - \frac{1-S^2}{S} V(u,v,w) + \frac{1+S^4}{S(1+S^2)} V(v,w,u)
        \biggr] \biggr\}\ +\ \cO(T^2)\,.
\esp
\label{OPE1111}
\ee

We match this expression to BSV's computation of the OPE in this channel~\cite{Basso2013aha}. At order $T^1$ only a single flux-tube excitation contributes; its contribution includes an integration over the excitation's rapidity $u$ and also involves its anomalous dimension (or energy) $\gamma(u)$, its momentum $p(u)$,
a measure factor $\mu(u)$, and the NMHV dressing functions $h$ and $\bar{h}$. Of these, $h$ and $\bar{h}$ can be given in closed form as 
\be
h(u) = \frac{2 x^+(u) x^-(u)}{a} \,, \qquad
\bar{h}(u) = \frac{1}{h(u)} \,,
\label{hhbar}
\ee
where
\be
x^\pm(u) = x(u\pm\tfrac{i}{2})
\label{xpmu}
\ee
is given in terms of the Zhukovsky variable defined in \eqn{Zhukovsky}, while $\gamma(u)$, $p(u)$, and $\mu(u)$ have perturbative expansions described in refs.~\cite{Basso2010in, Basso2013vsa}.

All together, the contribution of one gluonic excitation to the OPE is then
\be
\bsp
\cP^{(1111)} = 1 &
+ T F \int_{-\infty}^\infty 
\frac{du}{2\pi}\mu(u)(h(u)-1)e^{ip(u)\sigma-\gamma(u)\tau} \\
&+ \frac{T}{F} 
\int_{-\infty}^\infty 
\frac{du}{2\pi}\mu(u)(\bar{h}(u)-1)e^{ip(u)\sigma-\gamma(u)\tau} 
\,.
\esp
\label{OPE1111BSV}
\ee

Following ref.~\cite{Dixon2014iba}, we compute this integral as a sum of residues at $u=-im/2$ for positive integers $m$. Truncating the series in $m$ to a few hundred terms, we obtain an expansion in terms of $S=e^\sigma$, which we can then match to the expansion of an ansatz of HPLs in $S^2$. (Other methods for performing these sums are described in ref.~\cite{Papathanasiou2013uoa}.)  This expression in terms of HPLs can be compared with the $\cO(T)$ expansion of the ratio function.  The expansion of the transcendental functions $V$ and $\Vt$ is computed, as in ref.~\cite{Dixon2014iba}, from the differential equations method~\cite{Dixon2013eka}.

The $\chi_6\chi_1\chi_3\chi_4$ component has a simpler OPE at order $T^1$. All of the $R$-invariants vanish except for $(2)$ and $(5)$, which collapse to
\be
(2)\ =\ (5)\ =\ \frac{1}{\langle 6 1 3 4\rangle}
\ =\ \frac{e^{-\tau}}{2\cosh\sigma} \,.
\label{DCI6134}
\ee

Thus only the term multiplying $V(v,w,u)$ survives. This means that through $\cO(T)$ this component is remarkably simple, and is given by the following expression:
\be
\bsp
\mathcal{W}^{(6134)}&= \frac{e^{-\tau}}{2\cosh\sigma}
\sum_{L=0}^\infty \left(\frac{a}{2}\right)^{L} \sum_{n=0}^{L}
\tau^n F^{(L)}_n(\sigma) \,+\,\cO(e^{-2\tau})\\
&= \frac{T}{2\cosh\sigma}
\times V(v,w,u)|_{\cO(T^0)}\ +\ \cO(T^2) \,,
\esp
\label{OPE6134}
\ee
where the $F^{(L)}_n$ are given explicitly through three loops in appendix F of
ref.~\cite{Basso2013aha} and through six loops in ref.~\cite{Papathanasiou2013uoa}.

To check the $\Ord(T^{2})$ terms in the OPE, which receive contributions from two flux-tube excitations, we were assisted by Andrei Belitsky, who checked the $\chi_1^3\chi_4$ component in this limit using our expansions of the $V$ and $\Vt$ functions~\cite{BelitskyII}. For this component, $R$-invariants $(1)$ and $(4)$ vanish, while the behavior of the remaining components was detailed in ref.~\cite{BelitskyII}. In our variables, they behave as follows through $\Ord(T^{2})$:

\be
\bsp
(2)+(5)&=T \, \frac{1-S^2}{1+S^2}F^{1/2}
          - T^2 \, \left(\frac{S-2 S^3-S^5}{(1+S^2)^2}F^{3/2}+\frac{2 S+4 S^3}{(1+S^2)^2}F^{-1/2}\right)+\Ord(T^{3}) \,, \\
(3)+(6)&=(2)-(5)=(3)-(6)=T F^{1/2}-T^2 S F^{3/2}+\Ord(T^{3}) \,.
\esp
\ee
Belitsky has also checked the $\chi_1^2\chi_4^2$ component at $\Ord(T^{2})$ through four loops~\cite{BelitskyI,BelitskyPrivate}.

While the relevant expansions of $V$ and $\tilde{V}$ in the near-collinear limit are too lengthy to include in the text, in an ancillary file we include expressions for $V$ and $\tilde{V}$, as well as their cyclic permutations, expanded through $\Ord(T^{3})$.


\section{Multi-particle factorization}
\label{multiparticle}

In the limit that a three-particle momentum invariant goes on shell, the six-particle amplitude factorizes into a product of two four-particle amplitudes. For MHV amplitudes in supersymmetric theories this factorization is trivial, since at least one of the two resulting four-particle amplitudes is not MHV and thus the product vanishes. In the case of NMHV amplitudes, though, this factorization is nontrivial in some channels. For the limit $K^2=s_{345} \to 0$, where $K=k_3+k_4+k_5$, it behaves as follows~\cite{BernChalmers}:
\be
A_6^{\rm NMHV}(k_i)\ \mathop{\longrightarrow}^{s_{345} \to 0}\
A_4(k_6,k_1,k_2,K) \, \frac{F_6(K^2,s_{i,i+1})}{K^2} \, A_4(-K,k_3,k_4,k_5) \,,
\label{facts345}
\ee
where $F_6$ is the factorization function.

In terms of the cross-ratios, this limit corresponds to letting $u,w\to\infty$, with $u/w$ and $v$ held fixed. For the $R$-invariants, this entails picking out the pole as $s_{345} \to 0$. Only $R$-invariants $(1)$ and $(4)$ have poles in this limit, and their coefficients are equal. From \eqn{PVform}, we see that the factorization limit of the ratio function can be explored by considering the limit of $V(u,v,w)$ as $u,w\to\infty$.

We examined this limit through three loops in ref.~\cite{Dixon2014iba}.  We found that the function $U$ defined in \eqn{Udef}, rather than $V$, has a particularly simple limiting behavior.  In particular, in the factorization limit $U$ becomes a polynomial in $\ln(uw/v)$, with zeta-valued coefficients.  We have applied the same method as in ref.~\cite{Dixon2014iba} to take the limit of $U^{(4)}$, by iteratively working out the limiting behavior of its relevant coproducts, and fixing constants of integration using the line $(u,1,u)$ (see \sect{uu1subsection}).  We find that this simplicity of $U$ continues to be manifest at four loops, and the factorization limit of $U^{(4)}$ is given by:
\be
\bsp
U^{(4)}(u,v,w)|_{u,w\rightarrow\infty}\ =&\ \frac{1}{4}\zeta_4 \ln^4(uw/v)
 - (4\zeta_5 + 3\zeta_2\zeta_3) \ln^3(uw/v)
 + \left( \frac{3769}{32}\zeta_6
        + \frac{21}{4}\zeta_3^2 \right) \ln^2(uw/v) \\
& - \left( \frac{785}{8}\zeta_7
      + \frac{641}{4} \zeta_3 \zeta_4
      + \frac{191}{2} \zeta_2\zeta_5 \right)
   \ln(uw/v) \\
&
 + \frac{133}{4} \zeta_2 \zeta_3^2
 + \frac{289}{4}\zeta_3\zeta _5+ \frac{62629}{64}\zeta_8 \,.
\label{U4}
\esp
\ee
Note that the terms alternate strictly in sign from one power of $\ln(uw/v)$ to the next.  At a given power of $\ln(uw/v)$, they also alternate strictly from one loop order to the next.  The four loop limit~(\ref{U4}), as well as the analogous results from one to three loops~\cite{Dixon2014iba}, are in perfect agreement with a prediction based on integrability~\cite{BSVfact}.

Extracting the factorization function $F_6$ from this expression requires subtracting off the four-point amplitudes $A_4(k_6,k_1,k_2,K)$ and $A_4(-K,k_3,k_4,k_5)$, and adding back in the BDS-like ansatz that was subtracted off when defining $U$. Altogether, this results in the following formula for $F_6$ in terms of $U$ and quantities defined above in \eqns{M6hat}{BDSlikeAnsatz}, as previously presented in ref.~\cite{Dixon2014iba}:
\bea
[\ln F_6]^{(L)} &=& \frac{\gamma_K^{(L)}}{8\e^2L^2}
 \biggl( 1 + 2 \, \e \, L \, \frac{{\cal G}_0^{(L)}}{\gamma_K^{(L)}} \biggr)
  \biggl[ \biggl(\frac{(-s_{12})(-s_{34})}{(-s_{56})}\biggr)^{-L\e}
   + \biggl(\frac{(-s_{45})(-s_{61})}{(-s_{23})}\biggr)^{-L\e} \biggr]
\nonumber\\ &&\null\hskip0.0cm
  - \frac{\gamma_K^{(L)}}{8}\biggl[
     \frac{1}{2} \ln^2 \biggl(\frac{(-s_{12})(-s_{34})}{(-s_{56})}
        \bigg/ \frac{(-s_{45})(-s_{61})}{(-s_{23})}\biggr)
     + 6 \, \zeta_2 \biggr]
\nonumber\\ &&\null\hskip0.0cm
   + U^{(L)}(u,v,w)\bigr|_{u,w\to\infty}
   + \frac{f_2^{(L)}}{L^2} + C^{(L)} + {\cal O}(\e).
\label{lnF6L}
\eea
The limiting behavior of $U$ should also control the multi-particle factorization behavior of higher-point N$^k$MHV amplitudes~\cite{Dixon2014iba}.  It would be interesting to check this behavior once such amplitudes become available (or use this information as an aid in their construction).


\section{Quantitative behavior}
\label{quant}

In this section, we explore the ratio function quantitatively, plotting $V$ and $\Vt$ on a variety of lines and planes through the space of cross ratios.  We stay on the Euclidean branch in the positive octant, $u,v,w>0$, for which all the hexagon functions are real. On certain lines, these functions collapse to sums of well-known functions, such as HPLs. For another line, the diagonal line where $u=v=w$, we have series representations.  For faces of the unit cube, we have constructed the function space in a manner analogous to the full hexagon function construction --- see appendix~\ref{SP_basis} for the case where $w=1$.  We have used these constructions to obtain representations in terms of multiple polylogarithms whose arguments are the cross ratios.  We can then use the program {\sc GiNaC}~\cite{Bauer2000cp, Vollinga2004sn} to evaluate the functions numerically.  There are two other ``bulk'' regions where we have representations in terms of multiple polylogarithms using the $y_i$ variables.  These regions, called Regions I and II in ref.~\cite{Dixon2013eka}, are inside the unit cube and also have $\Delta(u,v,w) > 0$.  Although we won't plot the functions in these bulk regions in this paper, we provide the multiple polylog representations in ancillary files.


\subsection{The point $(1,1,1)$}

The first place we inspect the values of $V$ and $\Vt$ is the point where all the cross ratios are equal to one: $(u,v,w)=(1,1,1)$.  This point is our reference point for defining the constants of integration for all the irreducible hexagon functions:  We define them all to vanish there (except for $\Omega^{(2)}$ which was previously defined as a particular integral).  Also, the point $(1,1,1)$ is on the $\Delta=0$ surface, so all parity-odd hexagon functions (including $\Vt$) vanish there:
\be
\Vt^{(L)}(1,1,1) = 0 \qquad\hbox{for all $L$.}
\label{Vt_111}
\ee
However, $V$ is nonzero at this point.  The constant value of $V$ can be fixed via the collinear limits, or even the soft limits, which correspond to the point $(1,0,0)$, for example.  Then we fix $V$ along the line $(1,v,v)$, using the fact that it can be expressed here in terms of HPLs of the form $H_{\vec{w}}(v)$ with $w_i\in\{0,1\}$, as discussed in \sect{hexfnsubsection}.  Setting $v=1$,
we find that
\be
V^{(4)}(1,1,1) = 3 \, \zeta_2 \, \zeta_3^2 - 15 \, \zeta_3 \, \zeta_5
+ \frac{5051}{12} \, \zeta_8 - 3 \, \zeta_{5,3} \,.
\label{V4_111}
\ee
This value can be compared to previous results at lower loops:
\be
\bsp
V^{(1)}(1,1,1) & = - \zeta_2 \,, \\
V^{(2)}(1,1,1) & = 9 \, \zeta_4 \,, \\
V^{(3)}(1,1,1) &= - \frac{243}{4} \, \zeta_6 \,.
\esp
\label{V1V2V3_111}
\ee
Interestingly, odd zeta values first appear in $V(1,1,1)$ at four loops.  (A $(\zeta_3)^2$ term appears at three loops in $R_6^{(3)}(1,1,1)$ and $E^{(3)}(1,1,1)$, but it cancels in the ratio function.)


\subsection{The lines $(u,u,1)$ and $(u,1,u)$}
\label{uu1subsection}

When two of the cross ratios are equal and the remaining one is equal to unity, the hexagon functions collapse to HPLs.  On these lines, $\Delta=0$, so the parity-odd functions vanish.  For the parity-even functions, $E$ is simpler to express on these lines than $V$, so we present it instead.  Because it is symmetric in exchange of its first and third arguments, it suffices to give it on the lines $(u,u,1)$ and $(u,1,u)$.  We use the notation introduced in ref.~\cite{Dixon2014voa}, in which we expand all products of HPLs using the shuffle algebra in order to linearize the expression in terms of HPLs.  We then encode the HPL weight vectors $\vec w$, which consist entirely of $0$'s and $1$'s, as binary numbers, but written as a subscript in decimal. We track the length of the original weight vector with a superscript.  For example,
\be
H_1^u H_{2,1}^u = H_{1}^u H _{0,1,1}^u
= 3 H_{0,1,1,1}^u + H_{1,0,1,1}^u \to 3 h^{[4]}_7 + h^{[4]}_{11}\, .
\ee

In this notation, the parity-even functions are
\bea
E^{(1)}(u,u,1) &=& - \zeta_2 \,, \label{E1_uu1} \\
E^{(2)}(u,u,1) &=&
\frac{1}{2} \Big[ h^{[4]}_5 + h^{[4]}_{13} - 3 ( h^{[4]}_7 + h^{[4]}_{15} ) \Big]
- \zeta_2 \Big[ h^{[2]}_1 + h^{[2]}_3 \Big] + \frac{13}{2}\zeta_4 \,,
\label{E2_uu1} \\
E^{(3)}(u,u,1) &=& h^{[6]}_{21} + h^{[6]}_{53}
- 4 ( h^{[6]}_{23} + h^{[6]}_{55} ) - 5 ( h^{[6]}_{27} + h^{[6]}_{59} )
- 4 ( h^{[6]}_{29} + h^{[6]}_{61} ) + 10 ( h^{[6]}_{31} + h^{[6]}_{63} )
\nonumber\\
&&\null - \frac{1}{2} \zeta_2 \Big[
  5 ( h^{[4]}_5 + h^{[4]}_{13} ) - 19 ( h^{[4]}_7 + h^{[4]}_{15} ) \Big]
\nonumber\\
&&\null + \frac{21}{2} \zeta_4 \left[ h^{[2]}_1 + h^{[2]}_3 \right]
- \frac{235}{6}\zeta_6 + \zeta_3^2 \,,
\label{E3_uu1}
\eea
\bea
E^{(4)}(u,u,1) &=& \frac{1}{8} \Big[
- 18 ( h_{65}^{[8]} + h_{193}^{[8]} ) - 18 ( h_{67}^{[8]} + h_{195}^{[8]} )
- 18 ( h_{69}^{[8]} + h_{197}^{[8]} ) - 2 ( h_{71}^{[8]} + h_{199}^{[8]} )
\nonumber\\
&&\hskip0.3cm\null
- 18 ( h_{73}^{[8]} + h_{201}^{[8]} ) - 10 ( h_{75}^{[8]} + h_{203}^{[8]} )
- 10 ( h_{77}^{[8]} + h_{205}^{[8]} ) + 14 ( h_{79}^{[8]} + h_{207}^{[8]} )
\nonumber\\
&&\hskip0.3cm\null
- 21 ( h_{81}^{[8]} + h_{209}^{[8]} ) - 13 ( h_{83}^{[8]} + h_{211}^{[8]} )
+ 21 ( h_{85}^{[8]} + h_{213}^{[8]} ) - 107 ( h_{87}^{[8]} + h_{215}^{[8]} )
\nonumber\\
&&\hskip0.3cm\null
- 25 ( h_{89}^{[8]} + h_{217}^{[8]} ) - 161 ( h_{91}^{[8]} + h_{219}^{[8]} )
- 127 ( h_{93}^{[8]} + h_{221}^{[8]} ) + 225 ( h_{95}^{[8]} + h_{223}^{[8]} )
\nonumber\\
&&\hskip0.3cm\null
- 24 ( h_{97}^{[8]} + h_{225}^{[8]} ) - 8 ( h_{99}^{[8]} + h_{227}^{[8]} )
- 16 ( h_{101}^{[8]} + h_{229}^{[8]} ) + 16 ( h_{103}^{[8]} + h_{231}^{[8]} )
\nonumber\\
&&\hskip0.3cm\null
- 28 ( h_{105}^{[8]} + h_{233}^{[8]} ) - 156 ( h_{107}^{[8]} + h_{235}^{[8]} )
- 164 ( h_{109}^{[8]} + h_{237}^{[8]} ) + 348 ( h_{111}^{[8]} + h_{239}^{[8]} )
\nonumber\\
&&\hskip0.3cm\null
+ h_{113}^{[8]} + h_{241}^{[8]} + 25 ( h_{115}^{[8]} + h_{243}^{[8]} )
- 101 ( h_{117}^{[8]} + h_{245}^{[8]} ) + 411 ( h_{119}^{[8]} + h_{247}^{[8]} )
\nonumber\\
&&\hskip0.3cm\null
+ 41 ( h_{121}^{[8]} + h_{249}^{[8]} ) + 393 ( h_{123}^{[8]} + h_{251}^{[8]} )
+ 267 ( h_{125}^{[8]} + h_{253}^{[8]} ) - 525 ( h_{127}^{[8]} + h_{255}^{[8]} ) \Big]
\nonumber\\
&&\hskip0.0cm\null
+ \frac{1}{2} \zeta_2 \Big[ 2 ( h_{17}^{[6]} + h_{49}^{[6]} )
  + 2 ( h_{19}^{[6]} + h_{51}^{[6]} ) - 17 ( h_{21}^{[6]} + h_{53}^{[6]} )
  + 61 ( h_{23}^{[6]} + h_{55}^{[6]} )
\nonumber\\
&&\hskip0.9cm\null
  + 2 ( h_{25}^{[6]} + h_{57}^{[6]} ) + 80 ( h_{27}^{[6]} + h_{59}^{[6]} )
  + 61 ( h_{29}^{[6]} + h_{61}^{[6]} )  - 143 ( h_{31}^{[6]} + h_{63}^{[6]} ) \Big]
\nonumber\\
&&\hskip0.0cm\null
+ \frac{1}{4} \zeta_4 \Big[ 115 ( h_{5}^{[4]} + h_{13}^{[4]} )
                          - 429 ( h_{7}^{[4]} + h_{15}^{[4]} ) \Big]
+ \frac{3}{2} ( 5 \zeta_5 - 2 \zeta_2 \zeta_3 )
  \, ( h_{3}^{[3]} + h_{7}^{[3]} )
\nonumber\\
&&\hskip0.0cm\null
- 70 \zeta_6 ( h_{1}^{[2]} + h_{3}^{[2]} )
+ \frac{1}{2} \zeta_2 \zeta_3^2 - \frac{35}{2} \zeta_3 \zeta_5
+ \frac{36271}{144} \zeta_8 - \frac{3}{2} \zeta_{5,3} \,,
\label{E4_uu1}
\eea
\bea
E^{(1)}(u,1,u) &=& - 2 h^{[2]}_3 - \zeta_2 \,, \label{E1_u1u} \\
E^{(2)}(u,1,u) &=& \frac{1}{2} \Big[ h^{[4]}_5 - 3 h^{[4]}_7
  + 2 h^{[4]}_9 - 2 h^{[4]}_{11} - h^{[4]}_{13} + 15 h^{[4]}_{15} \Big]
\nonumber\\
&& - \zeta_2 \Big[ h^{[2]}_1 - 5 h^{[2]}_3 \Big]
+ \frac{13}{2} \zeta_4 \,, \label{E2_u1u} \\
E^{(3)}(u,1,u) &=& h^{[6]}_{21} - 4 h^{[6]}_{23} - 5 h^{[6]}_{27}
- 4 h^{[6]}_{29} + 10 h^{[6]}_{31} - 3 h^{[6]}_{33} - 2 h^{[6]}_{35} - 2 h^{[6]}_{37}
- 3 h^{[6]}_{39} - 2 h^{[6]}_{41} - 8 h^{[6]}_{43}
\nonumber\\
&&\hskip0.4cm\null
- 8 h^{[6]}_{45} + 8 h^{[6]}_{47} - 2 h^{[6]}_{49} - 3 h^{[6]}_{51}
- 7 h^{[6]}_{53} + 9 h^{[6]}_{55} - 3 h^{[6]}_{57} + 8 h^{[6]}_{59}
+ 4 h^{[6]}_{61} - 40 h^{[6]}_{63}
\nonumber\\
&&\null
- \frac{\zeta_2}{2} \Big[ 5 h^{[4]}_5 - 19 h^{[4]}_7 + 2 h^{[4]}_9
  - 22 h^{[4]}_{11} - 17 h^{[4]}_{13} + 55 h^{[4]}_{15} \Big]
\nonumber\\
&&\null
+ \frac{\zeta_4}{2} \Big[ 21 h^{[2]}_1 - 83 h^{[2]}_3 \Big]
- \frac{235}{6} \zeta_6 + \zeta_3^2 \,, \label{E3_u1u}
\eea
and
\bea
E^{(4)}(u,1,u) &=& \frac{1}{8} \Big[
  - 18 h^{[8]}_{65} - 18 h^{[8]}_{67} - 18 h^{[8]}_{69} - 2 h^{[8]}_{71} - 18 h^{[8]}_{73}
  - 10 h^{[8]}_{75} - 10 h^{[8]}_{77} + 14 h^{[8]}_{79} - 21 h^{[8]}_{81}
\nonumber\\&&\hskip0.4cm\null
  - 13 h^{[8]}_{83} + 21 h^{[8]}_{85} - 107 h^{[8]}_{87} - 25 h^{[8]}_{89}
  - 161 h^{[8]}_{91} - 127 h^{[8]}_{93} + 225 h^{[8]}_{95} - 24 h^{[8]}_{97}
\nonumber\\&&\hskip0.4cm\null
  - 8 h^{[8]}_{99} - 16 h^{[8]}_{101} + 16 h^{[8]}_{103} - 28 h^{[8]}_{105}
  - 156 h^{[8]}_{107} - 164 h^{[8]}_{109} + 348 h^{[8]}_{111} + h^{[8]}_{113}
\nonumber\\&&\hskip0.4cm\null
  + 25 h^{[8]}_{115} - 101 h^{[8]}_{117} + 411 h^{[8]}_{119} + 41 h^{[8]}_{121}
  + 393 h^{[8]}_{123} + 267 h^{[8]}_{125} - 525 h^{[8]}_{127}
\nonumber\\&&\hskip0.4cm\null
  + 120 h^{[8]}_{129} + 96 h^{[8]}_{131} + 88 h^{[8]}_{133} + 96 h^{[8]}_{135}
  + 88 h^{[8]}_{137} + 96 h^{[8]}_{139} + 88 h^{[8]}_{141} + 80 h^{[8]}_{143}
\nonumber\\&&\hskip0.4cm\null
  + 88 h^{[8]}_{145} + 96 h^{[8]}_{147} + 92 h^{[8]}_{149} + 84 h^{[8]}_{151}
  + 88 h^{[8]}_{153} + 80 h^{[8]}_{155} + 76 h^{[8]}_{157} + 100 h^{[8]}_{159}
\nonumber\\&&\hskip0.4cm\null
  + 78 h^{[8]}_{161} + 102 h^{[8]}_{163} + 86 h^{[8]}_{165} + 110 h^{[8]}_{167}
  + 74 h^{[8]}_{169} - 62 h^{[8]}_{171} - 78 h^{[8]}_{173} + 458 h^{[8]}_{175}
\nonumber\\&&\hskip0.4cm\null
  + 106 h^{[8]}_{177} + 130 h^{[8]}_{179} - 42 h^{[8]}_{181} + 654 h^{[8]}_{183}
  + 150 h^{[8]}_{185} + 686 h^{[8]}_{187} + 514 h^{[8]}_{189}
\nonumber\\&&\hskip0.4cm\null
  - 390 h^{[8]}_{191} + 114 h^{[8]}_{193} + 122 h^{[8]}_{195} + 114 h^{[8]}_{197}
  + 106 h^{[8]}_{199} + 114 h^{[8]}_{201} + 106 h^{[8]}_{203}
\nonumber\\&&\hskip0.4cm\null
  + 98 h^{[8]}_{205} + 122 h^{[8]}_{207} + 135 h^{[8]}_{209} + 151 h^{[8]}_{211}
  + 17 h^{[8]}_{213} + 553 h^{[8]}_{215} + 179 h^{[8]}_{217}
\nonumber\\&&\hskip0.4cm\null
  + 715 h^{[8]}_{219} + 581 h^{[8]}_{221} - 475 h^{[8]}_{223} + 126 h^{[8]}_{225}
  + 118 h^{[8]}_{227} + 114 h^{[8]}_{229} + 138 h^{[8]}_{231}
\nonumber\\&&\hskip0.4cm\null
  + 162 h^{[8]}_{233} + 538 h^{[8]}_{235} + 534 h^{[8]}_{237} - 546 h^{[8]}_{239}
  + 95 h^{[8]}_{241} + 119 h^{[8]}_{243} + 365 h^{[8]}_{245}
\nonumber\\&&\hskip0.4cm\null
  - 579 h^{[8]}_{247} + 79 h^{[8]}_{249} - 513 h^{[8]}_{251} - 267 h^{[8]}_{253}
  + 2205 h^{[8]}_{255} \Big]
\nonumber\\
&&\null
+ \frac{\zeta_2}{2} \Big[
  2 h^{[6]}_{17} + 2 h^{[6]}_{19} - 17 h^{[6]}_{21} + 61 h^{[6]}_{23}
  + 2 h^{[6]}_{25} + 80 h^{[6]}_{27} + 61 h^{[6]}_{29} - 143 h^{[6]}_{31}
  + 4 h^{[6]}_{33}
\nonumber\\&&\hskip0.4cm\null
  - 2 h^{[6]}_{35} - 2 h^{[6]}_{37} + 4 h^{[6]}_{39} + 84 h^{[6]}_{43}
  + 84 h^{[6]}_{45} - 180 h^{[6]}_{47} - 2 h^{[6]}_{49} + 4 h^{[6]}_{51}
  + 65 h^{[6]}_{53}
\nonumber\\&&\hskip0.4cm\null
  - 199 h^{[6]}_{55} + 4 h^{[6]}_{57} - 182 h^{[6]}_{59} - 121 h^{[6]}_{61}
  + 383 h^{[6]}_{63} \Big]
\nonumber\\
&&\null
+ \frac{\zeta_4}{4} \Big[
  115 h^{[4]}_5 - 429 h^{[4]}_7 + 20 h^{[4]}_9 - 524 h^{[4]}_{11}
  - 409 h^{[4]}_{13} + 1077 h^{[4]}_{15} \Big]
\nonumber\\
&&\null
+ \frac{1}{2} \left( 15 \zeta_5 - 6 \zeta_2 \zeta_3 \right)
    \Big[ h^{[3]}_3 + h^{[3]}_5 - 3 h^{[3]}_7 \Big]
- \frac{10}{3} \zeta_6 \left[ 21 h^{[2]}_1 - 79 h^{[2]}_3\right]
- 2\zeta_3^2 h^{[2]}_3
\nonumber\\
&&\null
+ \frac{1}{2} \zeta_2 \zeta_3^2 - \frac{35}{2} \zeta_3 \zeta_5
+ \frac{36271}{144} \zeta_8 - \frac{3}{2} \zeta_{5,3}\,.
\label{E4_u1u}
\eea
We provide an ancillary file containing these formulae, as well as the analogous ones for the remainder function.

The subscripts on the $h^{[m]}_i$ in these formulae are always odd, which means that the HPL weight vectors always end in 1.  This restriction enforces the condition that no branch cuts start at $u=1$.  On the line $(u,u,1)$, one can also see that there is a pairing of terms of the form $h^{[m]}_i + h^{[m]}_{i+2^{m-1}}$.  This pairing is due to the coproduct relation $E^u + E^{1-u} + E^v + E^{1-v} = 0$, which holds globally as a consequence of \eqns{Qbareven2}{Qbareven3}. On the line $(u,u,1)$, it implies that the $u$ derivative has the form, $dE(u,u,1)/du = E^u(u,u,1)/[u(1-u)]$, which in turn implies the pairing of HPLs of the form $H_{0,\vec{w}} + H_{1,\vec{w}}$, or equivalently $h^{[m]}_i + h^{[m]}_{i+2^{m-1}}$.

\begin{figure}
\begin{center}
\includegraphics[width=4.5in]{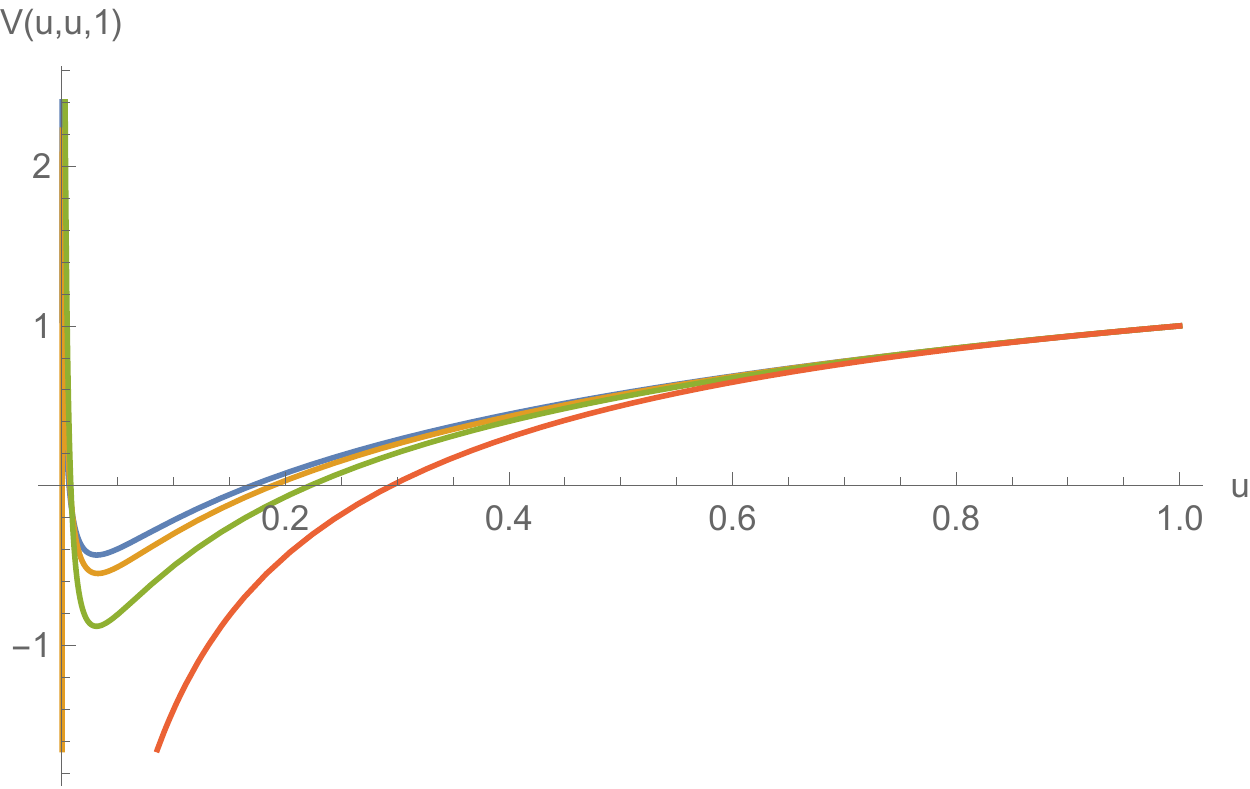}
\end{center}
\caption{$V^{(1)}(u,u,1)$, $V^{(2)}(u,u,1)$, $V^{(3)}(u,u,1)$, and $V^{(4)}(u,u,1)$ normalized to one at $(1,1,1)$. One loop is in red, two loops is in green, three loops is in yellow, and four loops is in blue.}
\label{fig:vuu1}
\end{figure}

\begin{figure}
\begin{center}
\includegraphics[width=4.5in]{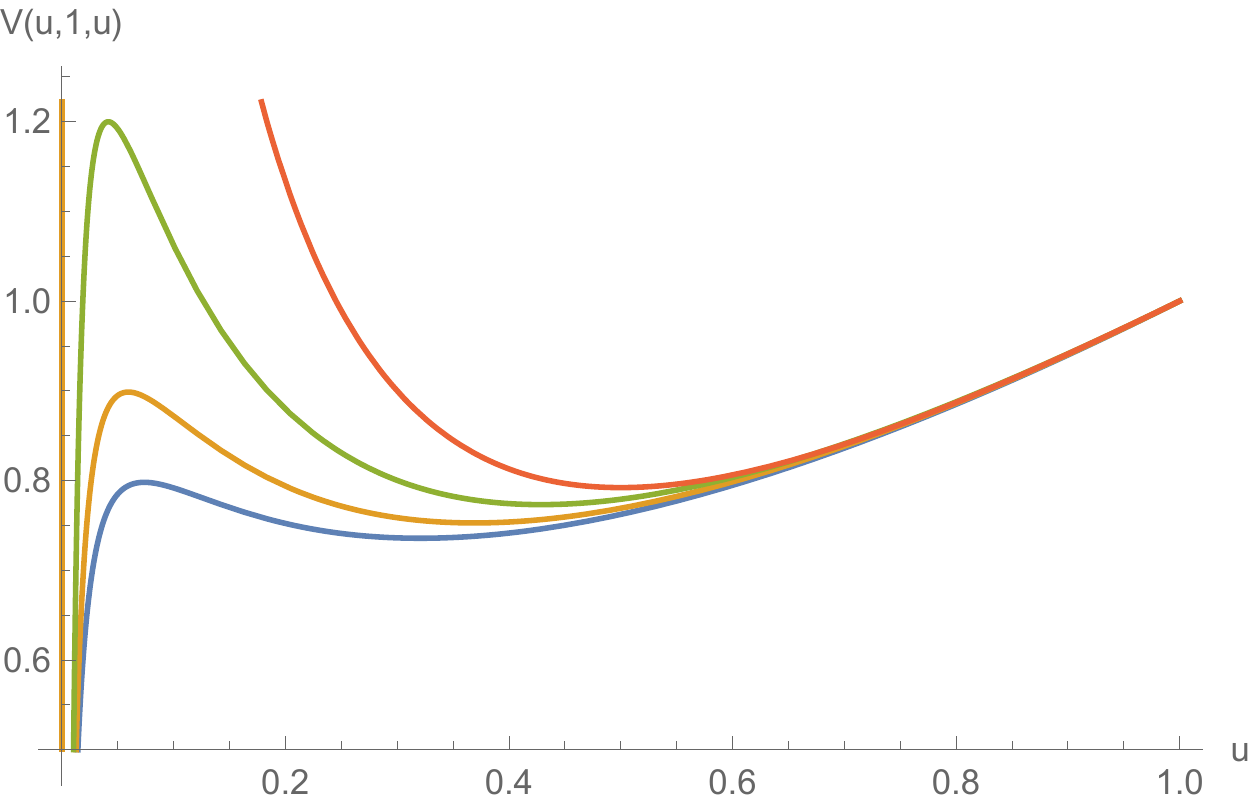}
\end{center}
\caption{$V^{(1)}(u,1,u)$, $V^{(2)}(u,1,u)$, $V^{(3)}(u,1,u)$, and $V^{(4)}(u,1,u)$ normalized to one at $(1,1,1)$. One loop is in red, two loops is in green, three loops is in yellow, and four loops is in blue.}
\label{fig:vu1u}
\end{figure}

We plot the behavior of $V$ on the lines $(u,u,1)$ and $(u,1,u)$ in figures~\ref{fig:vuu1} and~\ref{fig:vu1u}, respectively.  In both cases we plot the functions at each loop order, normalized so that they are all equal to unity at the point $(u,v,w)=(1,1,1)$. While these functions appear to have similar behavior at each loop order away from $u=0$, they do have dramatically varying $u\rightarrow 0$ limits, including oscillations at very small $u$.  In this limit, the curves in figure~\ref{fig:vu1u} approach the negatives of the corresponding curves in figure~\ref{fig:vuu1}.  That is, $V(u,1,u) \approx - V(u,u,1)$ as $u\to0$, which is a consequence of the collinear vanishing constraint~(\ref{collvanish}) if we also let $u\to0$, $v\to1$ in that relation.


\subsection{The lines $(u,1,1)$ and $(1,v,1)$}

The hexagon functions also collapse to the same class of HPLs on the lines where two of the three cross ratios are equal to one.  These lines are not on the $\Delta=0$ surface, so the parity-odd parts of the NMHV amplitude or ratio function do not automatically vanish.  However, on the line $(1,v,1)$, $\Et$ (or $\Vt$) vanishes due to its antisymmetry under $u\lr w$.  This vanishing also means that $\Et(u,1,1)$ is a physical quantity, because it is equal to $\Et(u,1,1) - \Et(1,u,1)$, which is a gauge-invariant difference of cyclic permutations.  Again, we preferentially present $E$ and $\Et$, rather than $V$ and $\Vt$, because they have somewhat simpler expressions.  Using the $u\lr w$ (anti-)symmetry, the functions we need to present are:
\bea
E^{(1)}(u,1,1) &=& - \frac{1}{2} h^{[2]}_3 - \zeta_2 \,, \label{E1_u11} \\
E^{(2)}(u,1,1) &=& \frac{1}{4} \Big[
  h^{[4]}_5 + h^{[4]}_9 +h^{[4]}_{11} + h^{[4]}_{13} + 3 h^{[4]}_{15} \Big]
  - \frac{1}{2} \zeta_2 \Big[ h^{[2]}_1 - 2 h^{[2]}_3 \Big]
  + \frac{13}{2} \zeta_4 \,, \label{E2_u11} \\
E^{(3)}(u,1,1)&=& - \frac{1}{8} \Big[
  - 4 h^{[6]}_{21} + h^{[6]}_{23} + h^{[6]}_{29} + 6 h^{[6]}_{33} + 6 h^{[6]}_{35}
  + 5 h^{[6]}_{37} + 6 h^{[6]}_{39} + 5 h^{[6]}_{41} + 5 h^{[6]}_{43} + 5 h^{[6]}_{45}
\nonumber\\&&\hskip0.4cm\null
  + 6 h^{[6]}_{47}
  + 6 h^{[6]}_{49} + 6 h^{[6]}_{51} + 6 h^{[6]}_{53} + 6 h^{[6]}_{55} + 6 h^{[6]}_{57}
  + 6 h^{[6]}_{59} + 6 h^{[6]}_{61} + 15 h^{[6]}_{63} \Big]
\nonumber\\
&&\null
- \frac{1}{4} \zeta_2 \Big[ 5 h^{[4]}_5 - h^{[4]}_7 + h^{[4]}_9 + h^{[4]}_{11}
  + 9 h^{[4]}_{15} \Big]
+ \frac{1}{4} \zeta_4 \Big[ 21 h^{[2]}_1 - 34 h^{[2]}_3 \Big]
\nonumber\\
&&\null - \frac{235}{6} \zeta_6 + \zeta_3^2 \,, \label{E3_u11}
\eea

\bea
E^{(4)}(u,1,1) &=& \frac{1}{16} \Big[
  - 18 h^{[8]}_{65} - 18 h^{[8]}_{67} - 18 h^{[8]}_{69} - 14 h^{[8]}_{71}
  - 18 h^{[8]}_{73} - 18 h^{[8]}_{75} - 14 h^{[8]}_{77} - 10 h^{[8]}_{79}
\nonumber\\&&\hskip0.4cm\null
  - 24 h^{[8]}_{81} - 24 h^{[8]}_{83} + 5 h^{[8]}_{85} - 20 h^{[8]}_{87}
  - 20 h^{[8]}_{89} - 20 h^{[8]}_{91} - 16 h^{[8]}_{93} - 5 h^{[8]}_{95}
  - 18 h^{[8]}_{97}
\nonumber\\&&\hskip0.4cm\null
  - 18 h^{[8]}_{99} - 18 h^{[8]}_{101} - 14 h^{[8]}_{103} - 24 h^{[8]}_{105}
  - 24 h^{[8]}_{107} - 20 h^{[8]}_{109} - 10 h^{[8]}_{111} - 14 h^{[8]}_{113}
\nonumber\\&&\hskip0.4cm\null
  - 14 h^{[8]}_{115} - 20 h^{[8]}_{117} - 9 h^{[8]}_{119} - 10 h^{[8]}_{121}
  - 10 h^{[8]}_{123} - 5 h^{[8]}_{125} + 60 h^{[8]}_{129} + 60 h^{[8]}_{131}
\nonumber\\&&\hskip0.4cm\null
  + 50 h^{[8]}_{133} + 54 h^{[8]}_{135} + 50 h^{[8]}_{137} + 50 h^{[8]}_{139}
  + 44 h^{[8]}_{141} + 48 h^{[8]}_{143} + 50 h^{[8]}_{145} + 50 h^{[8]}_{147}
\nonumber\\&&\hskip0.4cm\null
  + 46 h^{[8]}_{149} + 47 h^{[8]}_{151} + 44 h^{[8]}_{153} + 44 h^{[8]}_{155}
  + 41 h^{[8]}_{157} + 45 h^{[8]}_{159} + 45 h^{[8]}_{161} + 45 h^{[8]}_{163}
\nonumber\\&&\hskip0.4cm\null
  + 39 h^{[8]}_{165} + 42 h^{[8]}_{167} + 36 h^{[8]}_{169} + 36 h^{[8]}_{171}
  + 33 h^{[8]}_{173} + 39 h^{[8]}_{175} + 39 h^{[8]}_{177} + 39 h^{[8]}_{179}
\nonumber\\&&\hskip0.4cm\null
  + 33 h^{[8]}_{181} + 39 h^{[8]}_{183} + 36 h^{[8]}_{185} + 36 h^{[8]}_{187}
  + 36 h^{[8]}_{189} + 45 h^{[8]}_{191} + 54 h^{[8]}_{193} + 54 h^{[8]}_{195}
\nonumber\\&&\hskip0.4cm\null
  + 44 h^{[8]}_{197} + 50 h^{[8]}_{199} + 44 h^{[8]}_{201} + 44 h^{[8]}_{203}
  + 40 h^{[8]}_{205} + 46 h^{[8]}_{207} + 43 h^{[8]}_{209} + 43 h^{[8]}_{211}
\nonumber\\&&\hskip0.4cm\null
  + 34 h^{[8]}_{213} + 43 h^{[8]}_{215} + 39 h^{[8]}_{217} + 39 h^{[8]}_{219}
  + 39 h^{[8]}_{221} + 45 h^{[8]}_{223} + 48 h^{[8]}_{225} + 48 h^{[8]}_{227}
\nonumber\\&&\hskip0.4cm\null
  + 41 h^{[8]}_{229} + 47 h^{[8]}_{231} + 38 h^{[8]}_{233} + 38 h^{[8]}_{235}
  + 37 h^{[8]}_{237} + 46 h^{[8]}_{239} + 46 h^{[8]}_{241} + 46 h^{[8]}_{243}
\nonumber\\&&\hskip0.4cm\null
  + 40 h^{[8]}_{245} + 46 h^{[8]}_{247} + 45 h^{[8]}_{249} + 45 h^{[8]}_{251}
  + 45 h^{[8]}_{253} + 105 h^{[8]}_{255} \Big]
\nonumber\\
&&\null + \frac{\zeta_2}{8} \Big[
  4 h^{[6]}_{17} + 4 h^{[6]}_{19} - 25 h^{[6]}_{21} + 11 h^{[6]}_{23}
  + 4 h^{[6]}_{25} + 10 h^{[6]}_{27} + 11 h^{[6]}_{29} + 5 h^{[6]}_{31} + 4 h^{[6]}_{33}
\nonumber\\&&\hskip0.4cm\null
  + 4 h^{[6]}_{35} + h^{[6]}_{37} + 4 h^{[6]}_{39} + 3 h^{[6]}_{41} + 6 h^{[6]}_{43}
  + 6 h^{[6]}_{45} + 9 h^{[6]}_{47} + 6 h^{[6]}_{49} + 6 h^{[6]}_{51} + 9 h^{[6]}_{53}
\nonumber\\&&\hskip0.4cm\null
  + 6 h^{[6]}_{55} + 6 h^{[6]}_{57} + 9 h^{[6]}_{59} + 6 h^{[6]}_{61} + 60 h^{[6]}_{63}
  \Big]
\nonumber\\
&&\null + \frac{\zeta_4}{8} \Big[ 115 h^{[4]}_5 - 21 h^{[4]}_7 + 10 h^{[4]}_9
  + 10 h^{[4]}_{11} - 11 h^{[4]}_{13} + 186 h^{[4]}_{15} \Big]
\nonumber\\
&&\null + \frac{3}{8} \left( 5 \zeta_5 - 2 \zeta _2 \zeta_3 \right)
     \Big[ 2 h^{[3]}_3 + h^{[3]}_5 + h^{[3]}_7 \Big]
- \frac{\zeta_6}{24} \left[ 840 h^{[2]}_1 - 1373 h^{[2]}_3\right]
- \frac{1}{2}\zeta _3^2 h^{[2]}_3
\nonumber\\
&&\null + \frac{1}{2} \zeta_2 \zeta_3^2 - \frac{35}{2} \zeta_3 \zeta_5
+ \frac{36271}{144} \zeta_8 - \frac{3}{2} \zeta_{5,3} \,,
\label{E4_u11}
\eea
\bea
E^{(1)}(1,v,1) &=& - \frac{1}{2} h^{[2]}_3 - \zeta_2 \,,
\label{E1_1v1} \\
E^{(2)}(1,v,1) &=& \frac{1}{4} \Big[ h^{[4]}_5 + 3 h^{[4]}_{15} \Big]
- \frac{1}{2} \zeta_2 \Big[ h^{[2]}_1 - 3 h^{[2]}_3 \Big]
+ \frac{13}{2}\zeta_4 \,, \label{E2_1v1} \\
E^{(3)}(1,v,1) &=&
- \frac{1}{8} \Big[ - 4 h^{[6]}_{21} + h^{[6]}_{23} + h^{[6]}_{29} + h^{[6]}_{53}
  + 15 h^{[6]}_{63} \Big]
- \frac{1}{4}\zeta_2 \Big[ 5 h^{[4]}_5 - h^{[4]}_7 - h^{[4]}_{13}
  + 15 h^{[4]}_{15} \Big]
\nonumber\\
&&\null + \frac{1}{4} \zeta_4 \Big[ 21 h^{[2]}_1 - 55 h^{[2]}_3 \Big]
- \frac{235}{6} \zeta_6 + \zeta_3^2 \,,
\label{E3_1v1}
\eea
\bea
E^{(4)}(1,v,1) &=& \frac{1}{16} \Big[
  - 18 h^{[8]}_{65} - 18 h^{[8]}_{67} - 18 h^{[8]}_{69} - 14 h^{[8]}_{71}
  - 18 h^{[8]}_{73} - 18 h^{[8]}_{75} - 14 h^{[8]}_{77} - 10 h^{[8]}_{79}
  - 24 h^{[8]}_{81}
\nonumber\\&&\hskip0.4cm\null
  - 24 h^{[8]}_{83} + 5 h^{[8]}_{85} - 20 h^{[8]}_{87} - 20 h^{[8]}_{89}
  - 20 h^{[8]}_{91} - 16 h^{[8]}_{93} - 5 h^{[8]}_{95} - 18 h^{[8]}_{97}
  - 18 h^{[8]}_{99}
\nonumber\\&&\hskip0.4cm\null
  - 18 h^{[8]}_{101} - 14 h^{[8]}_{103} - 24 h^{[8]}_{105} - 24 h^{[8]}_{107}
  - 20 h^{[8]}_{109} - 10 h^{[8]}_{111} - 14 h^{[8]}_{113} - 14 h^{[8]}_{115}
\nonumber\\&&\hskip0.4cm\null
  - 20 h^{[8]}_{117} - 9 h^{[8]}_{119} - 10 h^{[8]}_{121} - 10 h^{[8]}_{123}
  - 5 h^{[8]}_{125} - 12 h^{[8]}_{161} - 12 h^{[8]}_{163} - 12 h^{[8]}_{165}
\nonumber\\&&\hskip0.4cm\null
  - 10 h^{[8]}_{167} - 18 h^{[8]}_{169} - 18 h^{[8]}_{171} - 16 h^{[8]}_{173}
  - 8 h^{[8]}_{175} - 12 h^{[8]}_{177} - 12 h^{[8]}_{179} - 18 h^{[8]}_{181}
\nonumber\\&&\hskip0.4cm\null
  - 10 h^{[8]}_{183} - 10 h^{[8]}_{185} - 10 h^{[8]}_{187} - 8 h^{[8]}_{189}
  - 8 h^{[8]}_{209} - 8 h^{[8]}_{211} - 14 h^{[8]}_{213} - 5 h^{[8]}_{215}
  - 8 h^{[8]}_{217}
\nonumber\\&&\hskip0.4cm\null
  - 8 h^{[8]}_{219} - 5 h^{[8]}_{221} - 6 h^{[8]}_{233} - 6 h^{[8]}_{235}
  - 6 h^{[8]}_{237} - 3 h^{[8]}_{245} + 105 h^{[8]}_{255} \Big]
\nonumber\\
&&\null + \frac{\zeta_2}{8} \Big[ 4 h^{[6]}_{17} + 4 h^{[6]}_{19}
  - 25 h^{[6]}_{21} + 11 h^{[6]}_{23} + 4 h^{[6]}_{25} + 10 h^{[6]}_{27}
  + 11 h^{[6]}_{29} + 5 h^{[6]}_{31} + 2 h^{[6]}_{41}
\nonumber\\&&\hskip0.4cm\null
  + 8 h^{[6]}_{43} + 8 h^{[6]}_{45} + 8 h^{[6]}_{47} + 9 h^{[6]}_{53}
  + 5 h^{[6]}_{55} + 6 h^{[6]}_{59} + 3 h^{[6]}_{61} + 105 h^{[6]}_{63} \Big]
\nonumber\\
&&\null + \frac{\zeta_4}{8} \Big[ 115 h^{[4]}_5 - 21 h^{[4]}_7 - 21 h^{[4]}_{13}
  + 333 h^{[4]}_{15} \Big]
+ \frac{3}{4} \left( 5 \zeta_5 - 2 \zeta_2 \zeta_3 \right)
    \Big[ h^{[3]}_3 + h^{[3]}_5 + h^{[3]}_7 \Big]
\nonumber\\
&&\null - \frac{\zeta_6}{24} \Big[ 840 h^{[2]}_1 - 2213 h^{[2]}_3 \Big]
- \frac{1}{2}\zeta_3^2 h^{[2]}_3
\nonumber\\
&&\null + \frac{1}{2} \zeta_2 \zeta_3^2 - \frac{35}{2} \zeta_3 \zeta_5
+ \frac{36271}{144} \zeta_8 - \frac{3}{2} \zeta_{5,3}\,,
\label{E4_1v1}
\eea
\bea
\Et^{(2)}(u,1,1) &=& \frac{1}{4} \Big[ h^{[4]}_9 + h^{[4]}_{11} + h^{[4]}_{13} \Big]
  - \frac{1}{2} \zeta_2 h^{[2]}_3 \,,
\label{Et2_u11}\\ 
\Et^{(3)}(u,1,1)&=& - \frac{1}{8} \Big[
  6  ( h^{[6]}_{33} + h^{[6]}_{35} + h^{[6]}_{39} + h^{[6]}_{47}
     + h^{[6]}_{49} + h^{[6]}_{51} + h^{[6]}_{55} + h^{[6]}_{57} + h^{[6]}_{59}
     + h^{[6]}_{61} )
\nonumber\\&&\hskip0.6cm\null
   + 5 ( h^{[6]}_{37} + h^{[6]}_{41} + h^{[6]}_{43} + h^{[6]}_{45} + h^{[6]}_{53} ) \Big]
\nonumber\\&&\hskip0.0cm\null
- \frac{1}{4} \zeta_2 \Big[
   h^{[4]}_{9} + h^{[4]}_{11} + h^{[4]}_{13} - 6 h^{[4]}_{15} \Big]
+ \frac{21}{4} \zeta_4  h^{[2]}_{3} \,,
\label{Et3_u11}
\eea
and
\bea
\Et^{(4)}(u,1,1)&=& \frac{1}{16} \Big[
40 h^{[8]}_{205}
   + 41 ( h^{[8]}_{157} + h^{[8]}_{229} ) + 43 ( h^{[8]}_{237} + h^{[8]}_{245} )
\nonumber\\&&\hskip0.6cm\null
   + 44 ( h^{[8]}_{141} + h^{[8]}_{153} + h^{[8]}_{155} + h^{[8]}_{189} + h^{[8]}_{197} + h^{[8]}_{201}
          + h^{[8]}_{203} + h^{[8]}_{221} + h^{[8]}_{233} + h^{[8]}_{235} )
\nonumber\\&&\hskip0.6cm\null
   + 45 ( h^{[8]}_{159} + h^{[8]}_{191} + h^{[8]}_{223} + h^{[8]}_{249} + h^{[8]}_{251} + h^{[8]}_{253} )
\nonumber\\&&\hskip0.6cm\null
   + 46 ( h^{[8]}_{149} + h^{[8]}_{185} + h^{[8]}_{187} + h^{[8]}_{207} + h^{[8]}_{239}
          + h^{[8]}_{241} + h^{[8]}_{243} + h^{[8]}_{247} )
\nonumber\\&&\hskip0.6cm\null
   + 47 ( h^{[8]}_{151} + h^{[8]}_{175} + h^{[8]}_{217} + h^{[8]}_{219} + h^{[8]}_{231} )
\nonumber\\&&\hskip0.6cm\null
   + 48 ( h^{[8]}_{143} + h^{[8]}_{213} + h^{[8]}_{215} + h^{[8]}_{225} + h^{[8]}_{227} )
   + 49 ( h^{[8]}_{173} + h^{[8]}_{183} )
\nonumber\\&&\hskip0.6cm\null
   + 50 ( h^{[8]}_{133} + h^{[8]}_{137} + h^{[8]}_{139} + h^{[8]}_{145} + h^{[8]}_{147} + h^{[8]}_{199} )
\nonumber\\&&\hskip0.6cm\null
   + 51 ( h^{[8]}_{165} + h^{[8]}_{177} + h^{[8]}_{179} + h^{[8]}_{181} + h^{[8]}_{209} + h^{[8]}_{211} )
   + 52 h^{[8]}_{167}
\nonumber\\&&\hskip0.6cm\null
   + 54 ( h^{[8]}_{135} + h^{[8]}_{169} + h^{[8]}_{171} + h^{[8]}_{193} + h^{[8]}_{195} )
   + 57 ( h^{[8]}_{161} + h^{[8]}_{163} ) + 60 ( h^{[8]}_{129} + h^{[8]}_{131} ) \Big]
\nonumber\\&&\hskip0.0cm\null
+ \frac{1}{8} \zeta_2 \Big[ 4 ( h^{[6]}_{33} + h^{[6]}_{35} + h^{[6]}_{39} )
  + h^{[6]}_{37} + h^{[6]}_{41} + h^{[6]}_{47} + h^{[6]}_{55}
  - 2 ( h^{[6]}_{43} + h^{[6]}_{45} )
\nonumber\\&&\hskip1.3cm\null
  + 6 ( h^{[6]}_{49} + h^{[6]}_{51} + h^{[6]}_{57} )
  + 3 ( h^{[6]}_{59} + h^{[6]}_{61} ) - 45 h^{[6]}_{63} \Big]
\nonumber\\&&\hskip0.0cm\null
+ \frac{1}{8} \zeta_4 \Big[ 10 ( h^{[4]}_{9} + h^{[4]}_{11} + h^{[4]}_{13} ) - 147 h^{[4]}_{15} \Big]
\nonumber\\&&\hskip0.0cm\null
- \frac{3}{8} ( 5 \zeta_5 - 2 \zeta_2 \zeta_3 ) ( h^{[3]}_{5} + h^{[3]}_{7} )
- 35 \zeta_6 h^{[2]}_{3} \,.
\label{Et4_u11}
\eea
We provide these formulae in the same ancillary file that contains the functions' values on the lines $(u,u,1)$ and $(u,1,u)$.

Actually, these functions are not all independent; they obey
\be
\Et(u,1,1) = E(u,1,1) - E(1,u,1).
\label{spur_u11}
\ee
This relation follows from the spurious pole constraint~(\ref{spurious}),
which holds for $E$ and $\Et$ as well as for $V$ and $\Vt$ because $R_6$
and $Y$ are totally symmetric.  However, there is an issue of choosing
the sign for the parity-odd function, or equivalently the choice of
$y_i$ versus $1/y_i$ as one approaches this limit.  If one lets
$u\to1$, $v\to u$, $w\to1$ in \eqn{spurious}, one obtains \eqn{spur_u11}.
On the other hand, if one lets $u\to u$, $v\to 1$, $w\to1$ in \eqn{spurious},
one obtains the same equation but with the opposite sign for $\Et(u,1,1)$.

The functions $\Et^{(L)}(u,1,1)$ have a relatively simple form because
$d\Et(u,1,1)/du$ has the form of $1/u$ times a pure function, with
no $1/(1-u)$ contribution.  Inspecting these terms in the $u$ derivative
in \eqn{dFu}, after taking the limit $(u,v,w)\to(u,1,1)$, we find that the
following linear combination of coproduct entries vanishes:
\be
\Et^{1-u}(u,1,1) + 2 \, \Et^{y_u}(u,1,1) - \Et^{y_v}(u,1,1) - \Et^{y_w}(u,1,1)
= 0.
\label{Etcoprodu11}
\ee
However, we have not yet been able to prove that this combination vanishes to all orders, for example as a consequence of the spurious-pole constraint and the $\bar{Q}$ relations.

\begin{figure}
\begin{center}
\includegraphics[width=4.5in]{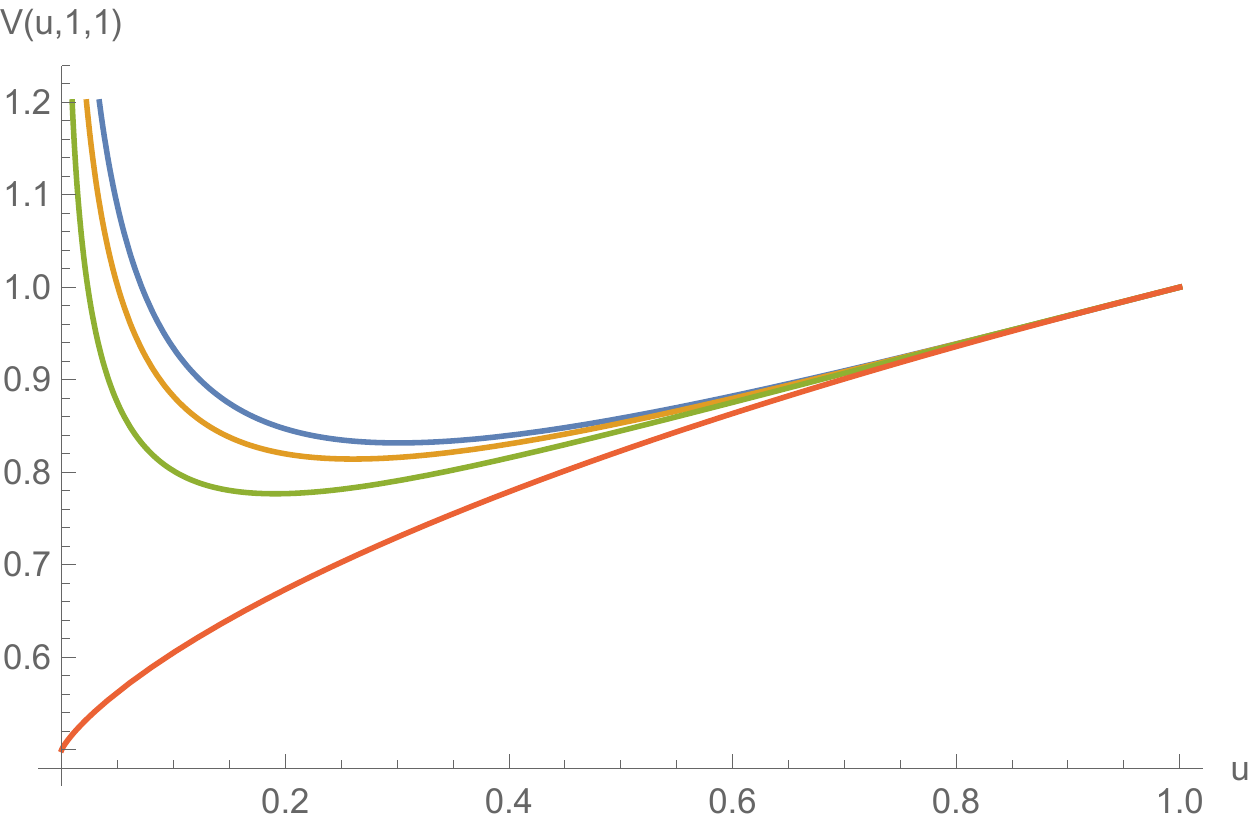}
\end{center}
\caption{$V^{(1)}(u,1,1)$, $V^{(2)}(u,1,1)$, $V^{(3)}(u,1,1)$, and $V^{(4)}(u,1,1)$ normalized to one at $(1,1,1)$. One loop is in red, two loops is in green, three loops is in yellow, and four loops is in blue.}
\label{fig:vu11}
\end{figure}

\begin{figure}
\begin{center}
\includegraphics[width=4.5in]{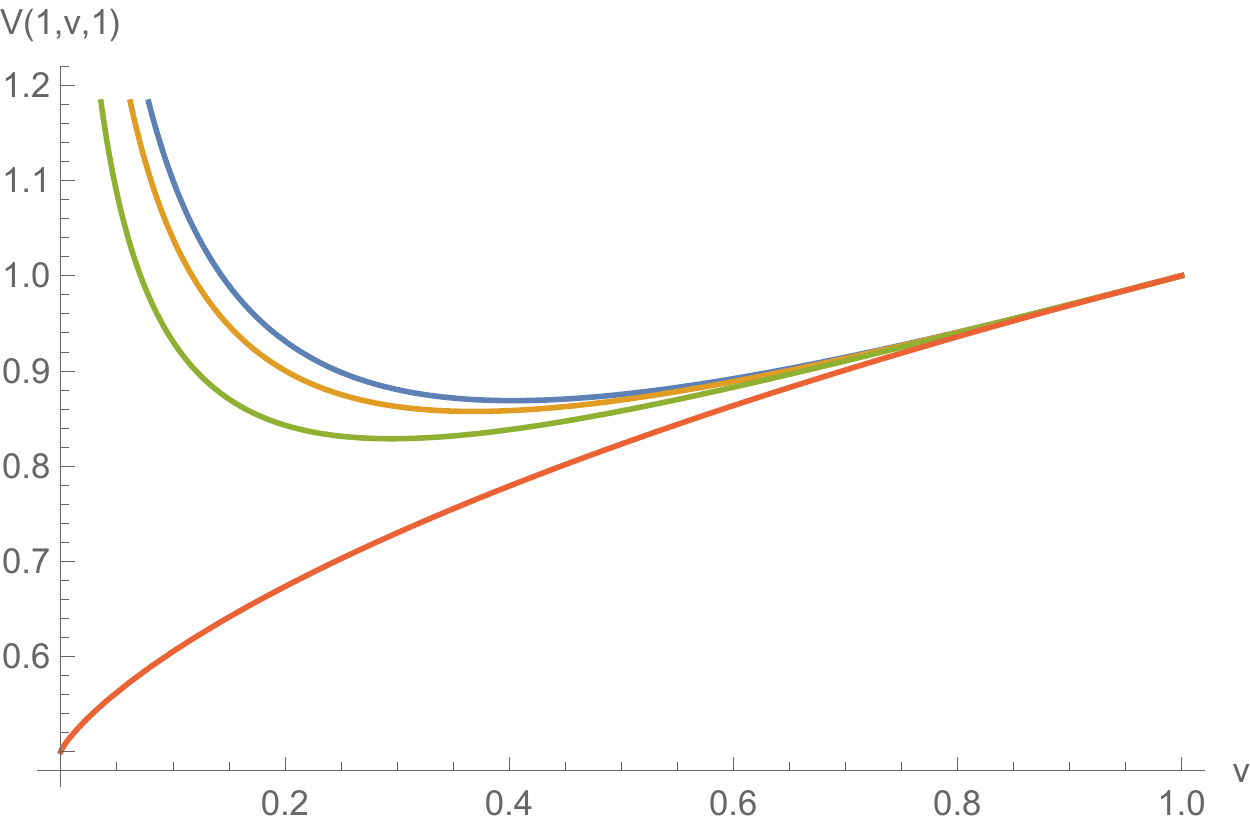}
\end{center}
\caption{$V^{(1)}(1,v,1)$, $V^{(2)}(1,v,1)$, $V^{(3)}(1,v,1)$, and $V^{(4)}(1,v,1)$ normalized to one at $(1,1,1)$. One loop is in red, two loops is in green, three loops is in yellow, and four loops is in blue.}
\label{fig:v1v1}
\end{figure}

Next we plot the functions $V$ and $\Vt$ on the lines $(u,1,1)$ and $(1,v,1)$. For $V(u,1,1)$ and $V(1,v,1)$, shown in figures~\ref{fig:vu11} and \ref{fig:v1v1}, respectively, we again normalize the plots so that each curve takes the value of unity at the point $(u,v,w)=(1,1,1)$. 


\begin{figure}
\begin{center}
\includegraphics[width=4.5in]{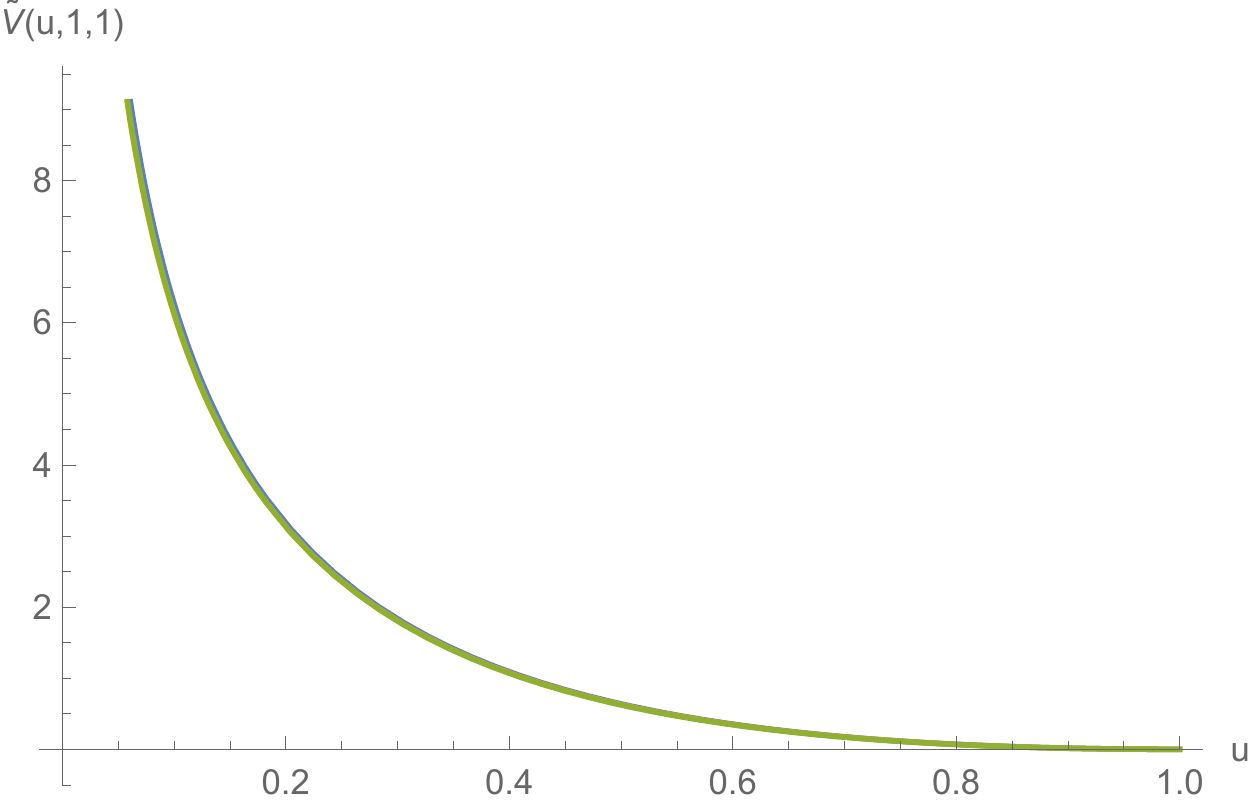}
\end{center}
\caption{$\Vt^{(2)}(u,1,1)$, $\Vt^{(3)}(u,1,1)$ and $\Vt^{(4)}(u,1,1)$ normalized so that the coefficient of the $\ln^2 u$ term in the $u\rightarrow 0$ limit is unity. Two loops is in green, three loops is in yellow, and four loops is in blue.}
\label{fig:vtu11}
\end{figure}

We cannot use this normalization for $\Vt(u,1,1)$, because this function vanishes at the point $(1,1,1)$. Instead, we normalize each loop order so that the coefficient of the $\ln^2 u$ term in the $u\rightarrow 0$ limit is equal to unity. As $u\rightarrow 0$, the functions (before normalization) behave as follows:
\bea
\Vt^{(2)}(u,1,1)|_{u\rightarrow 0}&=& - \frac{1}{8} \zeta_2 \ln^2 u
   - \frac{5}{16}\zeta_4 \,, \\
\Vt^{(3)}(u,1,1)|_{u\rightarrow 0} &=& \frac{47}{32}\zeta_4 \ln^2 u
   + \frac{343}{128} \zeta_6 - \frac{1}{4}\zeta_3^2 \,, \\
\Vt^{(4)}(u,1,1)|_{u\rightarrow 0} &=&
- \frac{13}{512} \zeta_4 \ln^4 u
+ \frac{1}{64} \left( 9 \zeta_5 - 2 \zeta_2 \zeta_3 \right) \ln^3 u
\nonumber\\
&&\null
- \frac{1}{768} \left( 8173 \zeta_6 + 48 \zeta_3^2 \right) \ln^2 u
+ \frac{1}{32} \left( 27 \zeta_2 \zeta_5 - 40 \zeta_3 \zeta_4 \right) \ln u
\nonumber\\
&&\null
- \frac{3}{8} \zeta_2 \zeta_3^2 + \frac{73}{16}\zeta_3 \zeta_5
- \frac{52217}{2560} \zeta_8 + \frac{33}{80} \zeta_{5,3} \,.
\eea
Note that when we use the normalization based on the $\ln^2 u$ coefficient, all three functions in figure~\ref{fig:vtu11} look almost identical!  This is quite surprising, because $\Vt^{(4)}(u,1,1)$ actually diverges like $\ln^4 u$ as $u\rightarrow 0$, while the lower-loop functions only diverge like $\ln^2 u$.  The coefficient in front of the $\ln^4 u$ divergence is apparently small enough that it does little to change the shape of $\Vt^{(4)}(u,1,1)$ over a large region of the $u$ line.

\subsection{The line $(u,u,u)$}

\begin{figure}
\begin{center}
\includegraphics[width=4.5in]{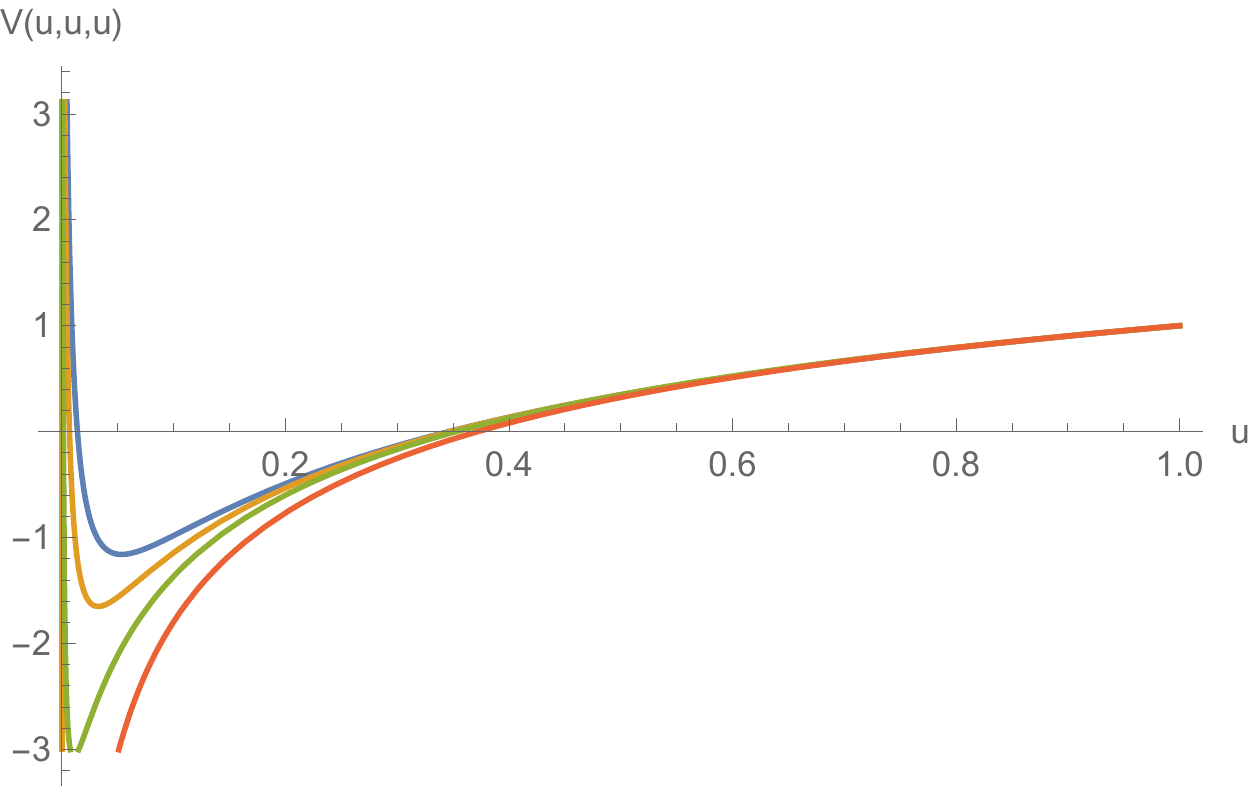}
\end{center}
\caption{$V^{(1)}(u,u,u)$, $V^{(2)}(u,u,u)$, $V^{(3)}(u,u,u)$, and $V^{(4)}(u,u,u)$ normalized to one at $(1,1,1)$. One loop is in red, two loops is in green, three loops is in yellow, and four loops is in blue.}
\label{fig:vuuu}
\end{figure}

Unlike the lines discussed above, the hexagon functions do not collapse to HPLs on the line where all of the cross ratios are equal.  Instead they become cyclotomic polylogarithms~\cite{Ablinger2011te}.  Using the differential equations that they obey, it is relatively straightforward to evaluate these functions in terms of series expansions, either around $u=0$, $u=1$ or $u=\infty$.  For the part of the line where $u<1/4$, we have an alternate representation of $V$ in terms of multiple polylogarithms.  That is because $\Delta(u,u,u)=(1-u)^2(1-4u)$ is positive for $u<1/4$, and this segment lies in the Region I defined in ref.~\cite{Dixon2013eka}.  On the whole line $(u,u,u)$, $\Vt$ vanishes by symmetry.  We plot $V(u,u,u)$ in figure~\ref{fig:vuuu}, normalized so that at each loop order the function has the value unity at the point $(1,1,1)$.

Among other uses, this line allows us to identify a place where the ratio function crosses zero, which is fairly stable with respect to the loop order. For each $L$, $V^{(L)}(u,u,u)$ crosses zero near $u=1/3$, although the exact point shifts slightly with the loop order. Denoting by $u_0^{(L)}$ the value of $u$ for which $V^{(L)}(u,u,u)$ equals zero, we have for this zero crossing,
\be
u_0^{(1)} = 0.372098\ldots, \quad
u_0^{(2)} = 0.352838\ldots, \quad
u_0^{(3)} = 0.347814\ldots, \quad
u_0^{(4)} = 0.346013\ldots. \quad
\label{zerocrossing}
\ee

The functions $V^{(L)}(u,u,u)$ oscillate as $u\rightarrow 0$, leading to additional zero crossings near the origin. In particular, $V^{(2)}(u,u,u)$ has a zero crossing near $0.0015$, while $V^{(3)}(u,u,u)$ crosses near $0.007$ and again near $1.3\times 10^{-6}$. $V^{(4)}(u,u,u)$ has three additional zero crossings, at roughly $0.014$, $0.000025$, and $7.2\times 10^{-10}$. This can be seen from the small-$u$ limits of these functions:
\bea
V^{(1)}(u,u,u) &\sim& \frac{1}{2} \, \ln^2 u 
+ \frac{1}{2} \, \zeta_2 \,, \label{V1_uuu_u0} \\
V^{(2)}(u,u,u) &\sim& \frac{1}{16} \, \ln^4 u
- \frac{3}{2} \, \zeta_2 \, \ln^2 u + \frac{1}{2} \, \zeta_3 \, \ln u
- \frac{53}{16} \, \zeta_4 \,, \label{V2_uuu_u0} \\
V^{(3)}(u,u,u) &\sim& \frac{1}{288} \, \ln^6 u
- \frac{41}{96} \, \zeta_2 \, \ln^4 u
+ \frac{1}{8} \, \zeta_3 \, \ln^3 u
+ \frac{419}{32} \, \zeta_4 \, \ln^2 u
- \Bigl( 2 \, \zeta_5 + \frac{3}{4} \, \zeta_2 \, \zeta_3 \Bigr) \, \ln u
\nonumber\\&&\null
+ \frac{2589}{128} \, \zeta_6 - \frac{1}{4} \, (\zeta_3)^2 \,,
\label{V3_uuu_u0}\\
V^{(4)}(u,u,u) &\sim& \frac{1}{9216}\,\ln^8 u
- \frac{43}{1152} \, \zeta_2 \, \ln^6 u
+ \frac{1}{96} \, \zeta_3 \, \ln^5 u
+ \frac{557}{96} \, \zeta_4 \,\ln^4 u
\nonumber\\&&\null
- \frac{1}{48} \Big( 23 \, \zeta_5 + 32 \, \zeta_2 \zeta_3 \Big) \, \ln^3 u 
- \frac{1}{256} \Big( 21971 \, \zeta_6 - 8 \, \zeta_3^2 \Big) \, \ln^2 u
\nonumber\\&&\null
+ \frac{1}{32} \Big( 300 \, \zeta_7 + 108 \, \zeta_2 \zeta_5
                   + 121 \,\zeta_3 \zeta_4 \Big) \, \ln u
\nonumber\\&&\null
- \frac{131867}{1024} \, \zeta_8
+ \frac{3}{8} \,\zeta_2 \zeta_3^2
+ \frac{11}{4}\,\zeta_3 \zeta_5 \,.
\label{V4_uuu_u0}
\eea
We note that the multiple zeta value $\zeta_{5,3}$ does not appear in this particular limit of the four-loop ratio function; nor did it appear in the same limit of the remainder function~\cite{Dixon2014voa}.  Its absence could be a hint that there might be a relatively simple description of this limit.


\subsection{Faces of the unit cube}
\label{cubefaces}

We can also examine $V$ and $\Vt$ on the faces of the unit cube in cross-ratio space. Here the functions do not collapse to HPLs, but they do still reduce to simpler bases of functions which can be readily treated numerically. There are two cases to consider: planes where one of the cross ratios goes to one, and planes where one of the cross ratios vanishes. We will consider each in turn.

First, we consider the plane where one of the cross ratios goes to one. For concreteness, we choose $w\rightarrow 1$, so the surface is $(u,v,1)$. This limit was discussed in section \ref{setupoverview}, where it was used to ensure the vanishing of spurious poles. Recall that in this limit, our symbol entries behave as follows:
\be
w\rightarrow 1 \,,
\quad y_u\rightarrow(1-w)\frac{u(1-v)}{(u-v)^2} \,,
\quad y_v\rightarrow\frac{1}{(1-w)}\frac{(u-v)^2}{v(1-u)} \,,
\quad y_w\rightarrow\frac{1-u}{1-v} \,.
\label{wto1yvars}
\ee
Thus in this limit our set of nine symbol letters reduces to the following five:
\be
\mathcal{S}_{w\rightarrow 1}\ =\ \{ u,\, v,\, 1-u,\, 1-v,\, u-v \} \,.
\ee

We cannot represent this function space solely with one-dimensional HPLs ($H_{\vec{w}}(u)$ and $H_{\vec{w}}(v)$ with $\vec{w}\in\{0,1\}$), due to the $u-v$ entry. However, it is relatively straightforward to express any function with these symbol letters in terms of Goncharov polylogarithms, which in turn can be evaluated numerically with {\sc GiNaC}~\cite{Bauer2000cp, Vollinga2004sn}.  (We could have used instead the 2dHPL functions introduced by Gehrmann and Remiddi~\cite{GehrmannRemiddi}.)

For $V(u,v,w)$, there are two distinct cases to consider. We can either let $v\rightarrow 1$, or let $w\rightarrow 1$. The $u\rightarrow 1$ case is related to the $w\rightarrow 1$ case by $u\leftrightarrow w$ symmetry.

\begin{figure}
\begin{center}
\includegraphics[width=4.5in]{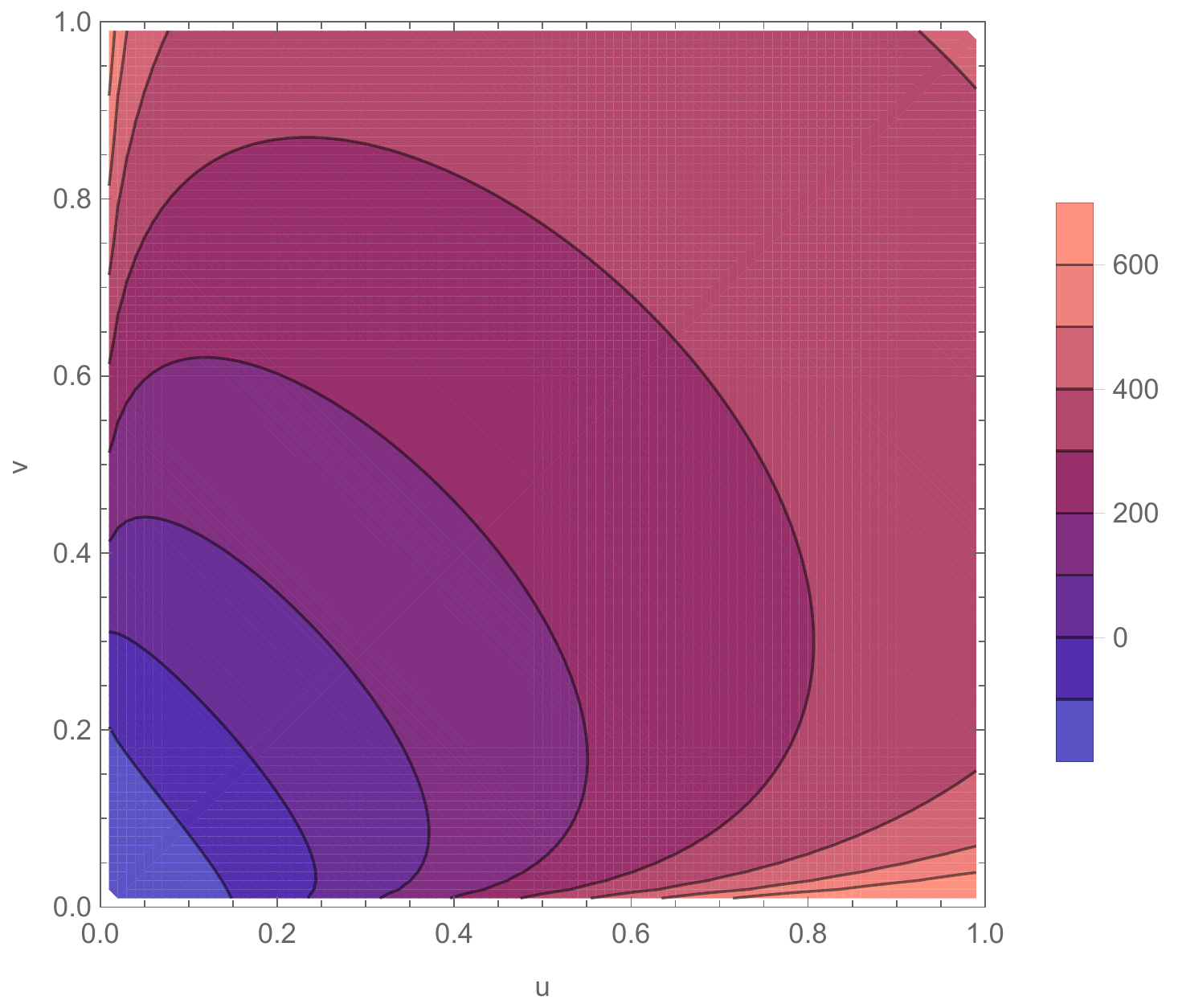}
\end{center}
\caption{$V^{(4)}(u,v,1)$ plotted in $u$ and $v$.}
\label{fig:vuv1}
\end{figure}

For the $w\rightarrow 1$ surface we find relatively simple behavior, shown in figure~\ref{fig:vuv1}. The function $V^{(4)}(u,v,1)$ is approximately symmetric under $u\lr v$.  It crosses zero around the line $u+v=0.3$, and increases as $u$ and $v$ increase. Since the function crosses zero on this surface, plotting ratios between $V$ at different loop orders is not especially informative, so here we plot only $V^{(4)}(u,v,1)$.

\begin{figure}
\begin{center}
\includegraphics[width=4.5in]{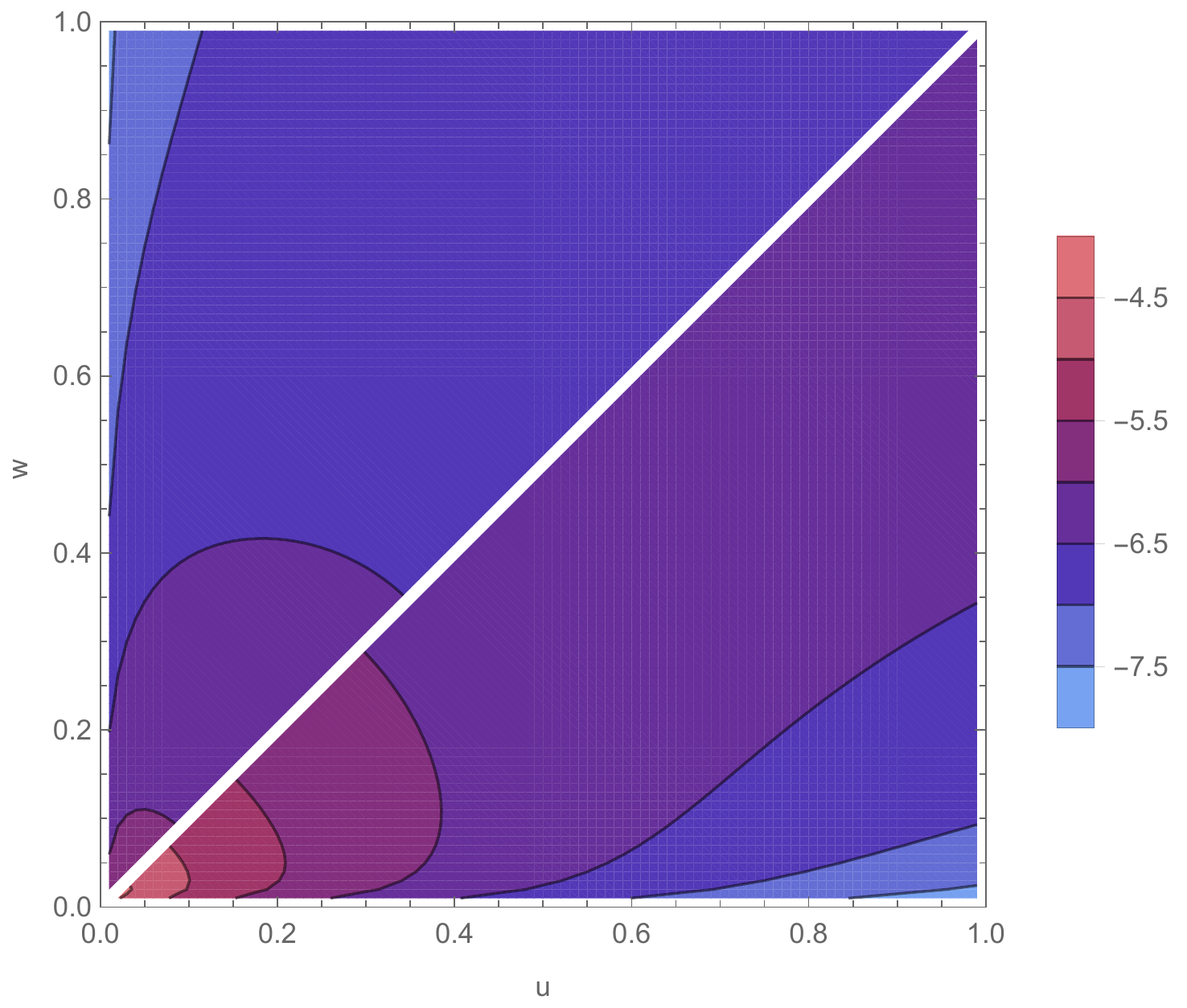}
\end{center}
\caption{Ratios of $V(u,1,w)$ between successive loop orders, plotted in $u$ and $w$. $V^{(4)}/V^{(3)}$ is in the top-left corner, while $V^{(3)}/V^{(2)}$ is in the bottom-right.}
\label{fig:vu1w}
\end{figure}

If we instead take $v\rightarrow 1$, the function $V(u,1,w)$ is exactly symmetric under exchange of $u$ and $w$. It also has uniform sign. Taking advantage of both of these properties, we show in figure~\ref{fig:vu1w} the ratios of $V^{(4)}(u,1,w)$ to $V^{(3)}(u,1,w)$ and $V^{(3)}(u,1,w)$ to $V^{(2)}(u,1,w)$ on the same plot. Here $V^{(4)}(u,1,w)/V^{(3)}(u,1,w)$ is plotted in the top-left corner, while $V^{(3)}(u,1,w)/V^{(2)}(u,1,w)$ is in the bottom-right. In both cases, the missing part of the plot is just the mirror image, due to $u\leftrightarrow w$ symmetry. We find that these inter-loop ratios are quite heavily constrained, staying between $-4$ and $-8$. Note in particular that $V^{(4)}(u,1,w)/V^{(3)}(u,1,w)$ is significantly flatter than $V^{(3)}(u,1,w)/V^{(2)}(u,1,w)$. This is encouraging; we expect the ratios to continue to become more constrained at higher loops due to the finite radius of convergence of the perturbative expansion.  In non-singular regions, we expect the inter-loop ratios to approach $-8$ at very large loop order~\cite{Dixon2014voa}.

We can also look at the parity-odd functions on this plane. Here, we need to make a choice to avoid ambiguity. As discussed in section \ref{setupoverview}, $\Vt$ has a ``gauge'' redundancy: we can add an arbitrary totally antisymmetric function to it without affecting the full ratio function. This ambiguity will have to be dealt with in order to present numerical results. Rather than fixing it in some arbitrary way, here we avoid the ambiguity altogether by taking differences of cyclic permutations of $\Vt(u,v,w)$.  Totally antisymmetric functions are cyclicly symmetric, so their contribution will cancel in these differences.  The full ratio function can be expressed only in terms of the cyclic differences, with no independent appearance of $\Vt$.

\begin{figure}
\begin{center}
\includegraphics[width=4.5in]{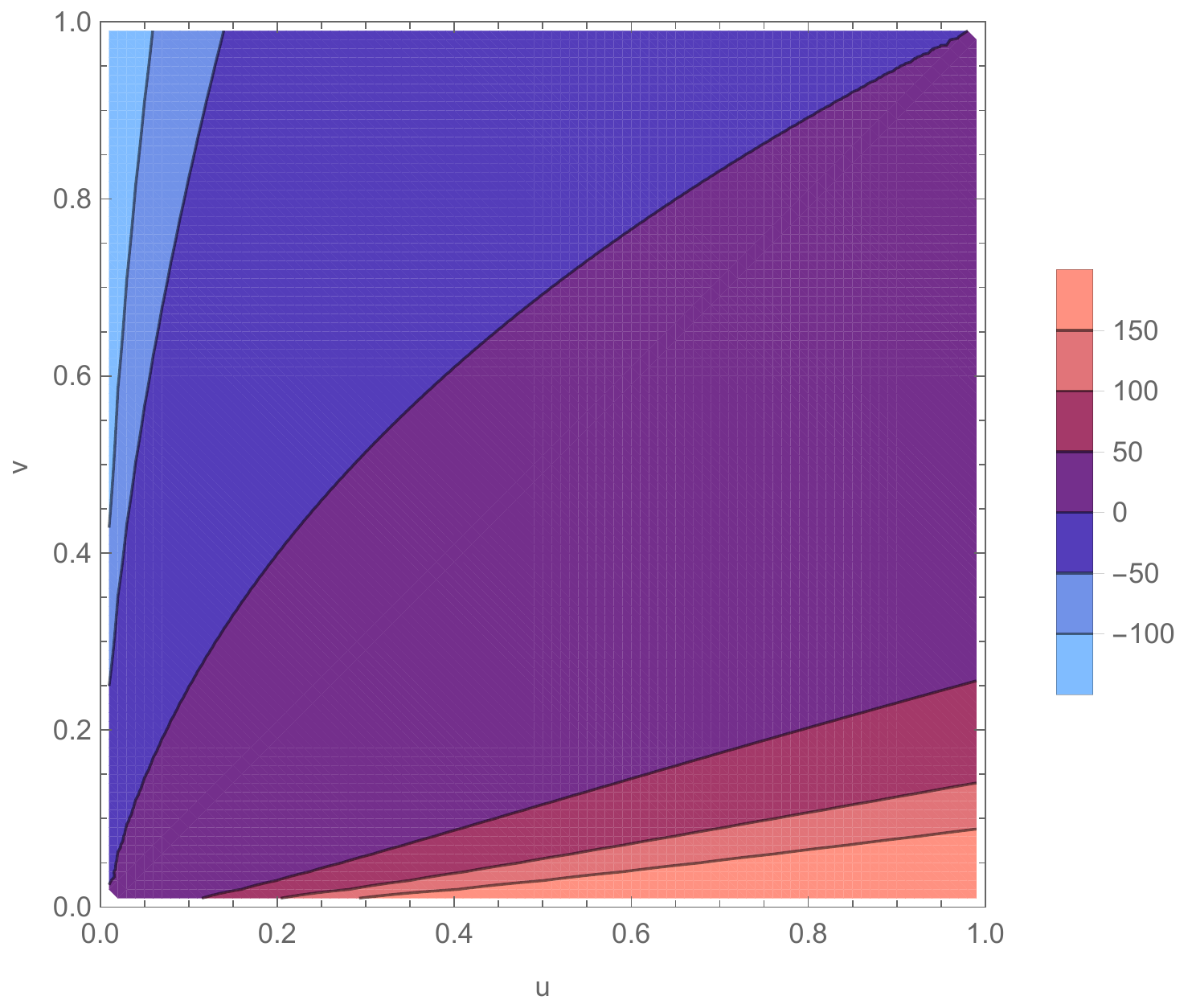}
\end{center}
\caption{$\Vt^{(4)}(v,1,u)-\Vt^{(4)}(1,u,v)$ plotted in $u$ and $v$.}
\label{fig:vtv1um1uv}
\end{figure}

\begin{figure}
\begin{center}
\includegraphics[width=4.5in]{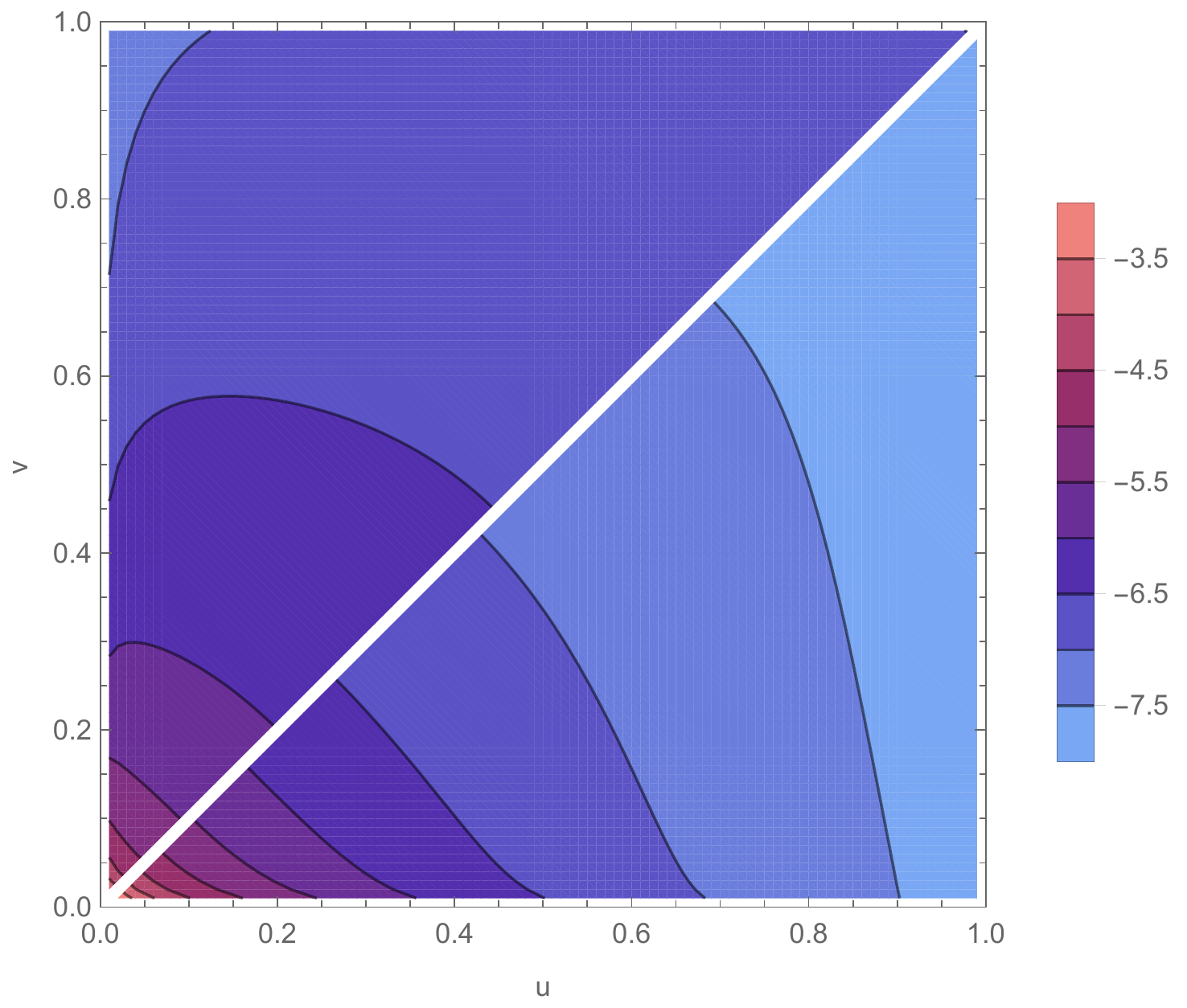}
\end{center}
\caption{Ratios of $\Vt(1,u,v)-\Vt(u,v,1)$ between successive loop orders, plotted in $u$ and $v$. $\Vt^{(4)}/\Vt^{(3)}$ is in the top-left corner, while $\Vt^{(3)}/\Vt^{(2)}$ is in the bottom-right.}
\label{fig:vt1uvmuv1}
\end{figure}

There are three such differences to consider, $\Vt(v,w,u)-\Vt(w,u,v)$, $\Vt(u,v,w)-\Vt(v,w,u)$, and $\Vt(w,u,v)-\Vt(u,v,w)$. Taking $w\rightarrow 1$, this gives us $\Vt(v,1,u)-\Vt(1,u,v)$, $\Vt(u,v,1)-\Vt(v,1,u)$, and $\Vt(1,u,v)-\Vt(u,v,1)$. Of these, $\Vt(v,1,u)-\Vt(1,u,v)$ and $\Vt(u,v,1)-\Vt(v,1,u)$ exchange under $u\leftrightarrow v$, while $\Vt(1,u,v)-\Vt(u,v,1)$ is symmetric under $u\leftrightarrow v$. As it turns out, $\Vt(v,1,u)-\Vt(1,u,v)$ crosses zero while $\Vt(1,u,v)-\Vt(u,v,1)$ does not. As such, we can plot these cyclic differences of $\Vt$ in the same format we used for $V$.  Figure~\ref{fig:vtv1um1uv} plots $\Vt^{(4)}(v,1,u)-\Vt^{(4)}(1,u,v)$, while figure~\ref{fig:vt1uvmuv1} shows the ratios $\left(\Vt^{(4)}(1,u,v)-\Vt^{(4)}(u,v,1)\right)/\left(\Vt^{(3)}(1,u,v)-\Vt^{(3)}(u,v,1)\right)$ and $\left(\Vt^{(3)}(1,u,v)-\Vt^{(3)}(u,v,1)\right)/\left(\Vt^{(2)}(1,u,v)-\Vt^{(2)}(u,v,1)\right)$ in the two panels separated by the diagonal line $u=v$. The latter plot again shows fairly constrained inter-loop ratios, varying between $-3$ and $-8$, and varying significantly less as the loop order increases.

Next, we consider the plane where one of the cross ratios goes to zero. For concreteness, take $v\rightarrow 0$. In this limit, as was also the case for the $w\rightarrow 1$ limit, the $y$ variables become rational functions of $u$, $v$, and $w$:
\be
y_u \rightarrow \frac{u}{1-w} \,, \qquad
y_v \rightarrow \frac{v(1-u)(1-w)}{(1-u-w)^2} \,, \qquad
y_w \rightarrow  \frac{w}{1-u} \,.
\label{vto0ylim}
\ee
(Equivalently, one could take the $y_i$ to the inverse of these values.)

In contrast to the limit $w\to1$, which is smooth (on the Euclidean branch), the hexagon functions can have logarithmically divergent behavior as $v\rightarrow 0$. As such, we expand all quantities in powers of $\ln v$. The coefficient of each power of $\ln v$ will then be a polylogarithmic function with symbol entries drawn from the following set:
\be
\mathcal{S}_{v\rightarrow 0}\ =\ \{ u,\, w,\, 1-u,\, 1-w,\, 1-u-w \} \,.
\ee
To plot these functions, we use a similar {\sc GiNaC}-based implementation to that used for the $w=1$ plane. Here there are two distinct regions to consider, due to the $1-u-w$ symbol entries. We can either consider $u+w>1$, or $u+w<1$. In general, these two regions require different implementations, which together can cover the whole positive quadrant $u,w>0$.  Here we just show results for the unit square.

\begin{figure}
\begin{center}
\begin{tabular}{>{\centering\arraybackslash} m{1cm} >{\centering\arraybackslash} m{2in} >{\centering\arraybackslash} m{2in}}
& $V^{(4)}(u,v,w)$ & $V^{(4)}(v,w,u)$\\
$\ln^4 v$ & \includegraphics[width=1.65in]{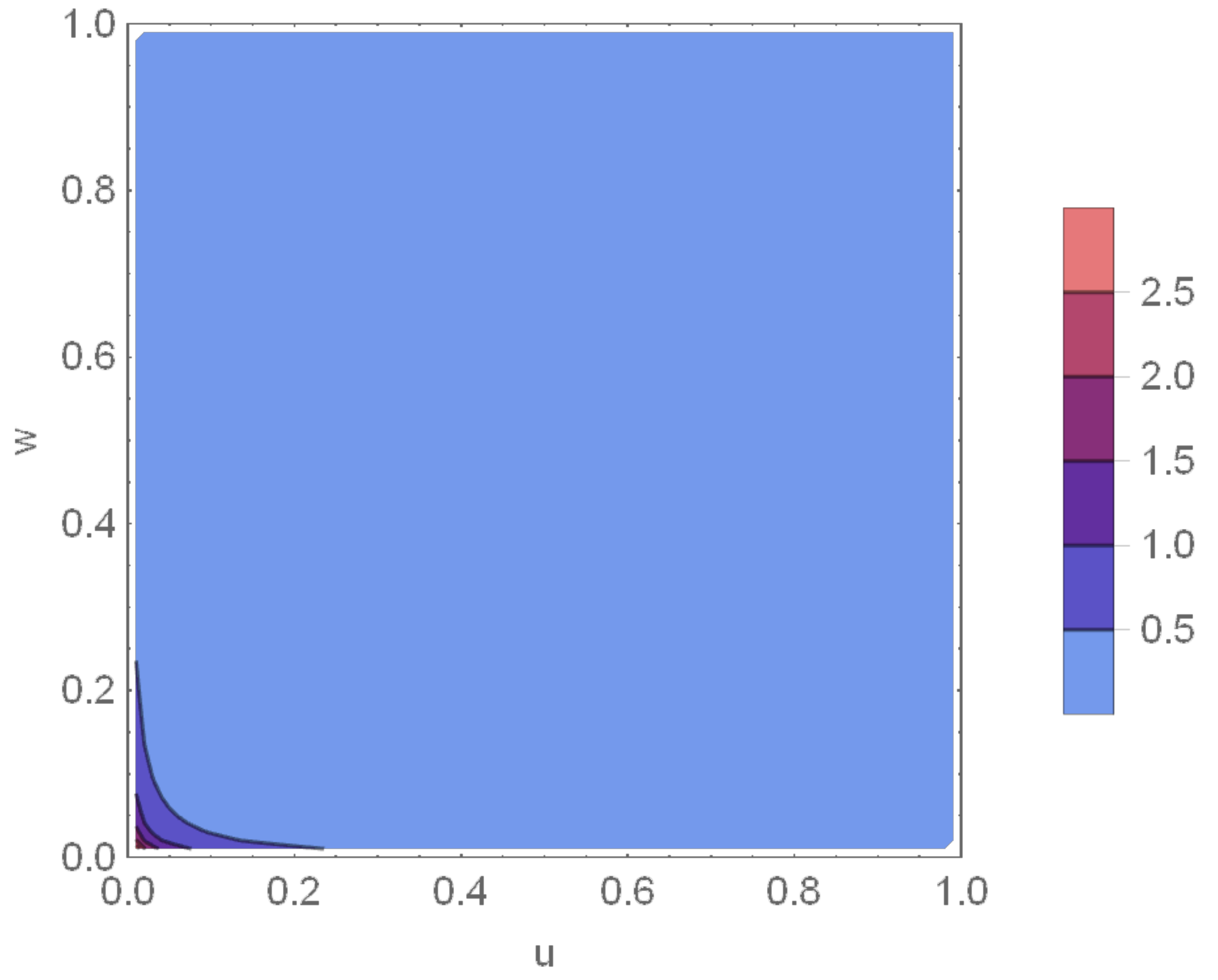} & \includegraphics[width=1.65in]{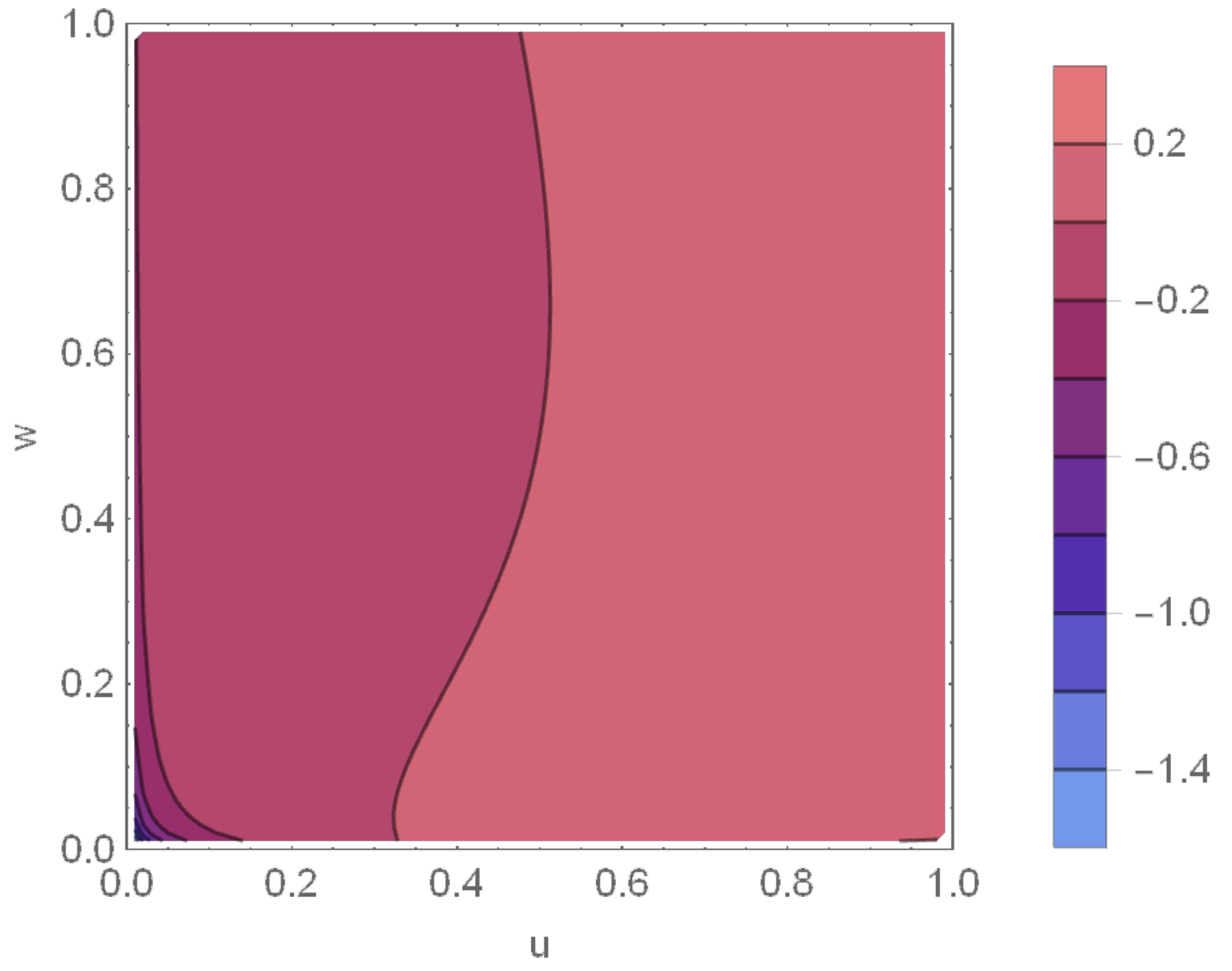}\\
$\ln^3 v$ & \includegraphics[width=1.65in]{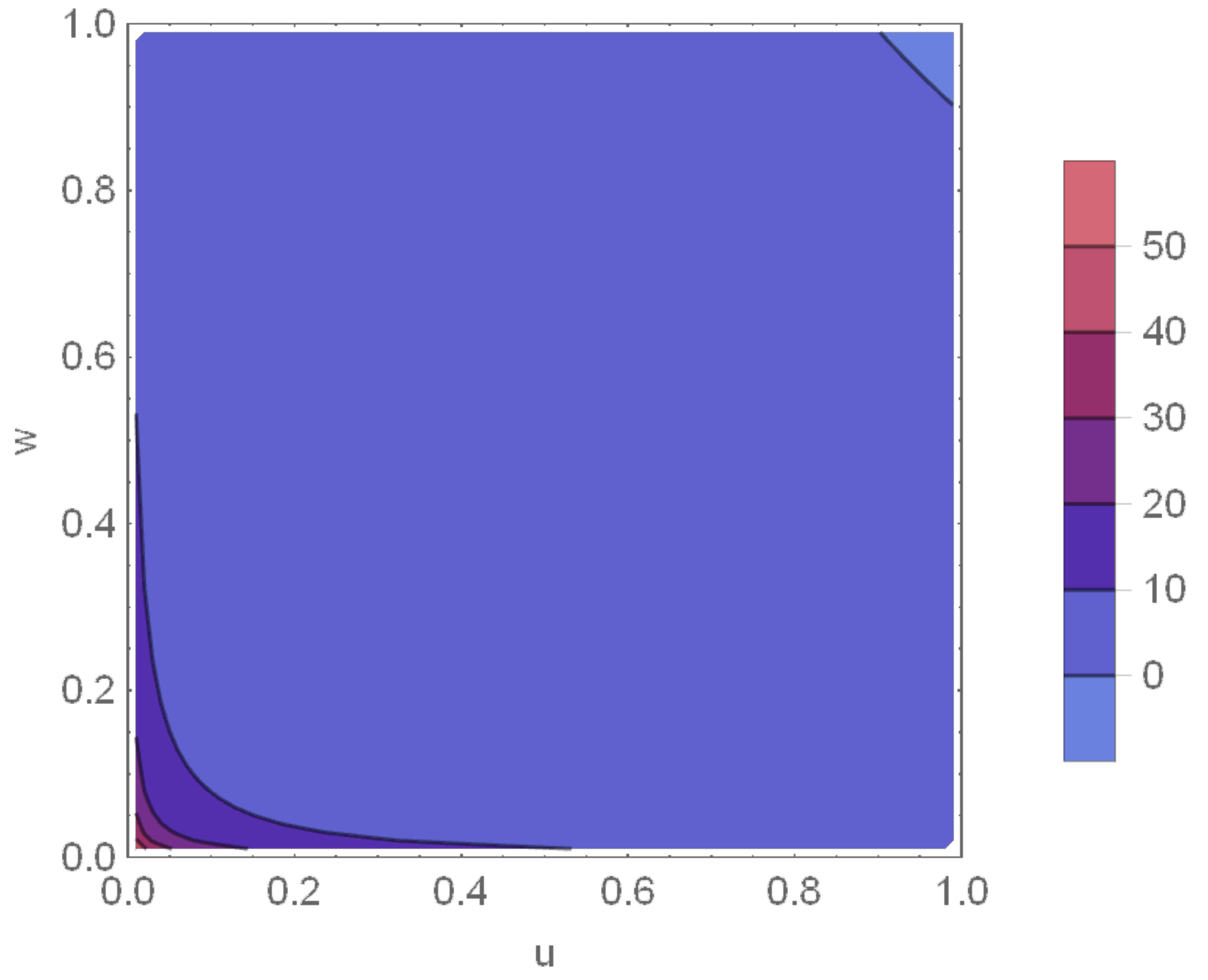} & \includegraphics[width=1.65in]{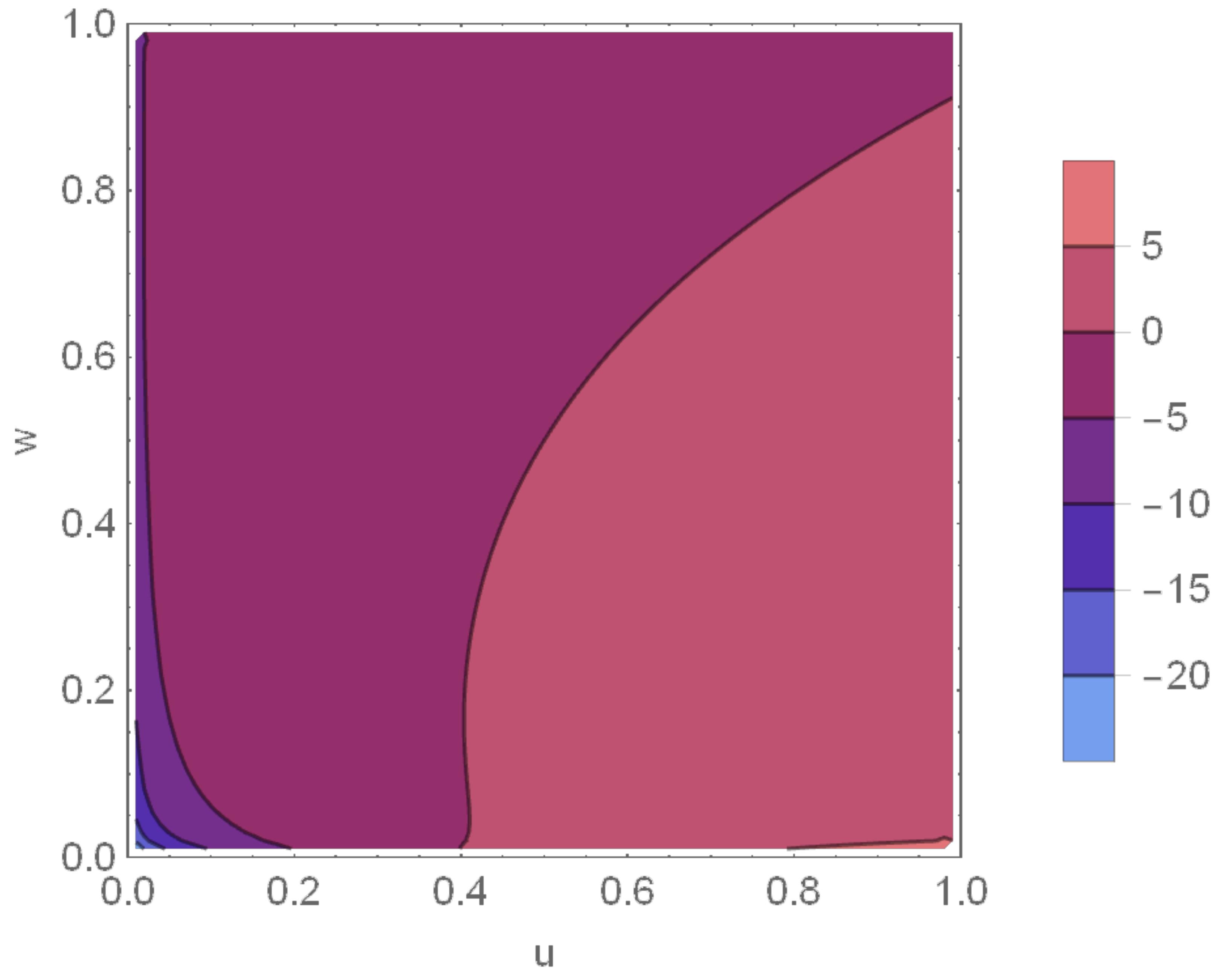}\\
$\ln^2 v$ & \includegraphics[width=1.65in]{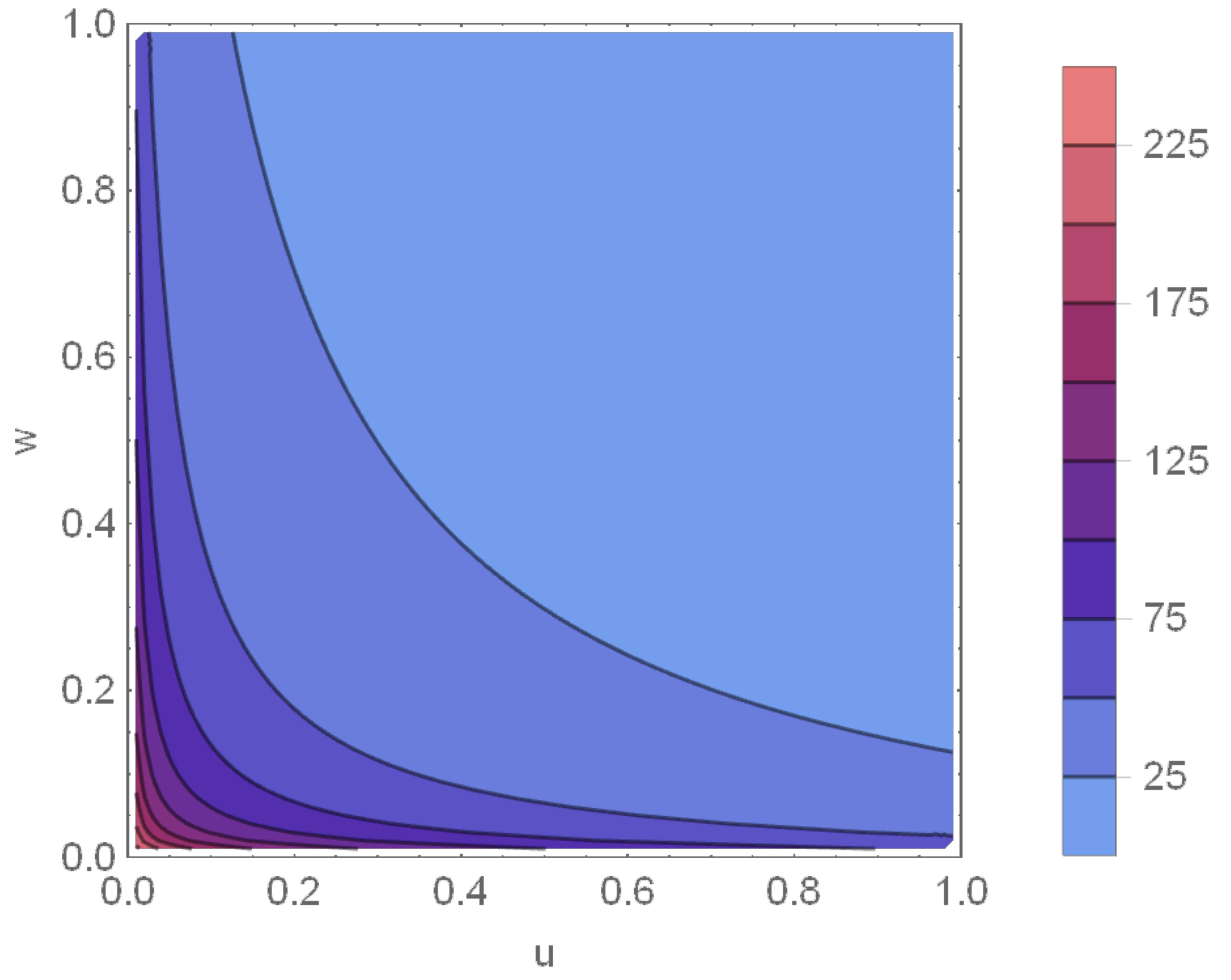} & \includegraphics[width=1.65in]{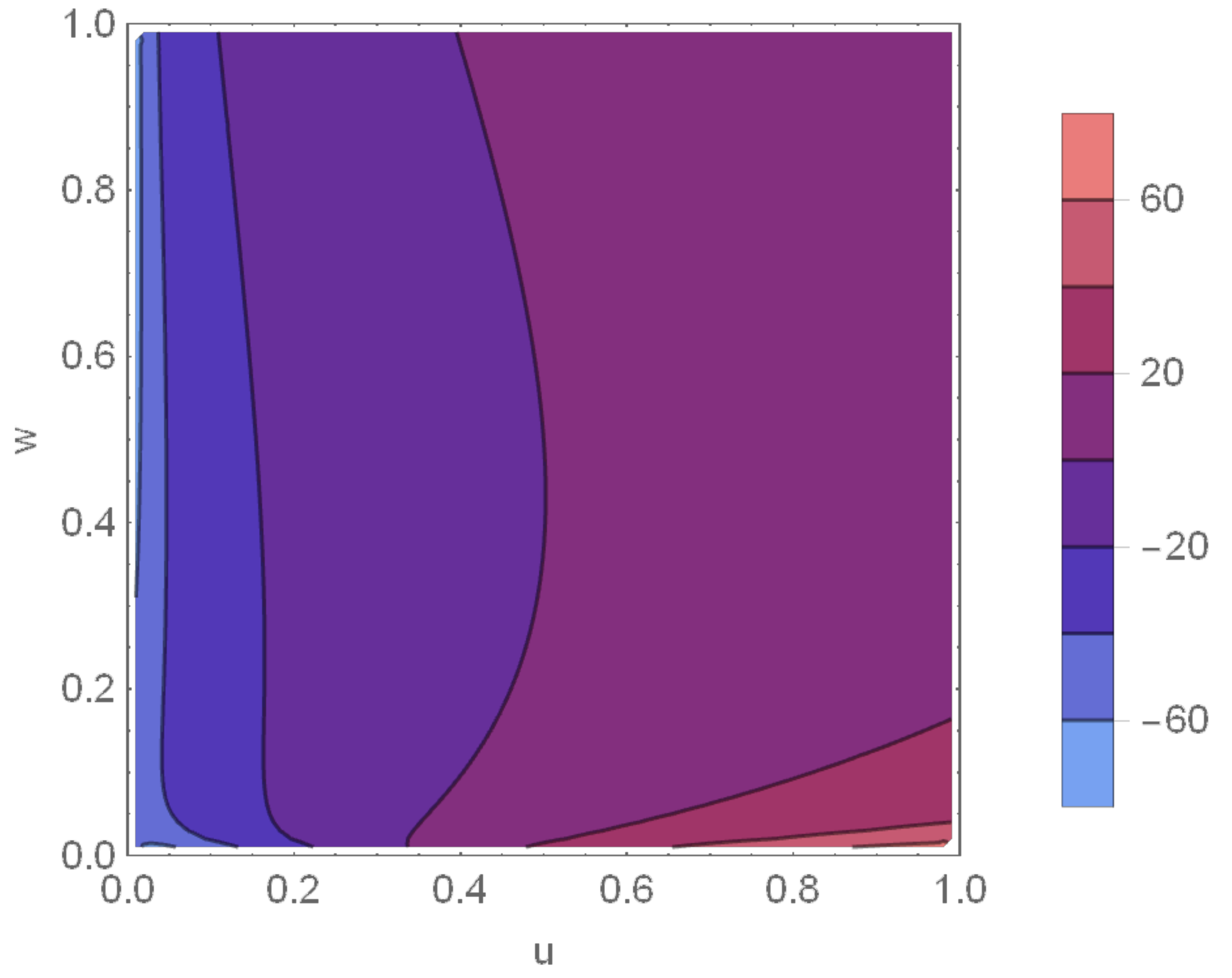}\\
$\ln^1 v$ & \includegraphics[width=1.65in]{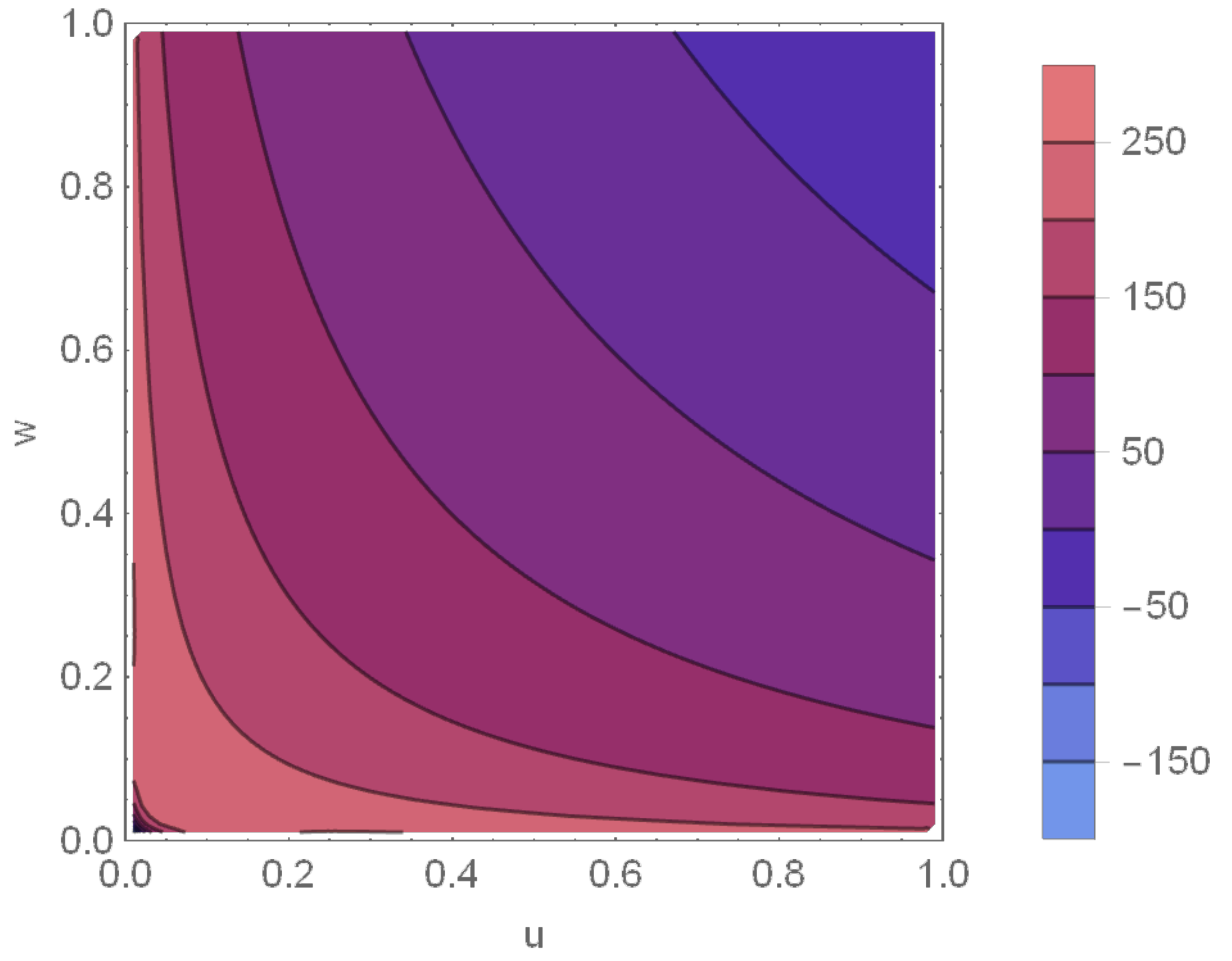} & \includegraphics[width=1.65in]{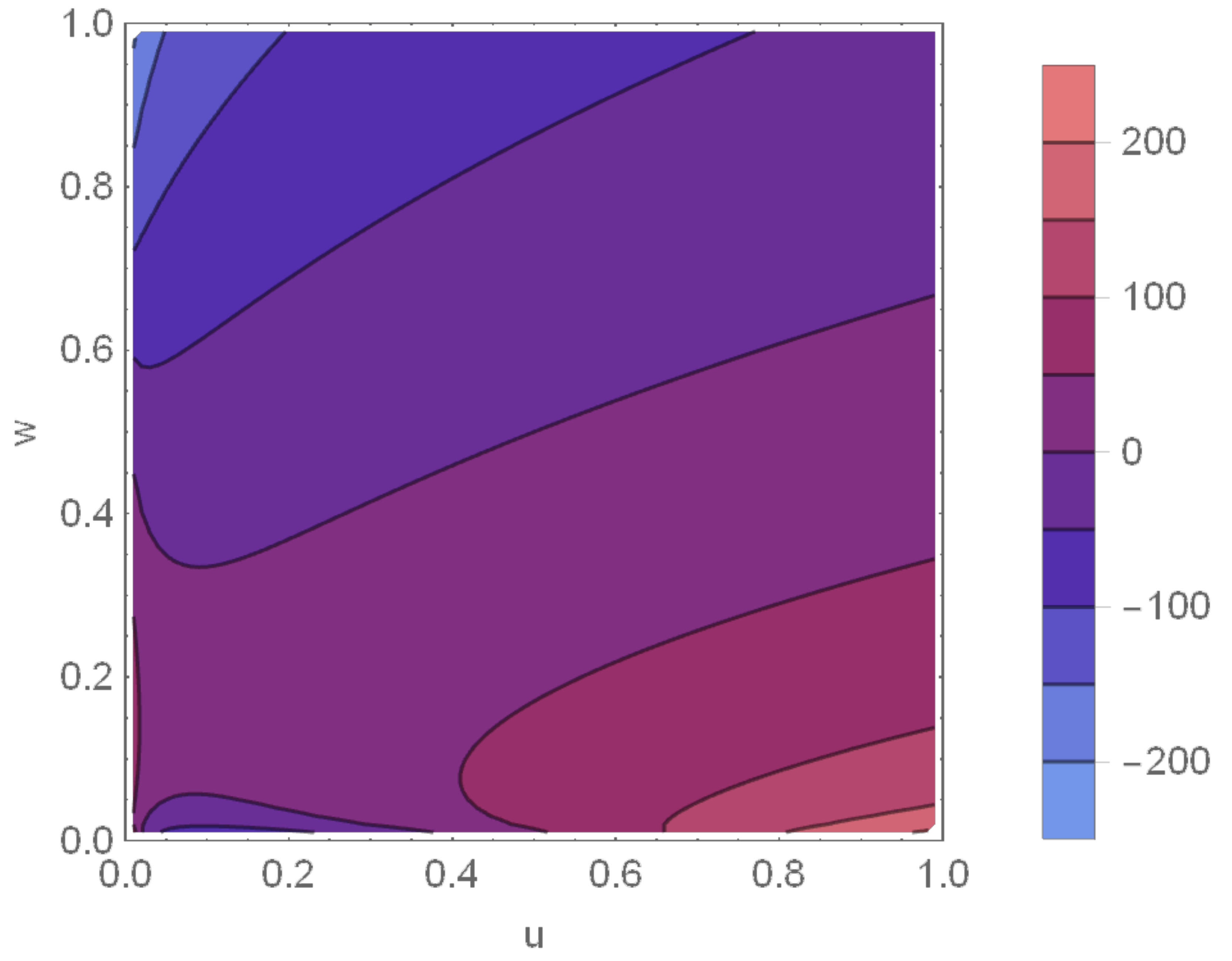}\\
$\ln^0 v$ & \includegraphics[width=1.65in]{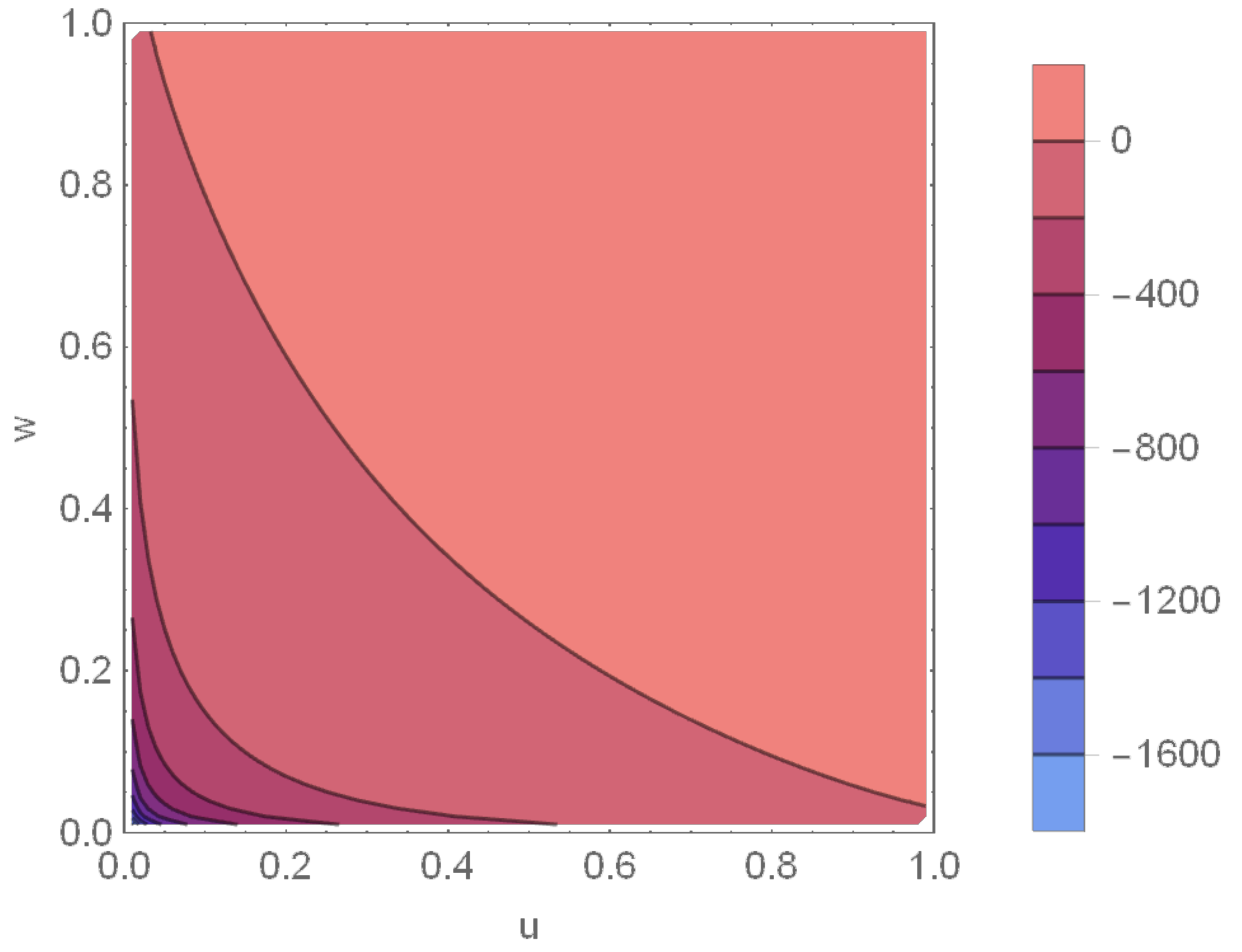} & \includegraphics[width=1.65in]{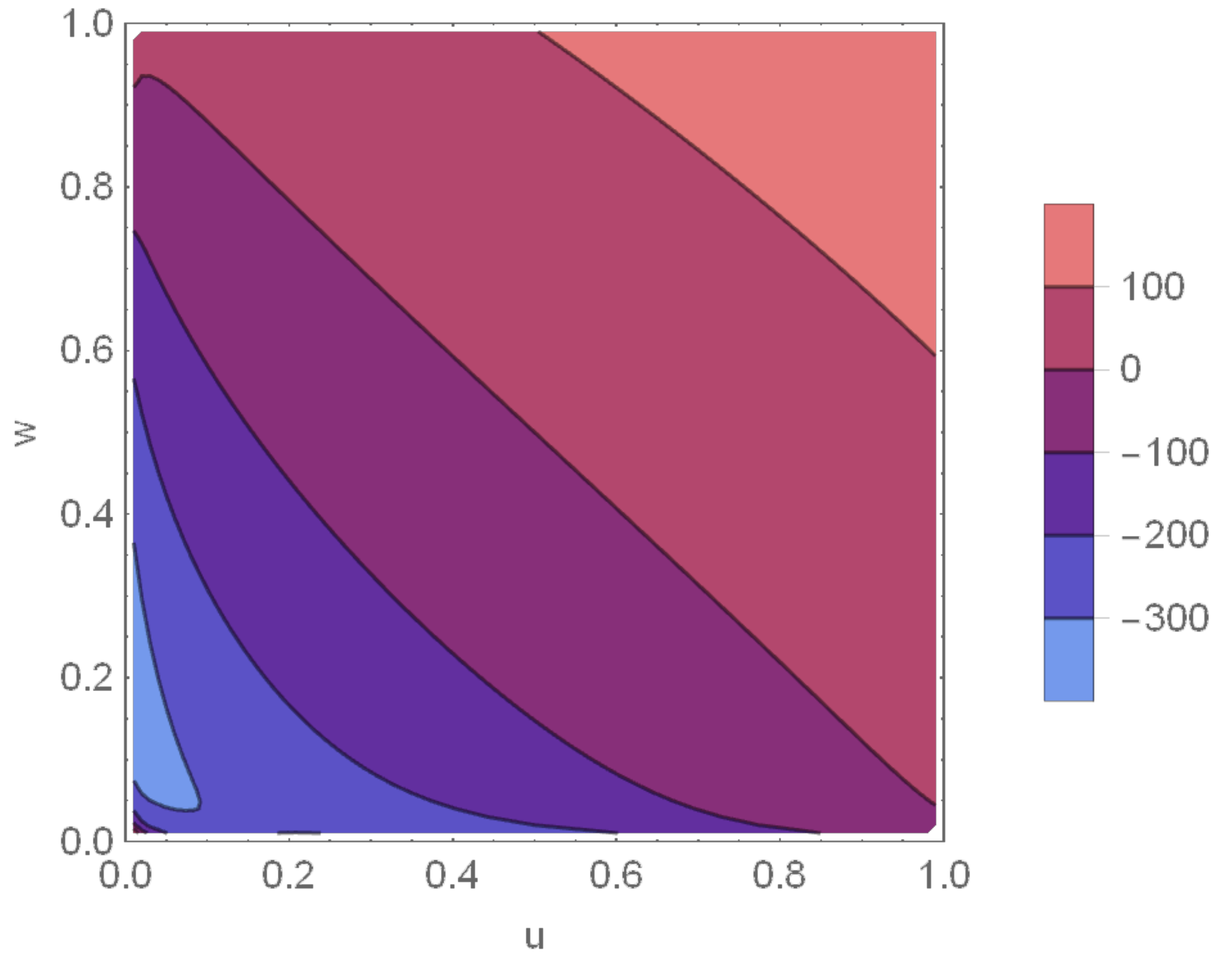}\end{tabular}
\end{center}
\caption{$V^{(4)}(u,v,w)$ and $V^{(4)}(v,w,u)$ plotted in the $v\rightarrow 0$ limit. The coefficient of each power of $\ln v$ is plotted separately.}
\label{fig:v4u0w}
\end{figure}

\begin{figure}
\begin{center}
\begin{tabular}{>{\centering\arraybackslash} m{1cm} >{\centering\arraybackslash} m{2in} >{\centering\arraybackslash} m{2in}}
& $\Vt^{(4)}(v,w,u)-\Vt^{(4)}(w,u,v)$ & $\Vt^{(4)}(u,v,w)-\Vt^{(4)}(v,w,u)$\\
$\ln^4 v$ & \includegraphics[width=1.65in]{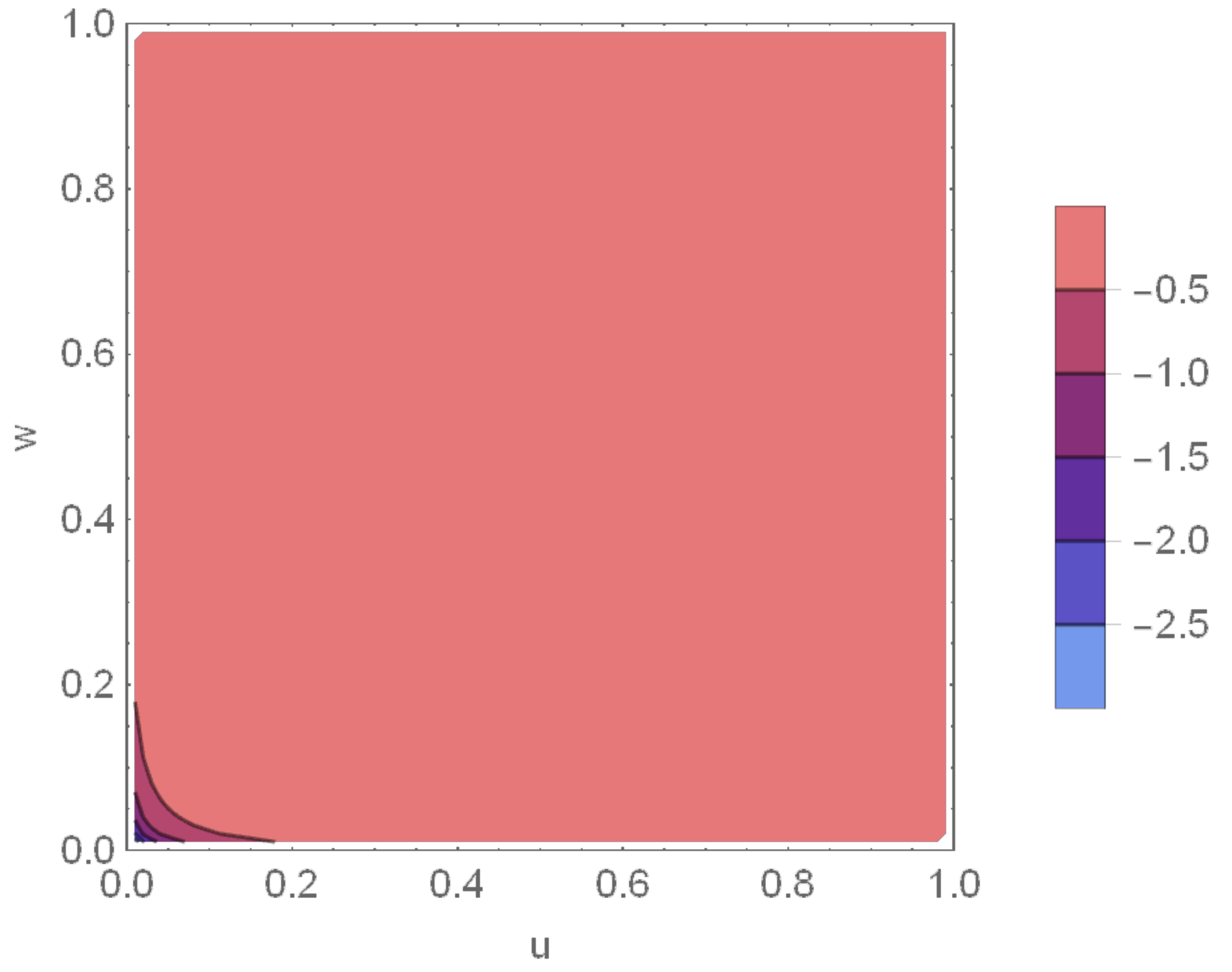} & \includegraphics[width=1.65in]{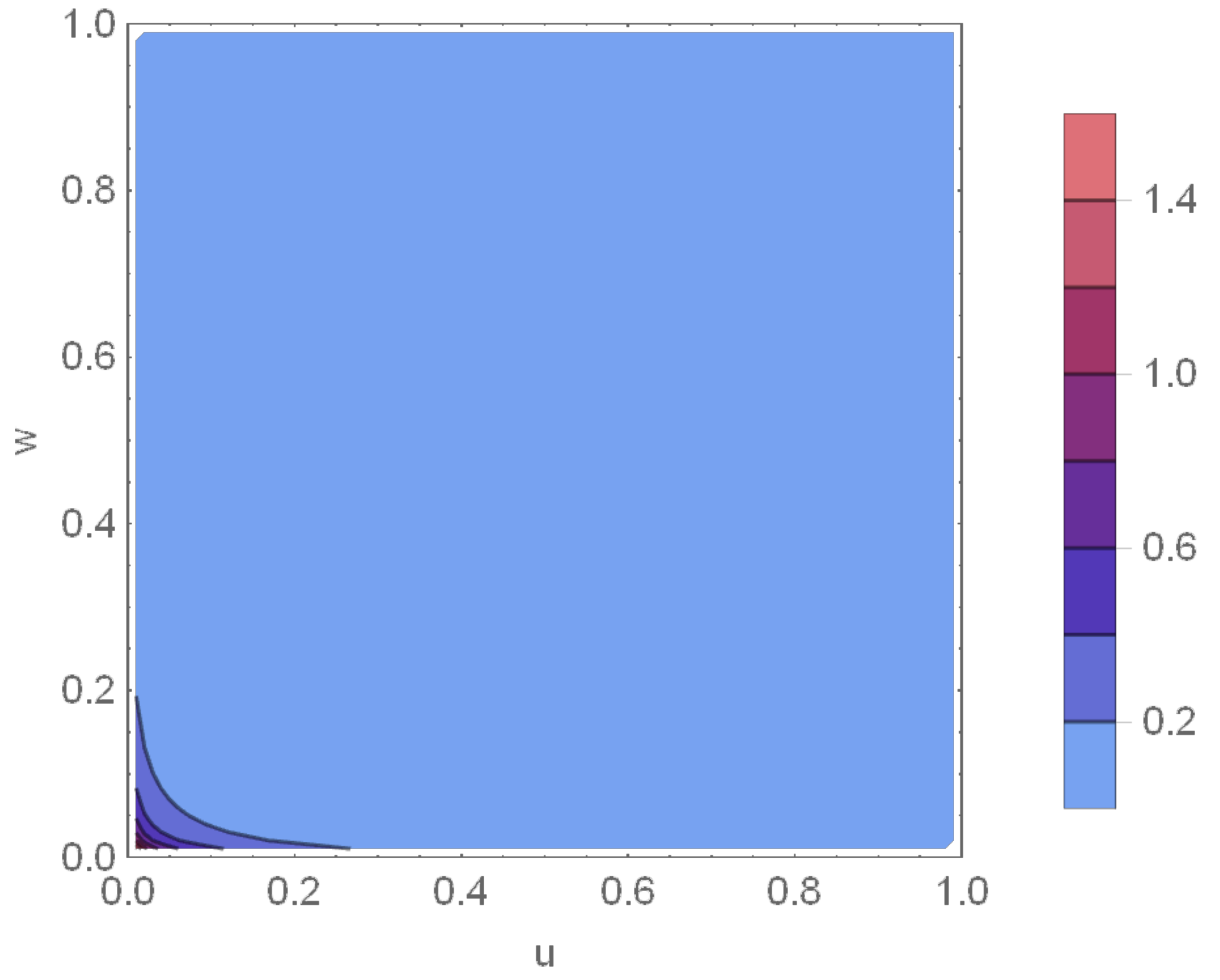}\\
$\ln^3 v$ & \includegraphics[width=1.65in]{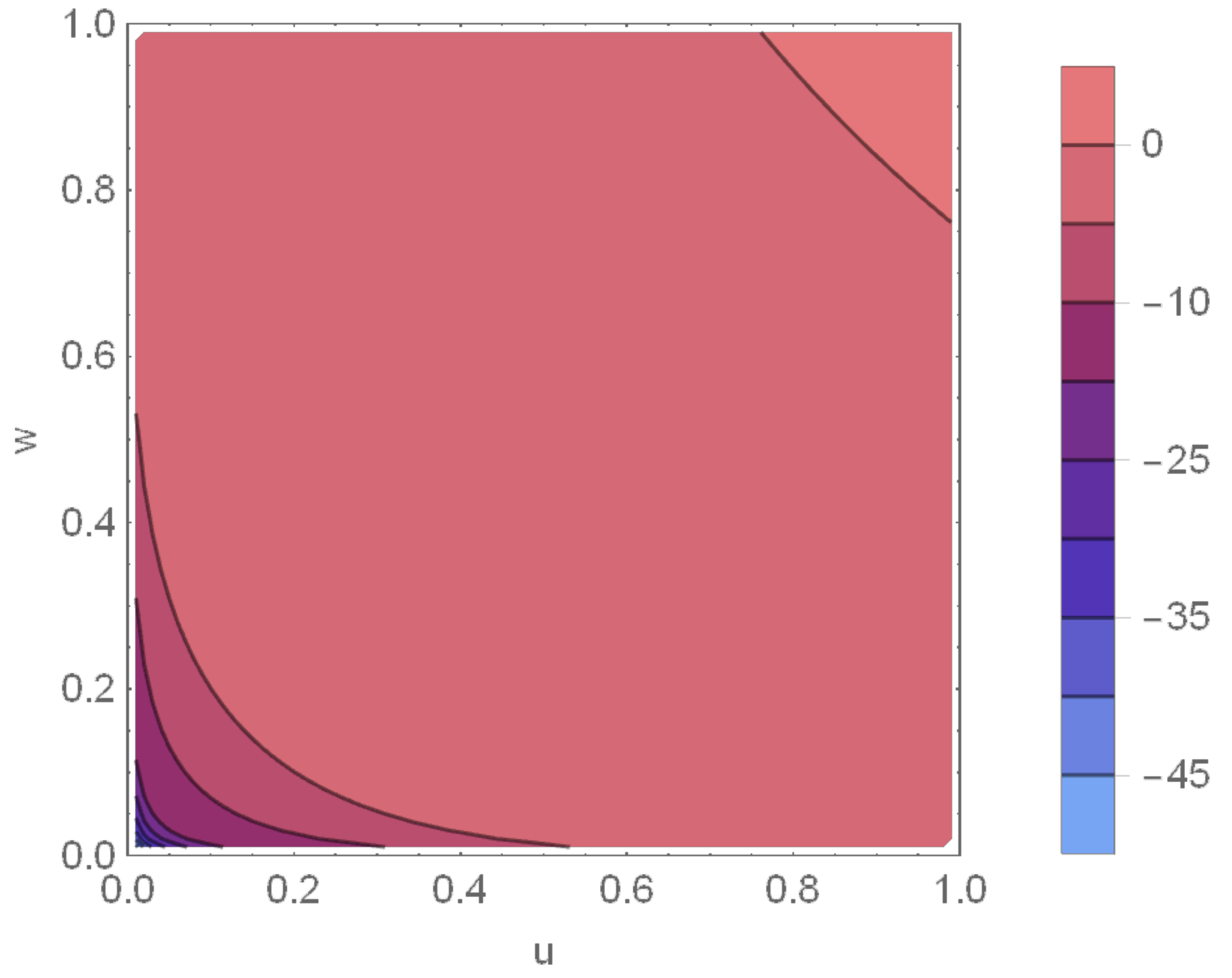} & \includegraphics[width=1.65in]{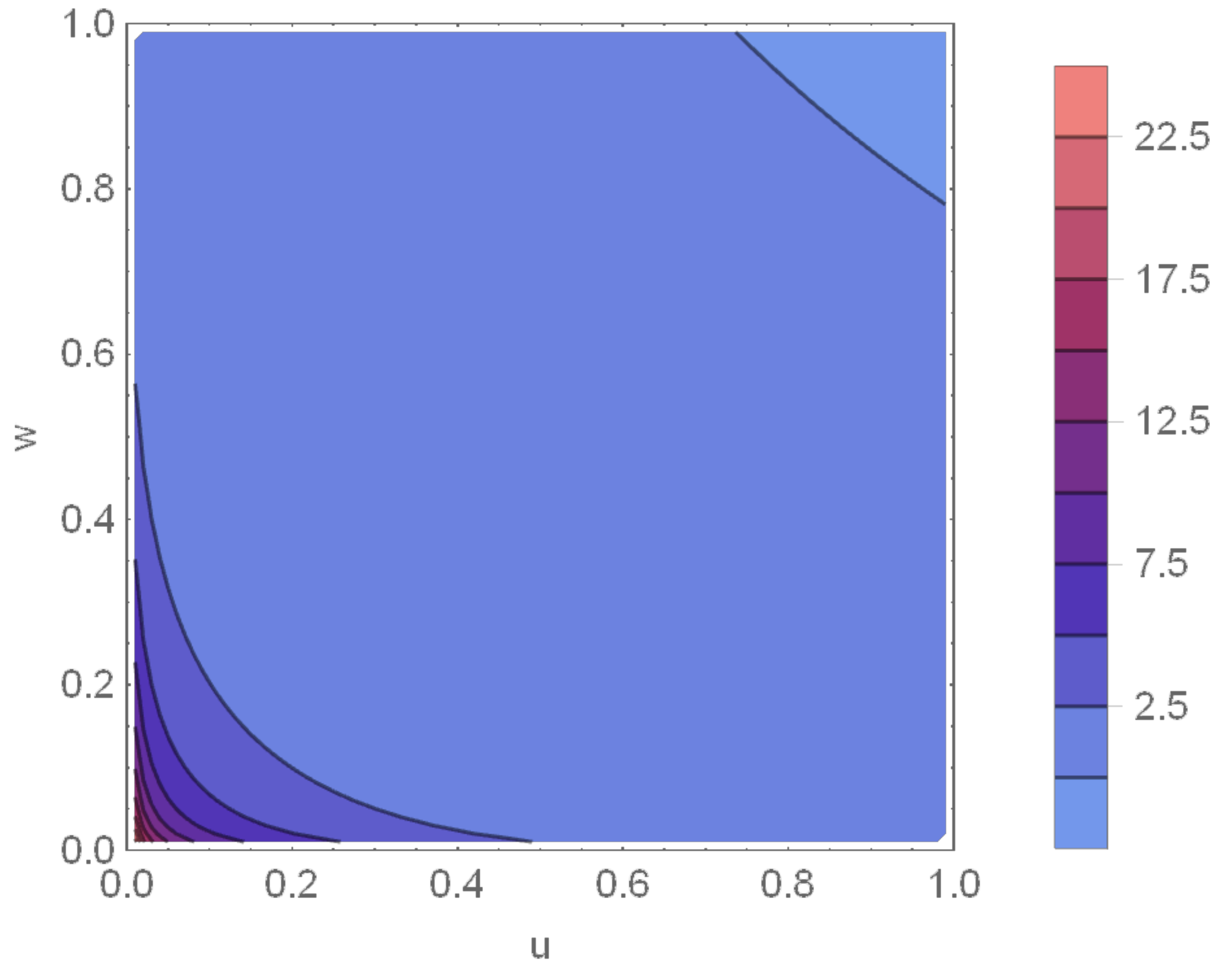}\\
$\ln^2 v$ & \includegraphics[width=1.65in]{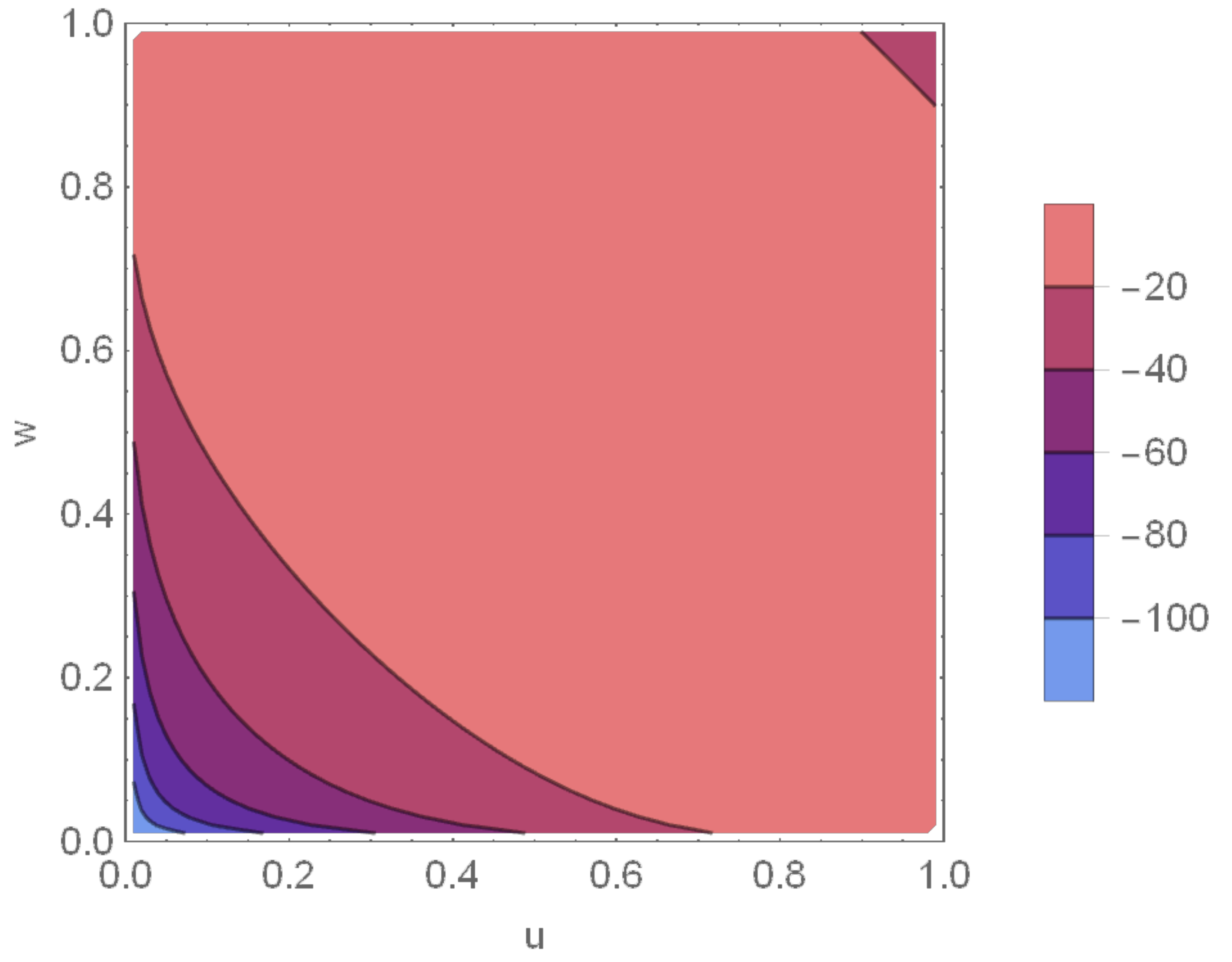} & \includegraphics[width=1.65in]{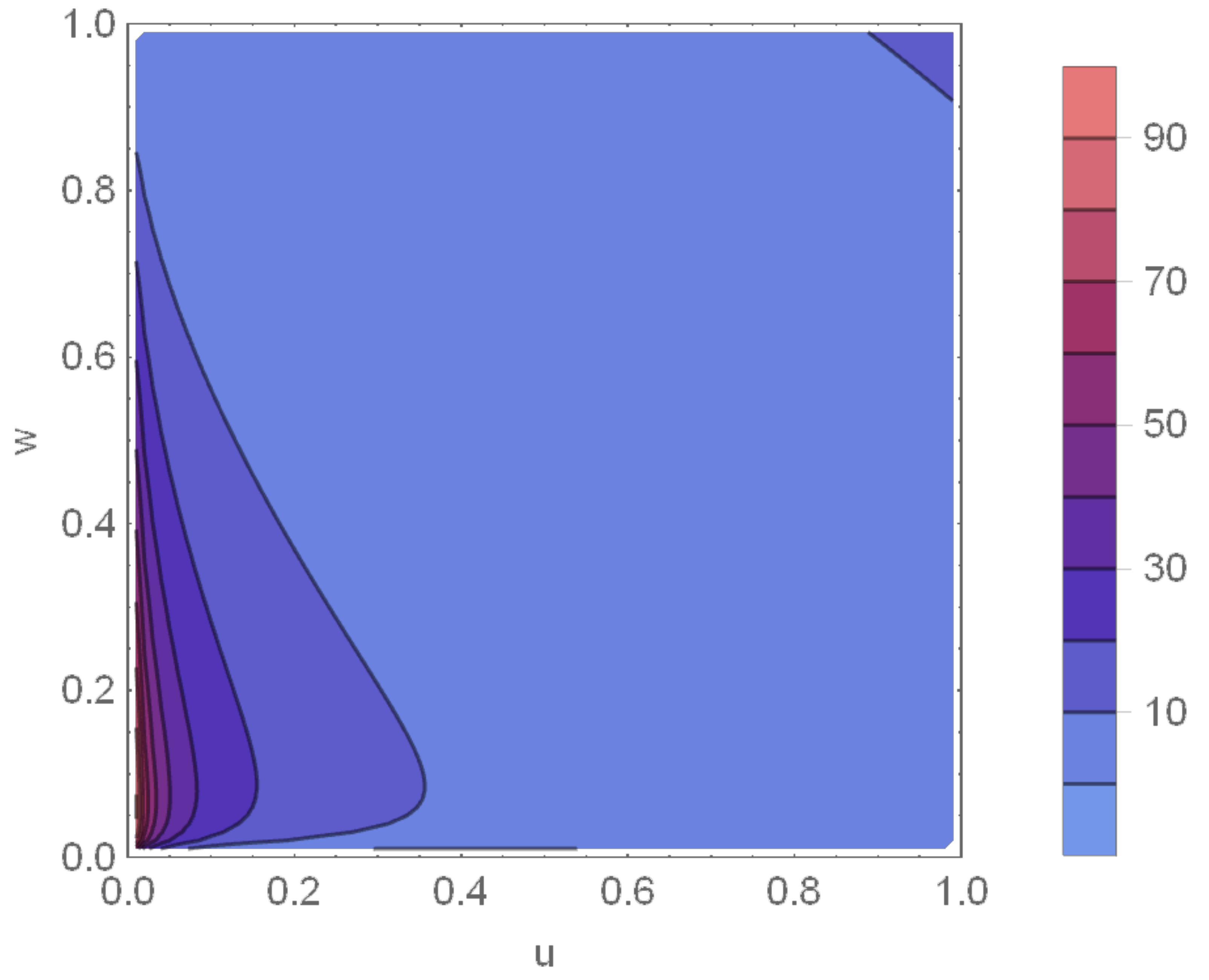}\\
$\ln^1 v$ & \includegraphics[width=1.65in]{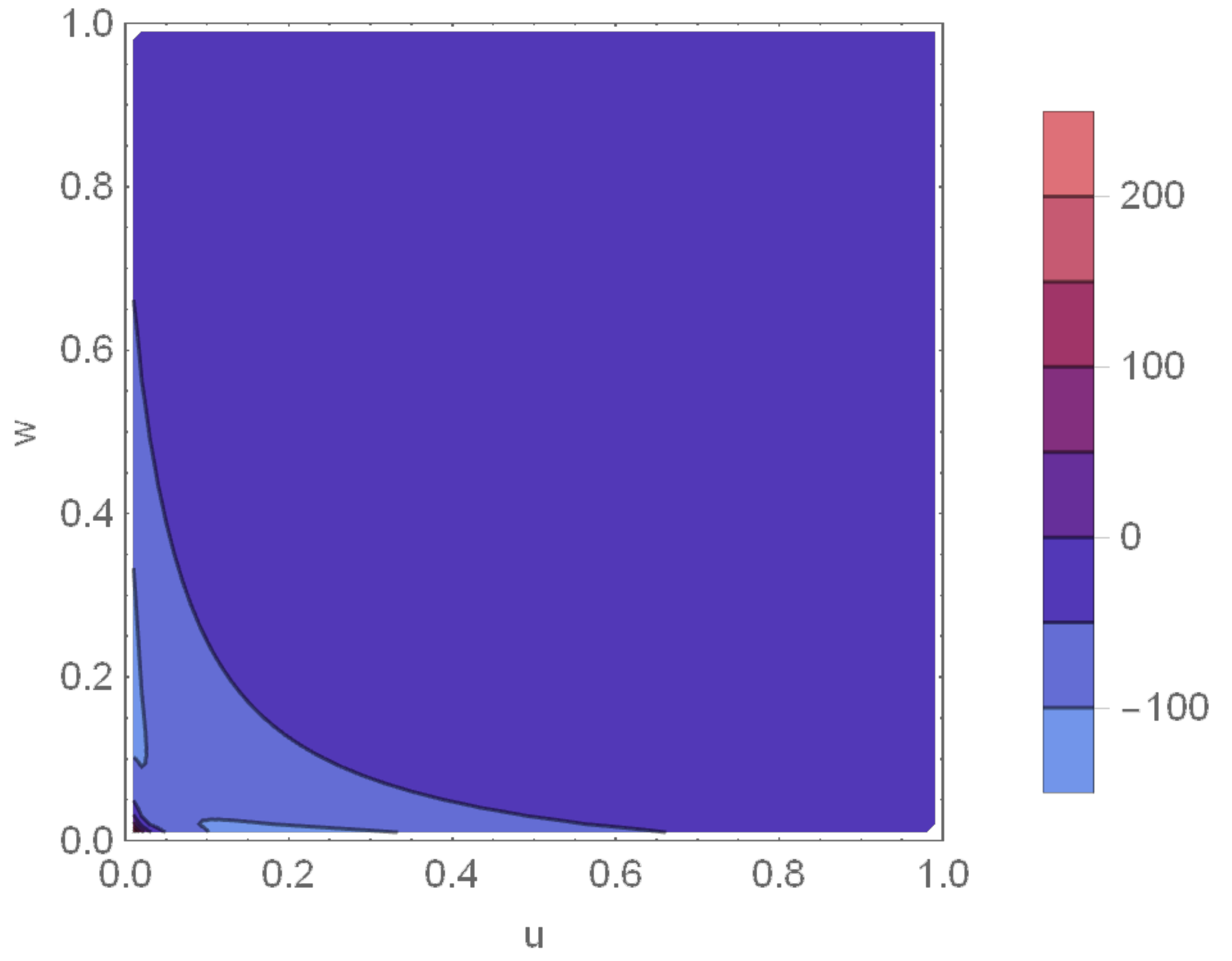} & \includegraphics[width=1.65in]{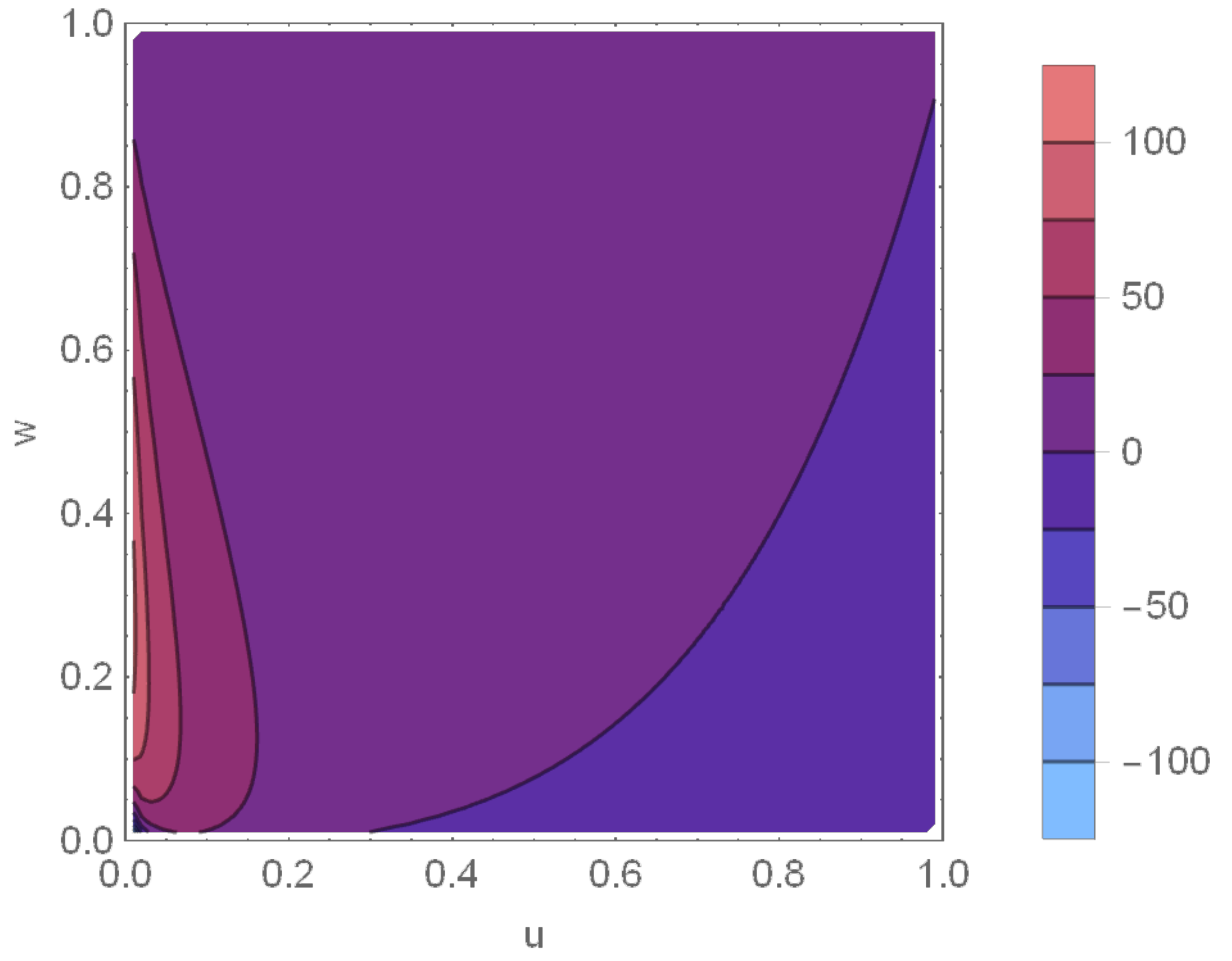}\\
$\ln^0 v$ & \includegraphics[width=1.65in]{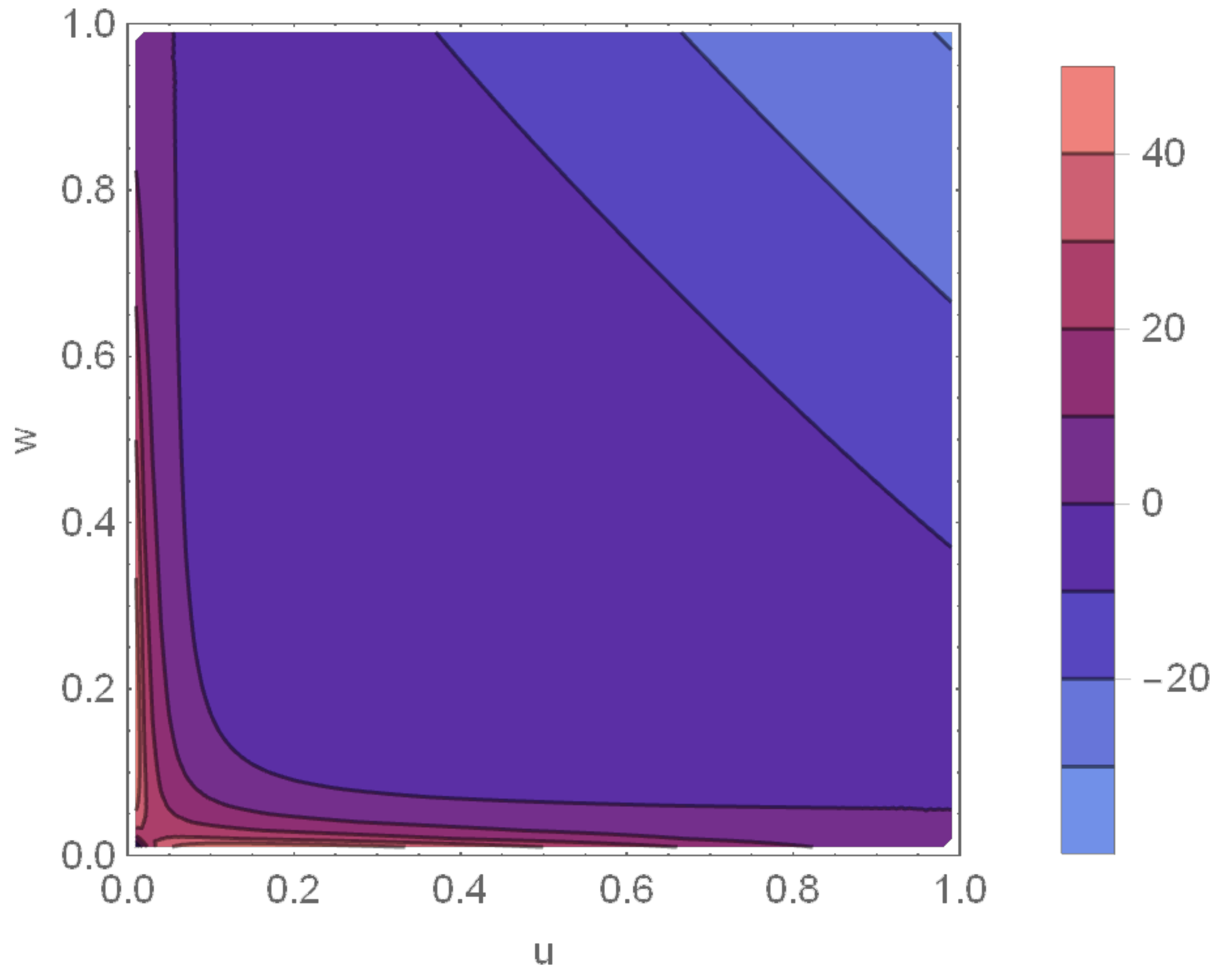} & \includegraphics[width=1.65in]{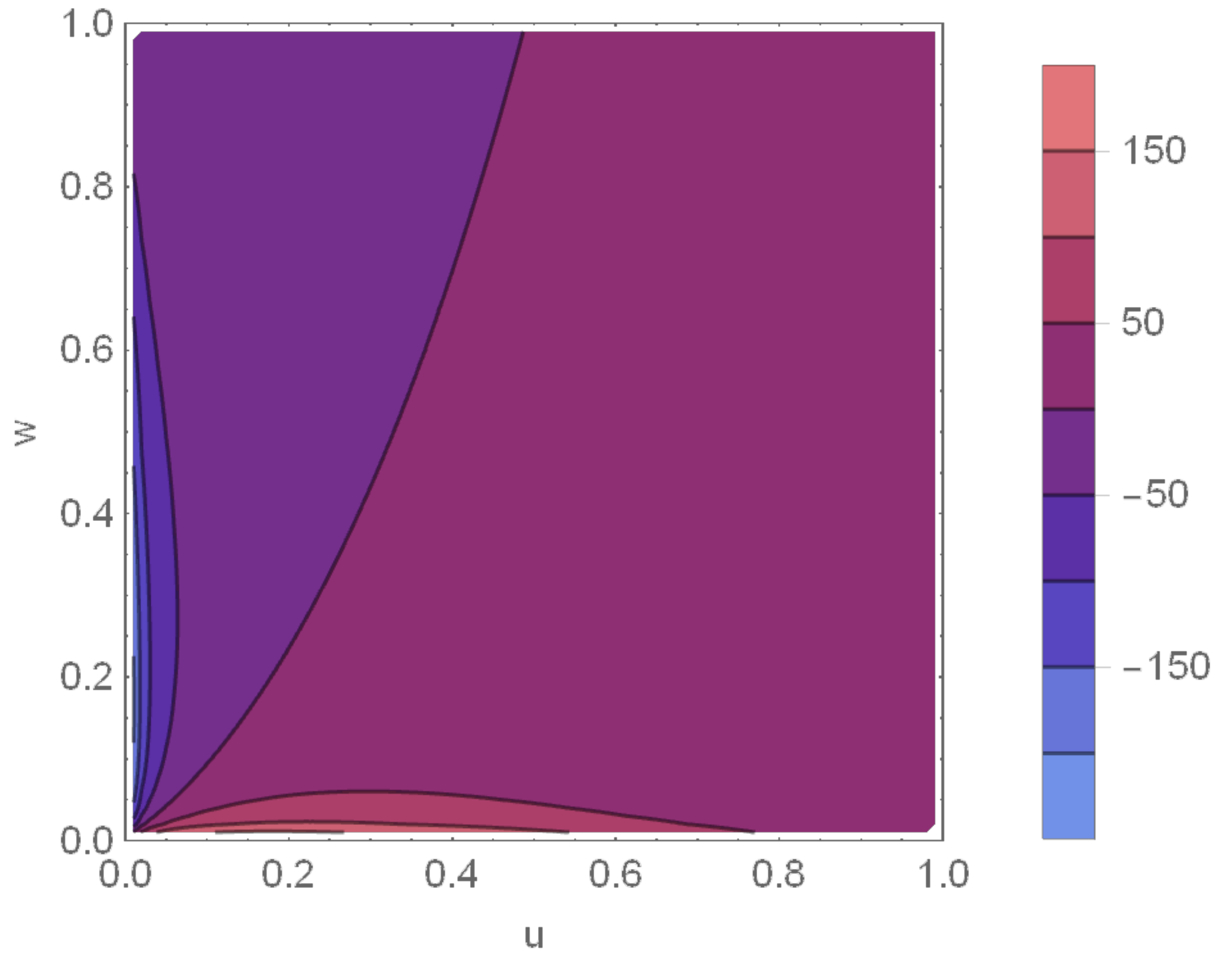}\end{tabular}
\end{center}
\caption{$\Vt^{(4)}(v,w,u)-\Vt^{(4)}(w,u,v)$ and $\Vt^{(4)}(u,v,w)-\Vt^{(4)}(v,w,u)$ plotted in the $v\rightarrow 0$ limit. The coefficient of each power of $\ln v$ is plotted separately.}
\label{fig:vt40wumwu0}
\end{figure}

In figure~\ref{fig:v4u0w} we plot the $v\rightarrow 0$ limit of the parity-even functions $V^{(4)}(u,v,w)$ and $V^{(4)}(v,w,u)$ in the left and right columns, respectively, for each of the coefficients of $\ln^k v$ that are nonvanishing, $k=0,1,2,3,4$.  (In general, $V^{(L)}$ and $\Vt^{(L)}$ have a maximum divergence of $\ln^L v$ at $L$ loops, at least for $L\leq4$.) Figure~\ref{fig:vt40wumwu0} plots the parity-odd functions $\Vt^{(4)}(v,w,u)-\Vt^{(4)}(w,u,v)$, and $\Vt^{(4)}(u,v,w)-\Vt^{(4)}(v,w,u)$.  The other possible arguments are related by permutations. In both figures, the left panels are exactly symmetric under the exchange $u\lr w$. Since the highest power of $\ln v$ in this limit increases with loop order there are no simple inter-loop ratios to show on this plane, which is why we plot only the four-loop functions.

\section{Conclusions and outlook}
\label{conclusions}

In this paper we have continued the hexagon function bootstrap program initiated in ref.~\cite{Dixon2011pw}. By calculating the six-particle NMHV ratio function through four loops, we have completed the description of six-point amplitudes at this loop order, complementing the earlier MHV results~\cite{Dixon2014voa}. We extended the basis of hexagon functions constructed in ref.~\cite{Dixon2013eka} to transcendental weight eight.  We used the weight-seven part of this basis to construct an ansatz for the $\{7,1\}$ component of coproduct of the NMHV coefficient functions $V$ and $\tilde{V}$, which we then constrained through a series of physical inputs.

The most powerful such input comes from the $\bar{Q}$ equation.  Previously, this equation was understood to imply a seven-final-entry condition.  We now understand that it actually leads to a much more powerful set of relations between different permutations of the functions $E(u,v,w)$ and $\Et(u,v,w)$~\cite{SimonSongPrivate}. This set of relations allowed us to work from an ansatz which, at four loops, had only 34 free parameters, and we could fix all but five of them by requiring the collinear limit of the ratio function to vanish.

The five remaining parameters were then fixed by appealing to the multi-Regge limit. By using an extension of the ansatz proposed in ref.~\cite{Lipatov2012gk} that we detailed in ref.~\cite{Dixon2014iba}, we used lower-loop information to predict the multi-Regge limit of the NMHV ratio function at leading-log and next-to-leading-log order. This allowed us to fix the remaining parameters in our ansatz. The terms in this limit that are of subleading logarithmic order have all been predicted using integrability~\cite{BCHS}.  They serve as a cross-check on our results.

With a unique result for the ratio function in hand, we proceeded to take the near-collinear limit and compare to the Pentagon OPE approach of Basso, Sever, and Vieira. Here we found perfect agreement with their published results~\cite{Basso2013aha} and those of Papathanasiou~\cite{Papathanasiou2013uoa} and Belitsky~\cite{BelitskyI,BelitskyII,BelitskyPrivate}. We also computed the multi-particle factorization limit, which takes a very simple form and agrees completely with integrability-based predictions~\cite{BSVfact}.

Plotting $V$ and $\tilde{V}$ over a variety of lines and planes through the space of cross ratios, we found its behavior to be broadly similar across loop orders. In particular, we observed that, outside of regions where the functions vanish and corners containing logarithmic divergences, the ratios between successive loop orders are fairly flat, and generally stay between $-4$ and $-8$.

Recently, based on investigation of the positive
Grassmannian~\cite{ArkaniHamed2012nw}, it was conjectured~\cite{AHCHTPrivate} that the remainder function ought to have uniform sign in a particular region of cross-ratio space. While this conjecture appears to be false near the origin for the remainder function, a similar conjecture seems to hold true for a bosonized version of the ratio function.  Using the ``data'' found in this paper through four loops, we will explore this conjecture in future work~\cite{DvHMTtoappear}.

Our new understanding of the $\bar{Q}$ equation has led to remarkably powerful constraints. After applying it, the number of free parameters remaining appears to only increase by around a factor of three at each loop order. If this trend continues, there should only be around a hundred unfixed parameters at five loops, comparable to the number that needed to be fixed for the four-loop MHV remainder function. This suggests that the five-loop ratio function may be well within reach. If so, it would be a great opportunity to see just how far the hexagon function program can extend.


\vskip0.5cm
\noindent {\large\bf Acknowledgments}
\vskip0.3cm

We are grateful to Benjamin Basso, Andrei Belitsky, Simon Caron-Huot, James Drummond, Song He, Yorgos Papathanasiou, Jeff Pennington, Amit Sever, Jara Trnka, and Pedro Vieira for many illuminating discussions. We particularly thank Simon and Song for enlightening us about the $\bar{Q}$ equation, Simon also for comments on the manuscript, and Jeff for explaining his Lyndon reduction algorithm. This research was supported by the US Department of Energy under contract DE--AC02--76SF00515 and grant DE--SC0011632, by the Walter Burke Institute, by the Gordon and Betty Moore Foundation through Grant No.~776 to the Caltech Moore Center for Theoretical Cosmology and Physics, by the Munich Institute for Astro- and Particle Physics (MIAPP) of the DFG cluster of excellence ``Origin and Structure of the Universe'', and by the Perimeter Institute for Theoretical Physics. Research at Perimeter Institute is supported by the Government of Canada through Industry Canada and by the Province of Ontario through the Ministry of Economic Development and Innovation. LD thanks Caltech, the Aspen Center for Physics and the NSF Grant \#1066293 for hospitality, as well as the Galileo Galilei Institute and the Perimeter Institute for Theoretical Physics.  MvH would like to thank the ICTP South American Institute for Fundamental Research for hospitality while this work was completed.

\vfill\eject


\appendix

\section{Hexagon function basis at weight six, seven and eight}
\label{hexagon_basis}

Building on the basis of hexagon functions through weight five introduced in ref.~\cite{Dixon2013eka}, we describe here a complete basis of hexagon functions at weight six, seven and eight. These functions can be defined in terms of their $\{n-1,1\}$ coproduct components, which for a generic hexagon function $F$ take the form
\be
\Delta_{n-1,1}(F) \equiv \sum_{i=1}^3 F^{u_i} \otimes \ln u_i + F^{1-u_i} \otimes \ln (1-u_i) + F^{y_i} \otimes \ln y_i,
\ee  
where the functions $\{F^{u_i}, F^{1-u_i},F^{y_i}\}$ uniquely determine the derivatives of $F$ to be
\bea
\frac{\partial F}{\partial u} \bigg |_{v,w} &=& \frac{F^u}{u} - \frac{F^{1-u}}{1-u} + \frac{1 - u - v - w}{u \sqrt{\Delta}} F^{y_u} + \frac{1 - u - v + w}{(1-u) \sqrt{\Delta}} F^{y_v} + \frac{1 - u + v - w}{(1-u) \sqrt{\Delta}} F^{y_w}, \nonumber\\
\sqrt{\Delta} y_u \frac{\partial F}{\partial y_u} &\bigg |_{y_v, y_w}& = (1 - u) (1 - v - w) F^u - u (1 - v) F^v - u (1 - w) F^w - u (1 - v - w) F^{1-u}   \nonumber\\ 
&&+ u v F^{1-v} + u w F^{1-w} + \sqrt{\Delta} F^{y_u} 
\eea
and the cyclic permutations of these formulae. We fix the overall integration constant of each function by stipulating that it vanish at the point $F(u,v,w) = F(1,1,1)$. The process of constructing such functions is described in ref.\ \cite{Dixon2013eka}, and proceeds analogously to the construction of the spurious pole functions which we cover in detail in appendix \ref{SP_basis}. The definitions of the basis functions in terms of their $\{n-1,1\}$ coproduct components are lengthy and unilluminating, so instead of writing them out explicitly we will only describe their formal properties; these definitions can be found in the ancillary files.

One way of organizing the space of irreducible hexagon functions is by the maximum number of times the $y_i$ variables appears in a single term of a function's symbol. Since this number is additive under function multiplication, it endows the space of hexagon functions with a grading which naturally separates parity-odd functions (those with odd numbers of $y$ entries) from parity-even functions (those with even numbers of $y$ entries). The dimension of the hexagon function space with each $y$-grading through weight eight is given in table \ref{tab:y_grading}. The hexagon function space also has an $S_3$ symmetry that acts by permuting the variables $u$, $v$, and $w$. We have selected basis functions that form orbits under this $S_3$ symmetry, and we only label one representative of each orbit, since the other members can be found by permuting the arguments of the representative.

\begin{table}[!t]
\centering
\begin{tabular}[t]{c || c c c c c c c}
\hline 
\hline
Weight & $ y^0 $ & $ y^1 $ & $ y^2 $ & $ y^3 $ & $ y^4 $ & $ y^5 $ & $ y^6 $ \\\hline
1 & 3 & - & - & - & - & - & - \\
2 & 3 & - & - & - & - & - & - \\
3 & 6 & 1 & - & - & - & - & - \\
4 & 9 & 3 & 3 & - & - & - & - \\
5 & 18 & 4 & 13 & 6 & - & - & - \\
6 & 27 & 4 & 27 & 29 & 18 & - & - \\
7 & 54 & 4 & 41 & 63 & 108 & 39 & - \\
8 & 90 & 4 & 50 & 108 & 306 & 238 & 114 \\
\hline
\hline
\end{tabular}
\caption{The dimension of the space of irreducible hexagon functions at each weight, graded by the maximum number of $y$ entries appearing in each function's symbol. The $y^0$ column counts one-dimensional HPLs, but the other columns are nontrivial.}
\label{tab:y_grading}
\end{table}

At weight six, we denote the $i^{\text{th}}$ odd function by $A_i(u,v,w)$ and the $i^{\text{th}}$ even function by $B_i(u,v,w)$. Up to the action of the $S_3$ symmetry there are 11 independent odd functions and 11 independent even functions. The size of each basis function's orbit is specified alongside its $y$-grading in table \ref{tab:w6_basis}. Similarly, we denote the weight seven odd functions by $C_i(u,v,w)$ and the weight seven even functions by $D_i(u,v,w)$. All 28 odd functions and 36 even functions fit into orbits of $S_3$, and these orbits are specified with the $y$-gradings of these functions in table \ref{tab:w7_basis}.
Finally, at weight eight there are 86 odd functions, denoted by $S_i(u,v,w)$, and 102 even functions, denoted by $T_i(u,v,w)$.  In table \ref{tab:w8_basis} we give their $S_3$ orbits and $y$-gradings.  We suppress the arguments of the basis functions in the tables.

\begin{table}[!t]
\centering
\begin{tabular}[t]{c || c | c | c | c}
\hline 
\hline
$S_3$ orbit & $ y^4$ & $ y^3 $ & $ y^2 $ & $ y^1 $ \\\hline
6-cycle & $B_1, B_2$ & $A_1, A_2$ & $B_5, B_6$ & - \\
3-cycle & $B_3, B_4$ & $A_3 \dots A_7$ & $B_7 \dots B_{11}$ & $A_{10}$  \\
singlet  & - & $A_8, A_9$ & - & $A_{11}$  \\
\hline
\hline
\end{tabular}
\caption{The weight-six hexagon basis functions organized by the size of their $S_3$ orbits and $y$-grading.}
\label{tab:w6_basis}
\end{table}

\begin{table}[!t]
\centering
\begin{tabular}[t]{c || c | c | c | c | c }
\hline 
\hline
$S_3$ orbit & $y^5$ & $y^4$ & $y^3$ & $y^2$ & $y^1$ \\\hline
6-cycle & $C_1 \dots C_3$ & $D_1 \dots D_{12}$ & $C_{11} \dots C_{15}$ & $D_{25},D_{26},D_{27}$ & - \\
3-cycle & $C_4 \dots C_{10}$ & $D_{13} \dots  D_{24}$ & $C_{16} \dots C_{26}$ &  $D_{28} \dots D_{34}$ & $C_{27}$  \\
singlet  & - & - & - & $D_{35}, D_{36}$ & $C_{28}$  \\
\hline
\hline
\end{tabular}
\caption{The weight-seven hexagon basis functions organized by the size of their $S_3$ orbits and $y$-grading.}
\label{tab:w7_basis}
\end{table}

\begin{table}[!t]
\centering
\begin{tabular}[t]{c || c | c | c | c | c | c}
\hline\hline $S_3$ orbit & $y^6$ & $y^5$ & $y^4$ & $y^3$ & $y^2$ & $y^1$
\\\hline
 6-cycle & $T_1 \dots T_{15} $ & $S_{1} \dots S_{26}$
   & $T_{16} \dots T_{54} $ & $S_{27} \dots S_{34}$ & $T_{55} \dots T_{58} $
   & - \\
 3-cycle & $T_{59} \dots T_{66} $ & $S_{35} \dots S_{61} $
   & $T_{67} \dots T_{89} $ & $S_{62} \dots S_{80} $ & $T_{90} \dots T_{97} $
   & $S_{81}$ \\
 singlet & - & $S_{82}$ & $T_{98}, T_{99}, T_{100} $
    & $S_{83}, S_{84}, S_{85} $ & $T_{101}, T_{102} $ & $S_{86}$
 \\ \hline \hline
\end{tabular}
\caption{The weight-eight hexagon basis functions organized by the size of their $S_3$ orbits and $y$-grading.}
\label{tab:w8_basis}
\end{table}


\section{$R_6^{(3)}$, $V^{(3)}$ and $\Vt^{(3)}$ in terms of weight-six basis}
\label{RVVt3inwt6basis}

The weight-six basis for the parity-odd sector includes functions $A_i(u,v,w)$,
$i=1,2,\ldots,11$, and for the parity-even sector, $B_i(u,v,w)$,
$i=1,2,\ldots,11$.  This basis allows us to write the three-loop remainder
and ratio functions directly, instead of in terms of their $\{5,1\}$
coproduct components, as was done
previously~\cite{Dixon2013eka,Dixon2014iba}.

Using the parity-even weight-six functions and the total symmetry of the
remainder function, we can write the three-loop result as
\bea
R_6^{(3)}(u,v,w) &=& R_{6,A}^{(3)}(u,v,w) + R_{6,A}^{(3)}(v,w,u) + R_{6,A}^{(3)}(w,u,v)
\nonumber\\ &&\hskip0cm\null
+ R_{6,A}^{(3)}(u,w,v) + R_{6,A}^{(3)}(v,u,w) + R_{6,A}^{(3)}(w,v,u)\,,
\label{R63A}
\eea
where
\bea
R_{6,A}^{(3)} \! &=& \!\frac{1}{128} \biggl\{
30 ( B_1 + B_2 ) + 18 B_3 + 14 B_4 - 2 ( B_5 + B_6 ) + 12 B_7
- 2 B_8 - 264 B_9 \! + \! 2 B_{10} \! - \! 24 B_{11}
\nonumber\\ &&\hskip0cm\null
 + 4 \ln(u/v) M_1 + \frac{128}{3} \ln(w/v) \Qep
 + [ - 400 H_{2}^u - 200 H_{2}^w + 10 \ln^2 u + 2 \ln^2 w
\nonumber\\ &&\hskip0cm\null
     + 204 \ln u \ln v - 412 \ln u \ln w + 408 \zeta_2 ]
    \, \Omega^{(2)}
 + \frac{104}{3} [\tPhi]^2
\nonumber\\ &&\hskip0cm\null
 - 96 H_{6}^u + 56 H_{5,1}^u - 6 H_{4,2}^u - 204 H_{4,1,1}^u - 44 H_{3,2,1}^u
 - 6 H_{3,1,2}^u + 168 H_{3,1,1,1}^u + 6 H_{2,2,1,1}^u
\nonumber\\ &&\hskip0cm\null
 - 210 H_{2,1,1,1,1}^u
 - 2 (H_{3}^u)^2 - 4 H_{3}^u H_{2,1}^u + 14 (H_{2,1}^u)^2
 + H_{2}^u \, [ - 376 H_{4}^u + 8 H_{3,1}^u + 594 H_{2,1,1}^u
\nonumber\\ &&\hskip0cm\null
   + 96 (H_{2}^u)^2 ]
 + \ln u \, [ 96 H_{5}^u - 56 H_{4,1}^u + 10 H_{3,2}^u + 168 H_{3,1,1}^u
           + 26 H_{2,2,1}^u - 144 H_{2,1,1,1}^u
\nonumber\\ &&\hskip0cm\null
           + H_{2}^u ( 380 H_{3}^u + 2 H_{2,1}^u ) ]
 + \ln^2 u \, [ - 26 H_{4}^u + 38 H_{3,1}^u - 60 H_{2,1,1}^u - 100 (H_{2}^u)^2 ]
\nonumber\\ &&\hskip0cm\null
 - 4 \ln^3 u ( H_{3}^u + 2 H_{2,1}^u ) + \frac{7}{4} \ln^4 u \, H_{2}^u
  - H_{3}^v \Bigl[ 338 H_{3}^u + 4 H_{2,1}^u + \frac{10}{3} \ln u H_{2}^u
            - \frac{7}{3} \ln^3 u \Bigr]
\nonumber\\ &&\hskip0cm\null
  + H_{2,1}^v \Bigl[ 690 H_{2,1}^u - \frac{34}{3} \ln u H_{2}^u
    - \frac{5}{3} \ln^3 u \Bigr]
 + H_{2}^v \Bigl[ - 760 H_{4}^u - 8 H_{3,1}^u + 1224 H_{2,1,1}^u
\nonumber\\ &&\hskip0cm\null
   + \frac{575}{2} (H_{2}^u)^2 - \frac{64}{3} H_{2}^u H_{2}^w
              + \ln u \Bigl( \frac{2314}{3} H_{3}^u + \frac{82}{3} H_{2,1}^u \Bigr)
              - \frac{610}{3} \ln^2 u H_{2}^u - \frac{5}{2} \ln^4 u \Bigr]
\nonumber\\ &&\hskip0cm\null
 + \ln v \Bigl[ - 8 H_{4,1}^u - 10 H_{3,2}^u + 120 H_{3,1,1}^u
             + 38 H_{2,2,1}^u - 48 H_{2,1,1,1}^u
             + H_{2}^u ( 4 H_{3}^u - 18 H_{2,1}^u )
\nonumber\\ &&\hskip0cm\null
             + \ln u  \Bigl( - 276 H_{4}^u - 320 H_{3,1}^u - 174 H_{2,1,1}^u
                       + 197 (H_{2}^u)^2 \Bigr)
             + 2 \ln^2 u ( - 30 H_{2,1}^u + 119 H_{3}^u )
\nonumber\\ &&\hskip0cm\null
             - 51 \ln^3 u H_{2}^u \Bigr]
 + \ln^2 v \Bigl[ 34 H_{4}^u - 10 H_{3,1}^u - 6 H_{2,1,1}^u - 9 (H_{2}^u)^2
             - \ln u \Bigl( \frac{403}{3} H_{3}^u + \frac{1837}{3} H_{2,1}^u \Bigr)
\nonumber\\ &&\hskip0cm\null
             - \frac{1687}{6} \ln^2 u H_{2}^u - 2 \ln^4 u \Bigr]
 - \ln^3 v \Bigl[ \frac{631}{3} \ln u H_{2}^u + \frac{103}{2} \ln^3 u \Bigr]
 + \frac{256}{3} \ln v H_{2}^v \ln u H_{2}^u
\nonumber\\ &&\hskip0cm\null
 + \ln w \Bigl[ H_{2}^v ( 16 H_{2,1}^u + 1018 \ln u H_{2}^u
                      + 201 \ln^3 u )
        + \ln v \Bigl( - 764 H_{4}^u - 164 H_{3,1}^u
\nonumber\\ &&\hskip0cm\null
                      + 846 H_{2,1,1}^u + 126 (H_{2}^u)^2
                       + \ln u ( 334 H_{3}^u + 582 H_{2,1}^u )
                       + \frac{1037}{2} \ln^2 u H_{2}^u + \frac{9}{2} \ln^4 u \Bigr)
\nonumber\\ &&\hskip0cm\null
           + \ln^2 v \Bigl( 114 H_{3}^u + 643 H_{2,1}^u + \frac{913}{2} \ln u H_{2}^u
                       + \frac{773}{12} \ln^3 u \Bigr) \Bigr]
\nonumber\\ &&\hskip0cm\null
 - \ln^2 w \Bigl[ 8 H_{2}^u H_{2}^v + \frac{335}{2} \ln^2 v H_{2}^u
             + \frac{457}{6} \ln^2 u \ln^2 v \Bigr]
\nonumber\\ &&\hskip0cm\null
 + \zeta_2 [ 788 H_{4}^u - 4 H_{3,1}^u - 1236 H_{2,1,1}^u + 220 (H_{2}^u)^2
             - 783 \ln u H_{3}^u + 4 \ln u H_{2,1}^u
\nonumber\\ &&\hskip0cm\null
             + 407 \ln^2 u H_{2}^u - \frac{5}{2} \ln^4 u
             + \ln v ( - 13 H_{3}^u + 2244 H_{2,1}^u + 297 \ln u H_{2}^u
                         - 39 \ln^3 u )
\nonumber\\ &&\hskip0cm\null
             + \ln^2 v ( 822 H_{2}^u + 314 \ln^2 u )
             + 858 H_{2}^u H_{2}^v
             - ( 400 H_{2}^u + 314 \ln^2 u ) \ln v \ln w ]
\nonumber\\ &&\hskip0cm\null
 + \zeta_3 [ 28 \ln u H_{2}^u + 6 \ln^3 u
             + \ln v ( - 40 H_{2}^u + 125 \ln^2 u )
             + 58 \ln u \ln v \ln w ]
\nonumber\\ &&\hskip0cm\null
 - \zeta_4 [ 5704 H_{2}^u + 980 \ln^2 u + 464 \ln u \ln v ]
 + \frac{41860}{9} \zeta_6 + \frac{64}{3} (\zeta_3)^2 \biggr\} \,,
\label{R63Adef}
\eea
and we dropped the arguments $(u,v,w)$ on $B_i$, $M_1$, $\Qep$, $\Omega^{(2)}$ 
and $\tPhi$ to save space.

The parity-even part of the three-loop ratio function is
\be
V^{(3)}(u,v,w)\ =\ V^{(3)}_A(u,v,w) + V^{(3)}_A(w,v,u),
\label{V3A} 
\ee
where
\bea
V^{(3)}_A(u,v,w) &=&
\frac{1}{128} \biggl\{
 - 42 B_1(u,v,w) - 38 B_1(v,w,u) - 54 B_1(w,u,v)
 - 38 B_2(u,v,w)
\nonumber\\ && \hskip0.0cm\null
 - 54 B_2(v,w,u) - 42 B_2(w,u,v)
 - 60 B_3(u,v,w) - 18 B_3(v,w,u)
\nonumber\\ && \hskip0.0cm\null
 - 40 B_4(u,v,w) - 20 B_4(v,w,u)
 + 2 B_5(u,v,w) + 2 B_5(v,w,u) + 10 B_5(w,u,v)
\nonumber\\ && \hskip0.0cm\null
 + 2 B_6(u,v,w) + 2 B_6(v,w,u) + 10 B_6(w,u,v)
 - 36 B_7(u,v,w) - 20 B_7(v,w,u)
\nonumber\\ && \hskip0.0cm\null
 + 4 B_8(u,v,w)
 + 816 B_9(u,v,w) + 348 B_9(v,w,u)
\nonumber\\ && \hskip0.0cm\null
 - 12 B_{10}(u,v,w) - 2 B_{10}(v,w,u)
 + 64 B_{11}(u,v,w) + 48 B_{11}(v,w,u)
\nonumber\\ && \hskip0.0cm\null
 + \ln(v/u) \Bigl[ 4 M_1(u,v,w)
              + \frac{128}{3} ( 5 \Qep(v,w,u) - \Qep(w,u,v) ) \Bigr]
\nonumber\\ && \hskip0.0cm\null
 + \ln(w/v) \Bigl[ 4 M_1(v,w,u)
              + \frac{128}{3} ( 5 \Qep(v,w,u) - \Qep(u,v,w) ) \Bigr]
\nonumber\\ && \hskip0.0cm\null
 + 2 \Bigl[ 292 ( H_2^u + H_2^v + H_2^w )
       - 6 \ln^2 u - 3 \ln^2 w - 5 \ln^2 v
       - 312 \ln u \ln v
\nonumber\\ && \hskip1.0cm\null
       + 298 \ln w \ln v + 316 \ln u \ln w
       - 596 \zeta_2 \Bigr] \Omega^{(2)}(u,v,w)
\nonumber\\ && \hskip0.0cm\null
 + 2 \Bigl[ 142 ( 2 H_2^u + H_2^v ) - 12 \ln^2 u - 3 \ln^2 v
       + ( 298 \ln v - 145 \ln w ) \ln u
\nonumber\\ && \hskip1.0cm\null
       - 286 \zeta_2 \Bigr] \Omega^{(2)}(w,u,v)
\nonumber\\ && \hskip0.0cm\null
 - 154 [\tPhi(u,v,w)]^2
\ +\ \hbox{pure HPLs} \biggr\} \,.
\label{V3Adef}
\eea
The pure HPL terms are quite lengthy, so we only present them
in an ancillary file.

The parity-odd part of the three-loop ratio function can be
presented here in its entirety,
\be
\Vt^{(3)}(u,v,w)\ =\ \Vt^{(3)}_A(u,v,w) - \Vt^{(3)}_A(w,v,u),
\label{Vt3A} 
\ee
where
\bea
\Vt^{(3)}_A(u,v,w) &=&
\frac{1}{128} \biggl\{
 - \frac{4}{3} \, A_1(u,v,w) - \frac{28}{3} \, A_1(v,w,u)
 + \frac{32}{3} \, A_1(w,u,v) + \frac{8}{3} \, A_2(u,v,w)
\nonumber\\ && \hskip0.3cm\null
 - \frac{28}{3} \, A_2(v,w,u) + \frac{20}{3} \, A_2(w,u,v) + 12 \, A_3(u,v,w)
 + 4 \, A_4(u,v,w) 
\nonumber\\ && \hskip0.3cm\null
 - 12 \, A_6(u,v,w) + 12 \, A_7(u,v,w) - 120 \, A_{10}(u,v,w)
\nonumber\\ && \hskip0.3cm\null
 - \frac{4}{3} \, \ln u \, H_1(u,v,w)
 - \frac{4}{3} \, ( 3 \, \ln w - \ln u - 2 \, \ln v ) \, H_1(v,w,u)
\nonumber\\ && \hskip0.3cm\null
 - \frac{23}{3} \, \ln u \, J_1(u,v,w)
 + \frac{1}{3} \, ( 3 \, \ln w - 13 \, \ln u + 10 \, \ln v ) \, J_1(v,w,u)
\nonumber\\ && \hskip0.3cm\null
 - 2 \, \Bigl[ 4 \, ( H_2^u + H_2^v + H_2^w )
       + 5 \, \ln^2u + 4 \, \ln^2w - 4 \, \ln u \, \ln w - 2 \, \ln u \, \ln v
\nonumber\\ && \hskip1.3cm\null
       + 3 \, \ln^2v - 12 \, \zeta_2 \Bigr] \, F_1(u,v,w)
\nonumber\\ && \hskip0.3cm\null
 + 2 \, ( \ln^2u - 2 \, \ln u \, \ln v ) \, F_1(v,w,u)
\nonumber\\ && \hskip0.3cm\null
 + 4 \, \Bigl[ 2 \, ( \ln u - \ln w ) \, H_2^u + 2 \, \ln^3u 
       - \ln^2u \, ( 3 \, \ln w + \ln v )
\nonumber\\ && \hskip1.3cm\null
       + 2 \, \ln u \, ( H_2^v + \ln^2v )
       - 26 \, \zeta_2 \, \ln u \Bigr] \, \tPhi(u,v,w) \biggr\} \,.
\label{Vt3Adef}
\eea

\newpage

\section{$R_6^{(4)}$, $V^{(4)}$ and $\Vt^{(4)}$ in terms of weight-eight basis}
\label{RVVt4inwt8basis}

Using the weight-eight basis, we can describe the four-loop quantities $R_6^{(4)}$, $V^{(4)}$ and $\Vt^{(4)}$ directly, instead of via their $\{7,1\}$ coproduct components.

First we present the four-loop remainder function $R_6^{(4)}$.  Because this function is totally symmetric in $(u,v,w)=(u_1,u_2,u_3)$, we can express it in terms of the weight-eight basis as,
\newcommand{\si}{\sigma}
\bea
R_6^{(4)} &=& \frac{1}{1024} \biggl\{
\sum_{\si\in S_3} \Bigl[
- 320 T_{1}^\si - 324 T_{2}^\si - 290 T_{3}^\si - 268 T_{4}^\si
- 252 T_{5}^\si - 292 T_{6}^\si - 248 T_{7}^\si
\nonumber\\&&\hskip2.0cm\null
- 252 T_{9}^\si - 248 T_{10}^\si - 248 T_{11}^\si - 272 T_{12}^\si
- 296 T_{13}^\si - 256 T_{14}^\si - 296 T_{15}^\si
\nonumber\\&&\hskip2.0cm\null
+ 4848 T_{16}^\si + 5268 T_{17}^\si - 4 T_{18}^\si - 4 T_{19}^\si
+ 1173 T_{20}^\si - 254 T_{21}^\si - 4 T_{22}^\si
\nonumber\\&&\hskip2.0cm\null
+ 12 T_{23}^\si + 312 T_{24}^\si + 292 T_{25}^\si
+ 24 T_{26}^\si + 252 T_{27}^\si + 8 T_{29}^\si + 4 T_{30}^\si
\nonumber\\&&\hskip2.0cm\null
+ \frac{725}{3} T_{31}^\si + 20 T_{32}^\si + 24 T_{33}^\si + 12 T_{34}^\si
+ \frac{1165}{3} T_{35}^\si + 724 T_{36}^\si + 4 T_{37}^\si
\nonumber\\&&\hskip2.0cm\null
+ 24 T_{38}^\si + 24 T_{39}^\si + 20 T_{40}^\si - 32 T_{41}^\si
- 48 T_{42}^\si - 32 T_{43}^\si + 4 T_{44}^\si
\nonumber\\&&\hskip2.0cm\null
- 16 T_{45}^\si - 48 T_{46}^\si - 16 T_{47}^\si + 40 T_{48}^\si
- 28 T_{49}^\si - 28 T_{50}^\si - 40 T_{51}^\si
\nonumber\\&&\hskip2.0cm\null
+ 16 T_{52}^\si + 20 T_{53}^\si + 20 T_{54}^\si
- 336 T_{55}^\si + 177 T_{57}^\si - 4 T_{58}^\si \Bigr]
\nonumber\\&&\hskip0.8cm\null
+ \sum_{\si\in Z_3} \Bigl[
- 200 T_{59}^\si - 128 T_{60}^\si - 136 T_{61}^\si - 132 T_{62}^\si
- 132 T_{64}^\si - 128 T_{65}^\si - 145 T_{66}^\si
\nonumber\\&&\hskip2.0cm\null
+ 2712 T_{67}^\si + 2520 T_{68}^\si - \frac{502}{3} T_{69}^\si - 114 T_{70}^\si
- \frac{122}{3} T_{71}^\si + \frac{2216}{3} T_{72}^\si
\nonumber\\&&\hskip2.0cm\null
+ 8 T_{73}^\si + 390 T_{74}^\si + 8 T_{75}^\si - 8 T_{76}^\si - 24 T_{77}^\si
- 8 T_{78}^\si + \frac{3827}{9} T_{80}^\si - 24 T_{81}^\si
\nonumber\\&&\hskip2.0cm\null
+ \frac{215}{6} T_{83}^\si - 160 T_{84}^\si + 20 T_{85}^\si - T_{86}^\si
- 4 T_{87}^\si + 2 T_{88}^\si - 116 T_{89}^\si
\nonumber\\&&\hskip2.0cm\null
+ \frac{11102}{3} T_{90}^\si + 197232 T_{91}^\si + 336 T_{92}^\si
- \frac{18465}{4} T_{93}^\si + \frac{12643}{3} T_{94}^\si
\nonumber\\&&\hskip2.0cm\null
- 79 T_{95}^\si + \frac{6113}{6} T_{96}^\si - \frac{3427}{6} T_{97}^\si \Bigr]
\nonumber\\&&\hskip0.8cm\null
- \frac{5741}{6} T_{100} + \frac{17467}{6} T_{101} - \frac{292661}{72} T_{102}
\nonumber\\&&\hskip0.8cm\null
+\ \hbox{products of lower weight functions} \biggl\} \,,
\label{R64_Abasis_T}
\eea
where $T_i^\si$ denotes a permuted version of $T_i \equiv T_i(u,v,w) = T_i(u_1,u_2,u_3)$,
namely
\be
T_i^\si\ \equiv\ T_i(u_{\si(1)},u_{\si(2)},u_{\si(3)}).
\ee
We sum over all six permutations
of the 6-cycle basis functions, $T_1,\ldots,T_{58}$, and over the three cyclic permutations of the 3-cycle ones, $T_{59},\ldots,T_{97}$. We have dropped the terms that are products of lower weight functions because they are very lengthy, but they are given in an ancillary file.

The parity-even part of the four-loop ratio function can be expressed in terms
of the same $T_i$ functions as
\be
V^{(4)}(u,v,w) = V_A^{(4)}(u,v,w) + V_A^{(4)}(w,v,u) + V_B^{(4)}(u,v,w) \,,
\label{V4_Abasis_T}
\ee
where
\bea
V_A^{(4)} \!\! &=& \!\! \frac{1}{1024} \biggl\{
  380 T_{1}^u + 620 T_{1}^v + 500 T_{1}^w
 + 596 T_{2}^u + 516 T_{2}^v + 396 T_{2}^w
 + 542 T_{3}^u + 440 T_{3}^v + 398 T_{3}^w
\nonumber\\&&\hskip-0.15cm\null
 + 376 T_{4}^u + 450 T_{4}^v + 446 T_{4}^w
 + 380 T_{5}^u + 434 T_{5}^v + 392 T_{5}^w
 + 436 T_{6}^u + 564 T_{6}^v + 394 T_{6}^w
\nonumber\\&&\hskip-0.15cm\null
 + 400 T_{7}^u +  414 T_{7}^v + 388 T_{7}^w
 + 422 T_{9}^u + 404 T_{9}^v + 396 T_{9}^w
 + 388 T_{10}^u + 426 T_{10}^v + 376 T_{10}^w
\nonumber\\&&\hskip-0.15cm\null
 + 392 T_{11}^u + 426 T_{11}^v + 376 T_{11}^w
 + 374 T_{12}^u + 464 T_{12}^v + 448 T_{12}^w
 + 404 T_{13}^u + 554 T_{13}^v + 446 T_{13}^w
\nonumber\\&&\hskip-0.15cm\null
 + 432 T_{14}^u + 406 T_{14}^v + 406 T_{14}^w
 + 554 T_{15}^u + 446 T_{15}^v + 404 T_{15}^w
 - 6984 T_{16}^u - 7584 T_{16}^v - 8604 T_{16}^w
\nonumber\\&&\hskip-0.15cm\null
 - 8347 T_{17}^u - 9102 T_{17}^v - 7576 T_{17}^w
 + 28 T_{18}^u + 4 T_{18}^v + 16 T_{18}^w
 + 16 T_{19}^u - 8 T_{19}^v + 4 T_{19}^w
\nonumber\\&&\hskip-0.15cm\null
 - \frac{7689}{4} T_{20}^u - \frac{22685}{12} T_{20}^v - 1852 T_{20}^w
 + 376 T_{21}^u + 403 T_{21}^v + 428 T_{21}^w
 + 16 T_{22}^u + 4 T_{22}^v + 28 T_{22}^w
\nonumber\\&&\hskip-0.15cm\null
 - 12 T_{23}^u - 26 T_{23}^v - 24 T_{23}^w
 - 482 T_{24}^u - 562 T_{24}^v - 388 T_{24}^w
 - 408 T_{25}^u - 434 T_{25}^v - 542 T_{25}^w
\nonumber\\&&\hskip-0.15cm\null
 + 12 T_{26}^u - 72 T_{26}^v - 24 T_{26}^w
 - 456 T_{27}^u - 394 T_{27}^v - 422 T_{27}^w
 + 40 T_{28}^u - 36 T_{28}^v - 36 T_{28}^w
\nonumber\\&&\hskip-0.15cm\null
 - 8 T_{29}^u - 32 T_{29}^v - 20 T_{29}^w
 - 4 T_{30}^u - 28 T_{30}^v - 16 T_{30}^w
 - \frac{2621}{6} T_{31}^u - \frac{605}{2} T_{31}^v - \frac{1219}{3} T_{31}^w
\nonumber\\&&\hskip-0.15cm\null
 - 6 T_{32}^u - 20 T_{32}^v - 26 T_{32}^w
 - 24 T_{33}^u - 10 T_{33}^v + 16 T_{33}^w
 - 26 T_{34}^u - 12 T_{34}^v - 24 T_{34}^w
\nonumber\\&&\hskip-0.15cm\null
 - \frac{2965}{6} T_{35}^u - \frac{1405}{2} T_{35}^v - 729 T_{35}^w
 - 1031 T_{36}^u - 1159 T_{36}^v - \frac{2537}{2} T_{36}^w
 - 4 T_{37}^u - 4 T_{37}^v - 4 T_{37}^w
\nonumber\\&&\hskip-0.15cm\null
 - 24 T_{38}^u - 96 T_{38}^v
 - 42 T_{39}^u + 12 T_{39}^v - 54 T_{39}^w
 + 2 T_{40}^u - 20 T_{40}^v - 90 T_{40}^w
\nonumber\\&&\hskip-0.15cm\null
 + 102 T_{41}^u + 32 T_{41}^v + 46 T_{41}^w
 + 120 T_{42}^u + 48 T_{42}^v + 18 T_{42}^w
 + 38 T_{43}^u + 32 T_{43}^v + 26 T_{43}^w
\nonumber\\&&\hskip-0.15cm\null
 - 4 T_{44}^u - 16 T_{44}^v + 8 T_{44}^w
 - 20 T_{45}^u + 16 T_{45}^v - 56 T_{45}^w
 + 80 T_{46}^u + 128 T_{46}^v + 48 T_{46}^w
\nonumber\\&&\hskip-0.15cm\null
 + 4 T_{47}^u + 16 T_{47}^v - 8 T_{47}^w
 - 40 T_{48}^u - 24 T_{48}^v - 96 T_{48}^w
 + 36 T_{49}^u + 38 T_{49}^v + 28 T_{49}^w
\nonumber\\&&\hskip-0.15cm\null
 + 28 T_{50}^u + 108 T_{50}^v + 42 T_{50}^w
 + 24 T_{51}^u + 96 T_{51}^v + 40 T_{51}^w
 - 16 T_{52}^u - 28 T_{52}^v - 22 T_{52}^w
\nonumber\\&&\hskip-0.15cm\null
 - 20 T_{53}^u - 26 T_{53}^v - 6 T_{53}^w
 - 10 T_{54}^u - 78 T_{54}^v - 20 T_{54}^w
 + 264 T_{55}^u + 756 T_{55}^v + 336 T_{55}^w
\nonumber\\&&\hskip-0.15cm\null
 + 3 T_{57}^u - 177 T_{57}^v - 102 T_{57}^w
 - 6 T_{58}^u - 2 T_{58}^v + 4 T_{58}^w
\nonumber\\&&\hskip-0.15cm\null
 + 200 T_{59}^u + 213 T_{60}^u + 190 T_{61}^u
 + 186 T_{62}^u + 186 T_{64}^u + 213 T_{65}^u
 + \frac{419}{2} T_{66}^u - 3468 T_{67}^u
\nonumber\\&&\hskip-0.15cm\null
 - \frac{8119}{2} T_{68}^u + 235 T_{69}^u + 204 T_{70}^u
 + 49 T_{71}^u - 1166 T_{72}^u - 44 T_{73}^u - 544 T_{74}^u
 - 20 T_{75}^u + 8 T_{76}^u
\nonumber\\&&\hskip-0.15cm\null
 + 48 T_{77}^u + 8 T_{78}^u - 18 T_{79}^u - \frac{23861}{36} T_{80}^u
 + 24 T_{81}^u + 22 T_{82}^u - \frac{190}{3} T_{83}^u
\nonumber\\&&\hskip-0.15cm\null
 + \frac{1291}{6} T_{84}^u - 20 T_{85}^u - \frac{7}{2} T_{86}^u
 + 4 T_{87}^u - 22 T_{88}^u + 202 T_{89}^u
 - 4999 T_{90}^u - 284328 T_{91}^u - 510 T_{92}^u
\nonumber\\&&\hskip-0.15cm\null
 + \frac{42173}{6} T_{93}^u - \frac{64501}{12} T_{94}^u + 79 T_{95}^u
 - \frac{34631}{24} T_{96}^u + \frac{4467}{4} T_{97}^u 
\nonumber\\&&\hskip-0.15cm\null
+ \ \hbox{products of lower weight functions} \biggr\} \,,
\label{V4A}
\eea
and
\bea
V_B^{(4)} &=& \frac{1}{1024} \biggl\{
   500 T_{59}^v + 193 T_{60}^v + 252 T_{61}^v
 + 244 T_{62}^v + 256 T_{64}^v + 193 T_{65}^v
 + 271 T_{66}^v - 5712 T_{67}^v
\nonumber\\&&\hskip0.8cm\null
 - 3922 T_{68}^v + \frac{1012}{3} T_{69}^v + 173 T_{70}^v
 + \frac{304}{3} T_{71}^v - 1215 T_{72}^v + 100 T_{73}^v
 - \frac{1561}{2} T_{74}^v + 28 T_{75}^v
\nonumber\\&&\hskip0.8cm\null
 - 100 T_{76}^v + 24 T_{77}^v - 28 T_{78}^v + 22 T_{79}^v
 - \frac{13825}{18} T_{80}^v + 144 T_{81}^v + 18 T_{82}^v
 - \frac{157}{6} T_{83}^v
\nonumber\\&&\hskip0.8cm\null
 + \frac{839}{3} T_{84}^v - 68 T_{85}^v + \frac{1}{2} T_{86}^v
 + 4 T_{87}^v + 30 T_{88}^v + 170 T_{89}^v
 - \frac{46967}{6} T_{90}^v - 367344 T_{91}^v
\nonumber\\&&\hskip0.8cm\null
 - 336 T_{92}^v + \frac{49109}{6} T_{93}^v - 9155 T_{94}^v + 364 T_{95}^v
 - \frac{8521}{4} T_{96}^v + \frac{1633}{3} T_{97}^v
\nonumber\\&&\hskip0.8cm\null
 + 12 T_{98}^u + 4 T_{99}^u + \frac{9155}{6} T_{100}^u
 - \frac{170141}{36} T_{101}^u + \frac{145829}{24} T_{102}^u
\nonumber\\&&\hskip0.8cm\null
+ \ \hbox{products of lower weight functions}  \biggr\} \,.
\label{V4B}
\eea
Here $T_i^u = T_i(u,v,w)$, $T_i^v = T_i(v,w,u)$, $T_i^w = T_i(w,u,v)$.
The 3-cycle functions $T_{59},\ldots,S_{97}$ are chosen to be symmetric
in their last two arguments, so for these functions the permutation $T_i(v,w,u)$
appears only in $V_B^{(4)}$.

Similarly, the parity-odd part of the four-loop ratio function can be expressed as
\be
\Vt^{(4)}(u,v,w) = \Vt_A^{(4)}(u,v,w) - \Vt_A^{(4)}(w,v,u) \,,
\label{Vt4_Abasis_S}
\ee
where
\bea
\Vt_A^{(4)} &=& \frac{1}{3072} \biggl\{
 - 300 S_{1}^u + 60 S_{1}^v + 240 S_{1}^w
 - 126 S_{2}^u + 18 S_{2}^v + 108 S_{2}^w
 + 156 S_{3}^u - 222 S_{3}^v + 66 S_{3}^w
\nonumber\\&&\hskip0.8cm\null
 - 40 S_{4}^u + 206 S_{4}^v - 166 S_{4}^w
 - 166 S_{5}^u - 112 S_{5}^v + 278 S_{5}^w
 - 976 S_{6}^u + 278 S_{6}^v + 698 S_{6}^w
\nonumber\\&&\hskip0.8cm\null
 + 44 S_{7}^u - 52 S_{7}^v + 8 S_{7}^w
 - 124 S_{8}^u + 224 S_{8}^v - 100 S_{8}^w
 + 48 S_{9}^u + 192 S_{9}^v - 240 S_{9}^w
\nonumber\\&&\hskip0.8cm\null
 + 720 S_{10}^u - 1110 S_{10}^v + 390 S_{10}^w
 + 178 S_{11}^u - 242 S_{11}^v + 64 S_{11}^w
 + 150 S_{12}^u - 150 S_{12}^v
\nonumber\\&&\hskip0.8cm\null
 - 196 S_{13}^u - 38 S_{13}^v + 234 S_{13}^w
 - 96 S_{14}^u - 18 S_{14}^v + 114 S_{14}^w
 + 78 S_{15}^u - 78 S_{15}^w
\nonumber\\&&\hskip0.8cm\null
 + 114 S_{16}^v - 114 S_{16}^w
 - 78 S_{17}^u - 18 S_{17}^v + 96 S_{17}^w
 - 122 S_{18}^u - 26 S_{18}^v + 148 S_{18}^w
\nonumber\\&&\hskip0.8cm\null
 - 122 S_{19}^u - 26 S_{19}^v + 148 S_{19}^w
 + 96 S_{20}^u + 18 S_{20}^v - 114 S_{20}^w
 - 454 S_{21}^u + 56 S_{21}^v + 398 S_{21}^w
\nonumber\\&&\hskip0.8cm\null
 + 12 S_{22}^u + 12 S_{22}^v - 24 S_{22}^w
 - 18 S_{23}^u + 96 S_{23}^v - 78 S_{23}^w
 + 114 S_{24}^u - 96 S_{24}^v - 18 S_{24}^w
\nonumber\\&&\hskip0.8cm\null
 - 166 S_{25}^u - 40 S_{25}^v + 206 S_{25}^w
 - 166 S_{26}^u - 40 S_{26}^v + 206 S_{26}^w
+ 396 S_{27}^u - 2664 S_{27}^v
\nonumber\\&&\hskip0.8cm\null
+ 2268 S_{27}^w
 + 2831 S_{28}^u - 259 S_{28}^v - 2572 S_{28}^w
 - 8 S_{29}^u - 146 S_{29}^v + 154 S_{29}^w
\nonumber\\&&\hskip0.8cm\null
 - \frac{215}{2} S_{30}^u + 218 S_{30}^v - \frac{221}{2} S_{30}^w
 - 20 S_{31}^u - 966 S_{31}^v + 986 S_{31}^w
\nonumber\\&&\hskip0.8cm\null
 - 136 S_{32}^u + 8 S_{32}^v + 128 S_{32}^w
 + 34 S_{33}^u - 8 S_{33}^v - 26 S_{33}^w
 + 1053 S_{34}^u - 1239 S_{34}^v + 186 S_{34}^w
\nonumber\\&&\hskip0.8cm\null
 + 126 S_{35}^u + 126 S_{36}^u - 1666 S_{38}^u + 228 S_{39}^u
+ 360 S_{40}^u + 712 S_{41}^u + 2843 S_{42}^u - 72 S_{43}^u
\nonumber\\&&\hskip0.8cm\null
+ 376 S_{45}^u- 153 S_{46}^u- 492 S_{47}^u + 610 S_{48}^u
+ 200 S_{49}^u - 846 S_{50}^u + 884 S_{52}^u - 462 S_{53}^u
\nonumber\\&&\hskip0.8cm\null
+ 27 S_{54}^u + 78 S_{55}^u + 114 S_{57}^u + 78 S_{58}^u
- 2313 S_{62}^u + 177 S_{63}^u - 3060 S_{64}^u
\nonumber\\&&\hskip0.8cm\null
+ \frac{14490793}{44} S_{65}^u + 81 S_{66}^u
- \frac{84153}{2} S_{67}^u + 2227 S_{68}^u + 20 S_{69}^u
+ 1354 S_{70}^u + \frac{1484251}{44} S_{71}^u
\nonumber\\&&\hskip0.8cm\null
+ \frac{1203}{2} S_{72}^u + \frac{657}{4} S_{73}^u - \frac{34985}{2} S_{75}^u
- 808 S_{76}^u + 62 S_{77}^u - \frac{28471}{4} S_{78}^u + 759 S_{79}^u
\nonumber\\&&\hskip0.8cm\null
+ 1065 S_{80}^u - 249048 S_{81}^u
\ + \ \hbox{products of lower weight functions} \biggr\} \,,
\label{Vt4A}
\eea
and $S_i^u = S_i(u,v,w)$, $S_i^v = S_i(v,w,u)$, $S_i^w = S_i(w,u,v)$.
Note that the singlet functions $S_{82},\ldots,S_{86}$ cannot appear in an
antisymmetric quantity such as $\Vt$.  Again, the 3-cycle functions
$S_{35},\ldots,S_{81}$ are chosen to be symmetric in their last two
arguments, so for these functions the permutation $S_i(v,w,u)$
cannot appear, and $S_i(w,u,v)$ is related by the $u\lr w$ exchange.
The products of lower weight functions for both $V^{(4)}$ and $\Vt^{(4)}$ are too lengthy to present here, but they are given in ancillary files.

\newpage

\section{Functions on the spurious pole surface $w=1$}
\label{SP_basis}

In \sect{cubefaces} we explored the behavior of the ratio function in the limit $w\to1$.  We also need to understand this limit in order to impose the spurious-pole constraint.  We call the functions that the hexagon functions approach in this limit \emph{spurious pole surface functions (SP functions)}.   Just as for the hexagon functions, the space of SP functions can be built up iteratively in the weight.  Because the construction is simpler than for the full hexagon function space, but contains the same essential ingredients, it may be useful for the reader to see it in some detail.\footnote{One can always use multiple polylogarithms, or the 2dHPLs of Gehrmann and Remiddi~\cite{GehrmannRemiddi} to describe this function space.  The main virtue of the construction described here, as with the hexagon function approach, is imposing the branch-cut condition at the beginning, which reduces the size of the space dramatically at high weights.}

The SP functions must have only physical branch cuts, and their symbol entries can only be drawn from the set of letters that appear in the $w \rightarrow 1$ limit of the hexagon function letters (\ref{uLetters}). These conditions translate to functions with symbols constructed out of the letters 
\be
\mathcal{S}_{w\rightarrow 1}\ =\ \{u,\, v,\, 1-u,\, 1-v,\, u-v \}\,,
\ee
with only $u$ and $v$ appearing in the first entry. The $\{n-1,1\}$ coproduct coponent of a generic SP function $f(u,v)$ of weight $n$ thus takes the form
\be
\Delta_{n-1,1}(f) \equiv f^{u} \otimes \ln u + f^{v} \otimes \ln v + f^{1-u} \otimes \ln (1-u) + f^{1-v} \otimes \ln (1-v) + f^{u-v} \otimes \ln (u-v) ,
\ee
where its derivatives are given by 
\bea
\frac{\partial f}{\partial u} \bigg |_{v} &=& \frac{f^u}{u} - \frac{f^{1-u}}{1-u} + \frac{f^{u-v}}{u - v} \,, \nonumber\\
\frac{\partial f}{\partial v} \bigg |_{u} &=& \frac{f^v}{v} - \frac{f^{1-v}}{1-v} - \frac{f^{u-v}}{u - v} \,.
\label{uvderiv}
\eea

We can take the $u$ and $v$ partial derivatives of a full hexagon function $F(u,v,w)$ using \eqn{dFu}, let $w \rightarrow 1$ in the rational prefactors, and compare with \eqn{uvderiv}.  This comparison relates the $\{n-1,1\}$ coproduct components for $F$ to the corresponding ones for the function $f(u,v)$ that it approaches on the $w=1$ surface:
\bea
f^{u} &=& F^u \pm F^{y_u}  \,, \nonumber\\
f^{v} &=& F^v \mp F^{y_v}  \,, \nonumber\\
f^{1-u} &=& F^{1-u} \mp F^{y_v} \pm F^{y_w} \,, \nonumber\\
f^{1-v} &=& F^{1-v} \pm F^{y_u} \mp F^{y_w} \,, \nonumber\\
f^{u-v} &=&  \mp 2 F^{y_u} \pm 2 F^{y_v} \,.
\label{matchw1coprod}
\eea
The overall sign ambiguity associated with the $F^{y_i}$ components simply reflects an ambiguity as to whether the limit~(\ref{wto1yvars}) holds, or the same limit with the $y_i$'s inverted, so it holds globally for all functions.  We note that ``coproduct matching relations'' like \eqn{matchw1coprod} provide a very useful way to collapse hexagon functions into functions on generic limiting surfaces, beyond the specific case of SP functions treated here.

We'll construct the irreducible part of the SP function space through weight three here, in order to illustrate the same methods used to construct hexagon functions.

At weight one, the only functions satisfying the branch-cut constraints are $\ln u$ and $\ln v$. Functions of higher weight $n$ can be constructed at the symbol level by requiring that their symbol satisfy an integrability condition.  This condition connects pairs of adjacent entries, and there are $n-1$ such conditions, one for each pair.  Imposing all these conditions ensures that the symbol can be integrated up to a single-valued function, or equivalently that partial derivatives acting on it commute.  However, integrability can also be imposed iteratively.  Suppose we have classified all functions with weight $n-1$.  Then we can construct an ansatz for the space of functions with weight $n$ by requiring that their derivatives are given by \eqn{uvderiv} (for the case of SP functions) where each of the coproduct entries $f^u, f^v, f^{1-u}, f^{1-v},f^{u-v}$ is a generic linear combination of weight $n-1$ functions.  Now we just need to impose integrability on the last two entries of the corresponding symbol.  At function level, this is a linear constraint on the $\{n-2,1,1\}$ coproduct entries $f^{x,y}$, which is a set of linear equations for the coefficients of $f^u, f^v, f^{1-u}, f^{1-v},f^{u-v}$, when they are expanded in terms of the weight $n-1$ functions.

On the spurious pole surface, the requirement that partial derivatives commute,
\be
\frac{\partial^2f}{\partial u \partial v}
\ =\ \frac{\partial^2f}{\partial v \partial u} \,,
\ee
gives rise to six relations between the $\{n-2,1,1\}$ coproduct entries of a weight $n$ function $f$:
\bea
f^{[u, 1 - v]} &=& 0, \nonumber \\
f^{[v, 1 - u]} &=& 0, \nonumber \\
f^{[u, u - v]} &=& f^{[v, u]}, \nonumber \\
f^{[v, u - v]} &=& f^{[u, v]}, \nonumber \\
f^{[1 - u, u - v]} &=& f^{[1 - v, 1 - u]},  \nonumber \\
f^{[1 - v, u - v]} &=& f^{[1 - u, 1 - v]},
\label{SP_integrability}
\eea
where the square brackets indicate that an antisymmetric combination of coproduct entries is being taken, $f^{[x,y]} \equiv f^{x,y} - f^{y,x}$.  These relations are the analogs of the relations~(\ref{integrabilityc}) for hexagon functions.

However, the relations~(\ref{SP_integrability}) don't completely exhaust the conditions we must impose on an SP function.  Note that transcendental constants of weight $n-1$ are in the kernel of the $\{n-2,1,1\}$ coproduct, so their coefficients remain undetermined by the above equations.  Some of these coefficients will lead to unwanted branch cuts for $f$, even if all of the $\{n-1,1\}$ coproducts $f^x$ have only the proper branch cuts.  We must also check the first derivatives of our candidate functions at particular locations, in order to make sure that they remain finite away from the allowed physical singularities ($u\to0$, $v\to0$). From \eqn{uvderiv} we see that we must inspect the lines $u=1$, $v=1$ and $u=v$, where the symbol letters (other than $u$ and $v$) vanish.  We must impose
\be
f^{1-u}|_{u=1}\ =\ f^{1-v}|_{v=1}\ =\ f^{u-v}|_{u=v}\ =\ 0,
\ee
which are the analogs for SP functions of \eqns{Fomuvanish}{Fyuvanish} for the hexagon functions.

After we have found the space of functions with good branch cuts, we remove the ones that are reducible, i.e.~products of lower weight functions, as well as the one-dimensional HPLs in $u$ and $v$.  The remaining irreducible functions can be classified by the discrete symmetry.  For hexagon functions this symmetry group includes parity and the $S_3$ symmetry permuting $(u,v,w)$.  For the SP functions, there is no parity; \eqn{matchw1coprod} shows that parity even and odd hexagon functions such as $F^u$ and $F^{y_u}$ combine to give SP functions. Also, the $S_3$ symmetry is broken to $S_2$, generated by the exchange $u\lr v$.

When we apply the integrability constraint, \eqn{SP_integrability}, at weight two we find, interestingly, that it already allows for the appearance of an irreducible function. (In the hexagon function case, the first irreducible function is $\tPhi$, at weight three.)  We choose to define this function's $\{1,1\}$ coproduct to be
\bea
\Delta_{1,1}\Big( \text{SP}^{(2)}_1(u,v) \Big) &=& - \, \ln u \otimes \ln (u-v) + \frac{1}{2} \ln u \otimes \ln v \nonumber\\ 
&\ & + \ln v \otimes \ln(u-v) - \frac{1}{2} \ln v \otimes \ln u \,,
\label{SP2_1_11}
\eea
so that it is antisymmetric under the exchange of $u$ and $v$.\footnote{$\ln(u-v)$ should be considered inert under this transformation.} 

No other integrable symbols at weight two involve the letter $u-v$.  We can see this easily from \eqn{SP_integrability}: The right-hand sides of the last two relations vanish for weight two because a first entry is never $1-u$ or $1-v$. Thus $f^{1-u,u-v} = f^{1-v,u-v} = 0$.   The third and fourth relations show that $f^{u,u-v} = - f^{v,u-v}$, which determines all the $(u-v)$-dependent terms up to an overall constant.  The rest of the space is spanned by products of HPLs in $u$ and $v$.

The derivative of $\text{SP}^{(2)}_1$ follows from \eqn{SP2_1_11}:
\be
\frac{\partial}{\partial u}\text{SP}^{(2)}_1(u,v) = -\frac{\ln v}{2u} + \frac{\ln v - \ln u}{u - v} \,.
\label{SP2_u_deriv}
\ee
It is indeed singular only in the $u \rightarrow 0$ limit.  At this weight, there would be no possibility of adding a transcendental constant to the (weight one) functions in the $\{n-1,1\}$ coproducts to fix such a singularity, had it been there.

We set the additive constant of all SP functions by requiring that they vanish in the limit $(u,v) \rightarrow (1,1)$. 

At weight three, there are four independent solutions to the integrability condition, besides the reducible space of HPLs and $\text{SP}^{(2)}_1$ times $\ln u$ or $\ln v$. These four irreducible solutions can be organized into two orbits of the $S_2$ group that permutes $u$ and $v$,
\be
\Big\{ \text{SP}^{(3)}_1(u,v), \text{SP}^{(3)}_1(v,u),
       \text{SP}^{(3)}_2(u,v), \text{SP}^{(3)}_2(v,u) \Big\} \,.
\label{fullSP3}
\ee
Each orbit is a two-cycle represented by one of the following functions,
defined by its $\{2,1\}$ coproduct:
\be
\bsp
\Delta_{2,1} \Big(\text{SP}^{(3)}_1(u,v) \Big) = -&H_2^u \otimes \ln(1 - v) + H_2^u \otimes \ln(u - v) - H_2^v \otimes \ln(u - v) \\
&+ \frac 1 2 \ln u \ln v  \otimes \ln(1 - v) + \text{SP}^{(2)}_1(u,v) \otimes \ln(1 - v) \,,
\esp
\label{SP3_1_21}
\ee
\be
\bsp
\Delta_{2,1} \Big(\text{SP}^{(3)}_2(u,v) \Big) =-& \frac 1 2 \ln u  \ln v  \otimes \ln v  + \text{SP}^{(2)}_1(u,v) \otimes \ln v  \\
&- 2 \ \text{SP}^{(2)}_1(u,v) \otimes \ln(u-v) \,.
\esp
\label{SP3_2_21}
\ee 
Note that, since each of these two-cycles represents two linearly independent SP functions, the dimension of the weight three irreducible space (four) is larger than the number of functions we have indexed (two).  Moreover, these definitions are relatively simple, compared to the number of terms required to specify each function's symbol. This feature becomes increasingly true as we move to higher weight.

Next we inspect the behavior of these functions at $u=1$, $v=1$ and $u=v$.
For $\text{SP}^{(3)}_1(u,v)$, \eqn{SP3_1_21} has no $\ln(1-u)$, so there can be no singularity as $u\to1$.  The singularity as $u\to v$ is cancelled because the first entry multiplying $\ln(u-v)$ is $H_2^u - H_2^v$, which vanishes in this limit.  The only subtlety is for $v\to1$, where we require, from setting $v=1$ in the first entry multiplying $\ln(1-v)$ in \eqn{SP3_1_21},
\be
\text{SP}^{(2)}_1(u,1) = H_2^u = {\rm Li}_2(1-u).
\ee
But this equation follows by evaluating the $u$ derivative \eqn{SP2_u_deriv} for $v=1$, and the fact that they match at $u=1$:
$\text{SP}^{(2)}_1(1,1) = 0 = {\rm Li}_2(0)$.
For the other weight three irreducible function, the only singularity that has to be checked is the limit $u\to v$, where the antisymmetry of $\text{SP}^{(2)}_1(u,v)$ ensures it.  So again at weight three, we do not need to add any transcendental constants (in this case only $\zeta_2$ would be expected) to the weight two functions appearing in the $\{2,1\}$ coproducts to fix the branch-cut behavior.  It turns out that such weight $n-1$ constants are never needed in the $\{n-1,1\}$ coproducts of SP functions.  (In contrast, they do appear in the coproducts of many hexagon functions, in order to enforce smoothness as $u_i\to1$.)

\begin{table}[!t]
\centering
\begin{tabular}[t]{c || c c c}
\hline 
\hline
Weight & two-cycles & symmetric & antisymmetric \\ \hline
2 & - & - & $\text{SP}^{(2)}_1$ \\
3 & $\text{SP}^{(3)}_1,\text{SP}^{(3)}_2$ & - & - \\
4 & $\text{SP}^{(4)}_1 \dots \text{SP}^{(4)}_5$ & - & $\text{SP}^{(4)}_6, \text{SP}^{(4)}_7 $ \\
5 & $\text{SP}^{(5)}_1 \dots \text{SP}^{(5)}_{16}$ & $\text{SP}^{(5)}_{17}, \text{SP}^{(5)}_{18}$ & $\text{SP}^{(5)}_{19}, \text{SP}^{(5)}_{20}$ \\
6 & $\text{SP}^{(6)}_1 \dots \text{SP}^{(6)}_{44}$ & $\text{SP}^{(6)}_{45}, \text{SP}^{(6)}_{46}, \text{SP}^{(6)}_{47}$ & $\text{SP}^{(6)}_{48} \dots \text{SP}^{(6)}_{54}$ \\
7 & $\text{SP}^{(7)}_1 \dots \text{SP}^{(7)}_{126}$ & $\text{SP}^{(7)}_{127} \dots \text{SP}^{(7)}_{138}$ & $\text{SP}^{(7)}_{139} \dots \text{SP}^{(7)}_{150}$ \\
\hline
\hline
\end{tabular}
\caption{The symmetry orbits of the SP basis functions through weight seven. The functional dependence on $u$ and $v$ has been suppressed. Upon exchange of $u$ and $v$, each two-cycle is sent to a linearly independent function within the SP function space.  Symmetric and antisymmetric functions are mapped back to themselves, with an overall sign change in the antisymmetric case.}
\label{tab:sp_permutation_symmetries}
\end{table}

A complete basis of SP functions through weight seven was constructed using this method, and can be found in an ancillary file. The symmetry properties of these basis functions under the permutation group $S_2$ are laid out in Table \ref{tab:sp_permutation_symmetries}.  We divide them into two-cycles, symmetric and antisymmetric functions.  Clearly one could form symmetric and antisymmetric combinations of each member of a two-cycle, but it is convenient to leave it as a two-cycle, in analogy to how we treat $S_3$ for the hexagon functions.  We introduced some explicitly symmetric functions into our basis starting at weight five. We provide another ancillary file which uses this SP basis to describe the ratio function and remainder function on the spurious pole surface through three loops.

\newpage

\end{document}